\documentclass[twoside,twocolumn,9pt]{article}
\usepackage{extsizes}
\usepackage[super,sort&compress,comma]{natbib} 
\usepackage[version=3]{mhchem}
\usepackage[left=1.5cm, right=1.5cm, top=1.785cm, bottom=2.0cm]{geometry}
\usepackage{balance}
\usepackage{mathptmx}
\usepackage{sectsty}
\usepackage{graphicx} 
\usepackage{lastpage}
\usepackage[format=plain,justification=justified,singlelinecheck=false,font={stretch=1.125,small,sf},labelfont=bf,labelsep=space]{caption}
\usepackage{float}
\usepackage{fancyhdr}
\usepackage{fnpos}
\usepackage[english]{babel}
\addto{\captionsenglish}{}
\usepackage{array}
\usepackage{droidsans}
\usepackage{charter}
\usepackage[T1]{fontenc}
\usepackage[usenames,dvipsnames]{xcolor}
\usepackage{setspace}
\usepackage[compact]{titlesec}
\usepackage{hyperref}

\usepackage{siunitx}
\DeclareSIUnit\angstrom{\text{Å}}
\usepackage{tabularray}
\UseTblrLibrary{booktabs}
\UseTblrLibrary{siunitx} 
\UseTblrLibrary{diagbox}
\usepackage{caption}
\usepackage{subcaption} %
\usepackage{enumitem} %
\usepackage{amssymb}

\usepackage[final]{changes}

\definecolor{cream}{RGB}{222,217,201}

\begin{document}

\pagestyle{fancy}
\thispagestyle{plain}
\fancypagestyle{plain}{
\renewcommand{\headrulewidth}{0pt}
}

\makeFNbottom
\makeatletter
\renewcommand\LARGE{\@setfontsize\LARGE{15pt}{17}}
\renewcommand\Large{\@setfontsize\Large{12pt}{14}}
\renewcommand\large{\@setfontsize\large{10pt}{12}}
\renewcommand\footnotesize{\@setfontsize\footnotesize{7pt}{10}}
\makeatother

\renewcommand{\thefootnote}{\fnsymbol{footnote}}
\renewcommand\footnoterule{\vspace*{1pt}%
\color{cream}\hrule width 3.5in height 0.4pt \color{black}\vspace*{5pt}} 
\setcounter{secnumdepth}{5}

\makeatletter 
\renewcommand\@biblabel[1]{#1}            
\renewcommand\@makefntext[1]%
{\noindent\makebox[0pt][r]{\@thefnmark\,}#1}
\makeatother 
\renewcommand{\figurename}{\small{Fig.}~}
\sectionfont{\sffamily\Large}
\subsectionfont{\normalsize}
\subsubsectionfont{\bf}
\setstretch{1.125} %
\setlength{\skip\footins}{0.8cm}
\setlength{\footnotesep}{0.25cm}
\setlength{\jot}{10pt}
\titlespacing*{\section}{0pt}{4pt}{4pt}
\titlespacing*{\subsection}{0pt}{15pt}{1pt}

\fancyfoot{}
\fancyfoot[LO,RE]{\vspace{-7.1pt}\includegraphics[height=9pt]{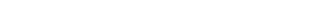}}
\fancyfoot[CO]{\vspace{-7.1pt}\hspace{13.2cm}\includegraphics{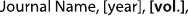}}
\fancyfoot[CE]{\vspace{-7.2pt}\hspace{-14.2cm}\includegraphics{head_foot/RF}}
\fancyfoot[RO]{\footnotesize{\sffamily{1--\pageref{LastPage} ~\textbar  \hspace{2pt}\thepage}}}
\fancyfoot[LE]{\footnotesize{\sffamily{\thepage~\textbar\hspace{3.45cm} 1--\pageref{LastPage}}}}
\fancyhead{}
\renewcommand{\headrulewidth}{0pt} 
\renewcommand{\footrulewidth}{0pt}
\setlength{\arrayrulewidth}{1pt}
\setlength{\columnsep}{6.5mm}
\setlength\bibsep{1pt}

\makeatletter 
\newlength{\figrulesep} 
\setlength{\figrulesep}{0.5\textfloatsep} 

\newcommand{\topfigrule}{\vspace*{-1pt}%
\noindent{\color{cream}\rule[-\figrulesep]{\columnwidth}{1.5pt}} }

\newcommand{\botfigrule}{\vspace*{-2pt}%
\noindent{\color{cream}\rule[\figrulesep]{\columnwidth}{1.5pt}} }

\newcommand{\dblfigrule}{\vspace*{-1pt}%
\noindent{\color{cream}\rule[-\figrulesep]{\textwidth}{1.5pt}} }

\makeatother

\twocolumn[
  \begin{@twocolumnfalse}
{\includegraphics[height=30pt]{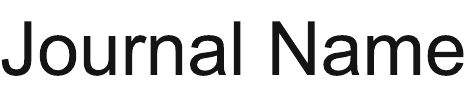}\hfill\raisebox{0pt}[0pt][0pt]{\includegraphics[height=55pt]{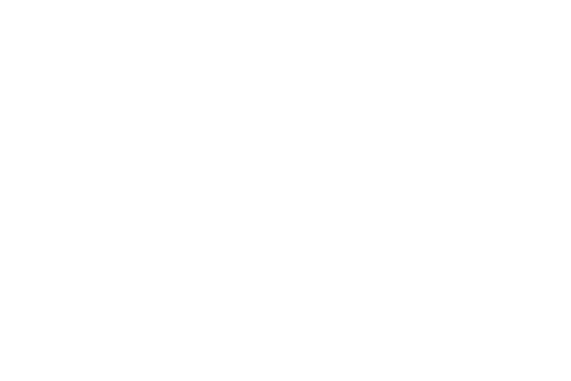}}\\[1ex]
\includegraphics[width=18.5cm]{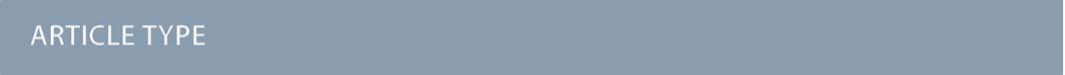}}\par
\vspace{1em}
\sffamily
\begin{tabular}{m{4.5cm} p{13.5cm} }

\includegraphics{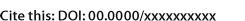} & \noindent\LARGE{\textbf{PoseBusters: 
AI-based docking methods fail to generate physically valid poses or generalise to novel sequences$^\dag$}} \\
\vspace{0.3cm} & \vspace{0.3cm} \\

 & \noindent\large{Martin Buttenschoen, Garrett M. Morris, and Charlotte M. Deane\textit{$^{\ddag}$}} \\

\includegraphics{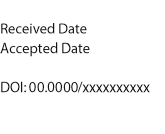} & \noindent\normalsize{
The last few years have seen the development of numerous deep learning-based protein-ligand docking methods. They offer huge promise in terms of speed and accuracy. However, despite claims of state-of-the-art performance in terms of crystallographic root-mean-square deviation (RMSD), upon closer inspection, it has become apparent that they often produce physically implausible molecular structures. It is therefore not sufficient to evaluate these methods solely by RMSD to a native binding mode. It is vital, particularly for deep learning-based methods, that they are also evaluated on steric and energetic criteria.
We present PoseBusters, a Python package that performs a series of standard quality checks using the well-established cheminformatics toolkit RDKit.
The PoseBusters test suite validates chemical and geometric consistency of a ligand including its stereochemistry, and the physical plausibility of intra- and intermolecular measurements such as the planarity of aromatic rings, standard bond lengths, and protein-ligand clashes. Only methods that both pass these checks and predict native-like binding modes should be classed as having ``state-of-the-art'' performance. We use PoseBusters to compare five deep learning-based docking methods (DeepDock, DiffDock, EquiBind, TankBind, and Uni-Mol) and two well-established standard docking methods (AutoDock Vina and CCDC Gold) with and without an additional post-prediction energy minimisation step using a molecular mechanics force field.
We show that both in terms of physical plausibility and the ability to generalise to examples that are distinct from the training data, no deep learning-based method yet outperforms classical docking tools. In addition, we find that molecular mechanics force fields contain docking-relevant physics missing from deep-learning methods. PoseBusters allows practitioners to assess docking and molecular generation methods and may inspire new inductive biases still required to improve deep learning-based methods, which will help drive the development of more accurate and more realistic predictions.
} \\

\end{tabular}

 \end{@twocolumnfalse} \vspace{0.6cm}

  ]

\renewcommand*\rmdefault{bch}\normalfont\upshape
\rmfamily
\section*{}
\vspace{-1cm}

\footnotetext{\textit{Department of Statistics, 24-29 St Giles', Oxford OX1 3LB, United Kingdom.%
}}

\footnotetext{\dag~Electronic Supplementary Information (ESI) available. See DOI: 00.0000/00000000.}

\footnotetext{\ddag~Corresponding author: \protect\url{deane@stats.ox.ac.uk}}

\section{Introduction}

Docking, an essential step in structure-based drug discovery \cite{deruyck2016molecular}, is the task of predicting the predominant binding modes of a protein-ligand complex given an experimentally solved or computationally modelled protein structure and a ligand structure \cite{morris2008molecular}.
The predicted complexes are often used in a virtual screening workflow to help select molecules from a large library of possible candidates \cite{lionta2014structurebased}; or directly by medicinal chemists to understand the binding mode and to decide whether a small molecule is a suitable drug candidate \cite{patrick2017drug}. 

Docking methods are designed with the understanding that binding is enabled by interactions between target and ligand structures but due to 
the complexity of this property methods tend to strike a balance between fast calculation and accuracy
\cite{patrick2017computers}.
Deep learning (DL) promises to disrupt the dominant design principle of classical docking software, and DL-based docking methods promise to unlock fast and accurate virtual screening for drug discovery.
To this end, a handful of different DL-based docking methods have already been proposed \cite{mendezlucio2021geometric, corso2023diffdock, stark2022equibinda, lu2022tankbinda, zhou2023unimol}.

Classical non-DL-based docking methods include within their search and scoring functions terms that help ensure chemical consistency and physical plausibility; for example
limiting the degrees of movement in the ligand %
to only the rotatable bonds in the ligand and 
including penalties if the protein and ligand clash \cite{jones1995molecular, trott2009autodock}. %
Some current DL-based docking methods, as we will show, still lack such key ``inductive biases'' resulting in the creation of unrealistic poses
despite obtaining root-mean-squared deviation (RMSD) values from the experimental binding mode that are less than the widely-used \qty{2}{\angstrom} threshold \cite{cole2005comparing}.
To assess such docking methods, an independent test suite is necessary to check the chemical consistency and physical plausibility alongside established metrics, such as the binding mode RMSD. 
Such a test suite would help the field to identify missing inductive biases required to improve DL-based docking methods, driving the development of more accurate and realistic docking predictions.

The problem of assessing the physical plausibility of docking predictions is akin to the structure validation of ligand data in the Protein Data Bank (PDB) \cite{berman2000protein, shao2022simplified}.
Structure validation assesses the agreement of the ligands bond lengths and angles with those observed in related chemical structures and the presence of steric clashes both within the ligand and between it and its surroundings \cite{shao2022simplified}.
While these tests were designed for users to select those ligand crystal structures which are likely to be correct \cite{shao2022simplified}, docking methods are evaluated on their ability to recover crystal structures so their output should pass the same physical plausibility tests.

Physical plausibility checks
are also part of some workflows for conformation generation \cite{hawkins2010conformer, friedrich2017benchmarking}. %
\citeauthor{friedrich2017benchmarking} use 
geometry checks performed by NAOMI \cite{urbaczek2011naomi} which measures---like the PDB tests mention above---the deviation from known optimal values for bond lengths and bond angles, and also
tests for divergences from the planarity of aromatic rings \cite{friedrich2017benchmarking}.

In addition to physical checks, chemical checks are also needed 
\cite{warren2012essential}. 
Chemical checks proposed for checking PDB structures include the identification of mislabelled stereo assignment, inconsistent bonding patterns, missing functional groups, and unlikely ionisation states \cite{warren2012essential}. 
The problem of checking chemical plausibility
has also come up in \emph{de novo} molecule generation, where \citeauthor{brown2019guacamol} proposed a test suite including checks for the chemical validity of any proposed molecule \cite{brown2019guacamol}. For docking, the focus is less on stability and synthetic accessibility of a molecular structure as it is hoped that these have been tested prior to attempting docking, but more on chemical consistency and physical realism of the predicted bound conformation.

Some comparisons of docking methods have included additional metrics based on volume overlap \cite{warren2006critical} or protein-ligand interactions \cite{ciancetta2014alternative} to supplement pose accuracy-based metrics such as RMSD of atomic positions and run time measurements, 
but the majority of comparisons of docking methods are predominantly based on binding mode RMSD \cite{cole2005comparing, onodera2007evaluations, plewczynski2011can, bolcato2019can}. 

The current standard practice of comparing docking methods based on RMSD-based metrics alone also extends to the introduction papers of recent new methods. 
The five DL-based docking methods we test in this paper \cite{mendezlucio2021geometric, corso2023diffdock, stark2022equibinda, lu2022tankbinda, zhou2023unimol} all claim better performance than standard docking methods
but these claims rest entirely on RMSD. None of these methods test their outputs for physical plausibility. 

In this paper we present PoseBusters, a test suite that is designed to identify implausible conformations and ligand poses.
We used PoseBusters to evaluate the predicted ligand poses generated by the five DL-based docking methods (DeepDock\cite{mendezlucio2021geometric}, DiffDock\cite{corso2023diffdock}, EquiBind\cite{stark2022equibinda}, TankBind\cite{lu2022tankbinda}, and Uni-Mol\cite{zhou2023unimol}) and two standard non-DL-based docking methods (AutoDock Vina\cite{trott2009autodock} and Gold\cite{verdonk2003improved}). These poses were generated by re-docking the cognate ligands of the 81 protein-ligand crystal complexes in the Astex Diverse set\cite{hartshorn2007diverse} and \replaced{308}{428} ligands of the protein-ligand crystal complexes in the PoseBusters Benchmark set, a new set of complexes released from 2021 onwards, into their cognate receptor crystal structures.
On the commonly-used Astex Diverse set, the DL-based docking method DiffDock appears to perform best in terms of RMSD alone but when taking physical plausibility into account, Gold and AutoDock Vina perform best. On the PoseBusters Benchmark set, a test set that is harder because it contains only complexes that the DL methods have not been trained on, Gold and AutoDock Vina are the best methods in terms of RMSD alone and when taking physical plausibility into account or when proteins with novel sequences are considered. The DL-based methods make few valid predictions for the unseen complexes. 
Overall, we show that no DL-based method yet outperforms standard docking methods when consideration of physical plausibility is taken into account. The PoseBusters test suite will enable DL method developers to better understand the limitations of current methods, ultimately resulting in more accurate and realistic predictions.

\section{Methods}
Five DL-based and two classical docking methods were used to re-dock known ligands into their respective proteins and the predicted ligand poses were evaluated with the PoseBusters test suite. 
The following section describes the docking methods, the data sets, and the PoseBusters test suite for checking physicochemical consistency and structural plausibility of the generated poses.  

\subsection{Docking methods}
The selected five DL-based docking methods \cite{mendezlucio2021geometric, corso2023diffdock, stark2022equibinda, lu2022tankbinda, zhou2023unimol} cover a wide range of DL-based approaches for pose prediction. Table~\ref{tab:methods} lists the methods and their publications.
In order to examine the ability of standard non-DL-based methods to predict accurate chemically and physically valid poses, we also included the well-established docking methods AutoDock Vina \cite{forli2016computational} and Gold \cite{jones1997development}.

\begin{table}
\centering
\caption{Selected DL-based docking methods. The selection includes five methodologically different DL-based docking methods published over the last two years.}
\label{tab:methods}
\begin{tabular}{l l l l l l}
\toprule
Method & Authors & Date & Search space \\ %
\midrule
DeepDock \cite{mendezlucio2021geometric} & \citeauthor{mendezlucio2021geometric} & Dec 2021 & pocket \\
DiffDock \cite{corso2023diffdock} & \citeauthor{corso2023diffdock} & Feb 2023 & blind \\ %
EquiBind \cite{stark2022equibinda} & \citeauthor{stark2022equibinda} & Feb 2022 & blind \\ %
TANKBind \cite{lu2022tankbinda} & \citeauthor{lu2022tankbinda}  & Oct 2022 & blind \\ %
Uni-Mol \cite{zhou2023unimol} & \citeauthor{zhou2023unimol} & Feb 2023 & pocket \\ %
\bottomrule
\end{tabular}%
\end{table}%

The five DL-based docking methods can be summarised as follows. Full details of each can be found in their respective references. 
DeepDock \cite{mendezlucio2021geometric} learns a statistical potential based on the distance likelihood between ligand heavy atoms and points of the mesh of the surface of the binding pocket. 
DiffDock \cite{corso2023diffdock} uses equivariant graph neural networks in a diffusion process for blind docking. 
EquiBind \cite{stark2022equibinda} applies equivariant graph neural networks for blind docking.
TankBind \cite{lu2022tankbinda} is a blind docking method that uses a trigonometry-aware neural network for docking in each pocket predicted by a binding pocket prediction method. 
Uni-Mol \cite{zhou2023unimol} carries out docking with SE3-equivariant transformers. 
All five DL-based docking methods are trained on subsets of the PDBbind General Set \cite{liu2017forging} as detailed in Table~\ref{tab:methods_training_data}. DeepDock is trained on v2019 and the other four are trained on v2020. \added{It should be noted that we used the DL models as trained by the respective authors without further hyperparameter tuning.}

\begin{table}
\centering
\caption{Data sets used to train the selected five machine learning-based docking methods. All five DL-based methods were trained on subsets of the PDBBind 2020 General Set.}
\label{tab:methods_training_data}
\begin{tblr}{l X[1,l]}
\toprule
Method & Training and validation set \\ 
\midrule
DeepDock                     
& PDBBind 2019 General Set without complexes included in CASF-2016 or those that fail pre-processing---\num{16367} complexes
\\
{DiffDock,\\ EquiBind}        
& PDBbind 2020 General Set keeping complexes published before 2019 and without those with ligands found in test set---\num{17347} complexes
\\
TankBind                     
& PDBbind 2020 General Set keeping complexes published before 2019 and without those failing pre-processing---\num{18755} complexes
\\
Uni-Mol                      
& PDBBind 2020 General Set without complexes where protein sequence identity (MMSeq2) with CASF-2016 is above 40\% and ligand fingerprint similarity is above 80\%---\num{18404} complexes
\\
\bottomrule
\end{tblr}
\end{table}

The docking protocols that were used to generate predictions with each method and the software versions used are given in section~S1 of the Supplementary Information. 
Table~\ref{tab:binding_site_definitions} lists the search space definitions that we used for each method. DeepDock and Uni-Mol require the definition of a binding site while DiffDock, EquiBind, and TankBind are `blind' docking methods that search over the entire protein. 
We used the default search spaces for \replaced{DiffDock, DeepDock, EquiBind, and TankBind}{the DL-based methods} but larger than default search spaces for AutoDock Vina\replaced{,}{ and} Gold\added{, and Uni-Mol} such that they are more comparable with the blind docking \deleted{DL-based} methods. 
SI~Figure~S1 shows the search spaces for one example protein-ligand complex.
\added{We show results for Uni-Mol across a range of binding site definitions starting from their preferred definition of all residues with an atom within \qty{6}{\angstrom} of a heavy atom of the crystal ligand. Under this tight pocket definition Uni-Mol performs better than any of the blind docking methods (SI~Figure~S21).}

\begin{table}
\centering
\caption{Search spaces of the docking methods used. 
}
\begin{tblr}{width=\linewidth, colspec={lX[1,l]}}
\toprule
Method & Search space \\
\midrule
\SetCell[c=2]{l} \emph{Classical docking methods} \\
Gold         & Sphere of radius \qty{25}{\angstrom} centered on the geometric centre of the crystal ligand heavy atoms \\
Vina         & Cube with side length \qty{25}{\angstrom} centered on the geometric centre of crystal ligand heavy atoms \\
\SetCell[c=2]{l} \emph{DL-based docking methods} \\
DeepDock     & Protein surface mesh nodes within \qty{10}{\angstrom} of any crystal ligand atom \\
Uni-Mol      & Protein residues within \replaced{\qty{8}{\angstrom}}{\qty{6}{\angstrom}} of any crystal ligand heavy atom \\
\SetCell[c=2]{l} \emph{DL-based blind docking methods} \\
DiffDock	 & Entire crystal protein \\
EquiBind	 & Chains of crystal protein which are within \qty{10}{\angstrom} of any crystal ligand heavy atom \\
TankBind	 & Pockets identified by P2Rank\cite{krivak2018p2rank} \\
\bottomrule
\end{tblr}
\label{tab:binding_site_definitions}
\end{table}

\subsection{The PoseBusters test suite}
\label{sec:methods_testing_testsuite}

The PoseBusters test suite is organised into three groups of tests. 
The first checks chemical validity and contains tests for the chemical validity and consistency relative to the input.
The second group checks intramolecular properties and tests for the ligand geometry and the ligand conformation's energy computed using the universal force field (UFF)\cite{rappe1992uff}.
The third group considers intermolecular interactions and checks for protein-ligand and ligand-cofactor clashes. Descriptions of all the tests PoseBusters performs in the three sections are listed in Table~\ref{tab:methods_tests}.
Molecule poses which pass all tests in PoseBusters are `PB-valid'. 

For evaluating docking predictions, PoseBusters requires three input files: an SDF file containing the re-docked ligands, an SDF file containing the true ligand(s), and a PDB file containing the protein with any cofactors. 
The three files are loaded into RDKit molecule objects with the sanitisation option turned off. 

\begin{table*}
\centering
\caption{Description of the checks used in the PoseBusters test suite.}
\label{tab:methods_tests}
\begin{tblr}{width=\linewidth,colspec={X[1,l]X[2.5,l]}}
\toprule
Test name & Description \\
\midrule
\addlinespace   
\SetCell[c=2]{l} \bfseries Chemical validity and consistency & \\
\addlinespace
File loads
& The input molecule can be loaded into a molecule object by RDKit. 
\\
Sanitisation
& The input molecule passes RDKit's chemical sanitisation checks. 
\\
Molecular formula %
& The molecular formula of the input molecule is the same as that of the true molecule. 
\\
Bonds %
& The bonds in the input molecule are the same as in the true molecule.
\\
Tetrahedral chirality %
& The specified tetrahedral chirality in the input molecule is the same as in the true molecule.
\\
Double bond stereochemistry %
& The specified double bond stereochemistry in the input molecule is the same as in the true molecule.
\\
\addlinespace
\SetCell[c=2]{l} \bfseries Intramolecular validity & \\
\addlinespace
Bond lengths %
& The bond lengths in the input molecule are within  \num{0.75} of the lower and  \num{1.25} of the upper bounds determined by distance geometry. 
\\
Bond angles %
& The angles in the input molecule are within  \num{0.75} of the lower and  \num{1.25} of the upper bounds determined by distance geometry. 
\\
Planar aromatic rings
& All atoms in aromatic rings with 5 or 6 members are within \qty{0.25}{\angstrom} of the closest shared plane.
\\
Planar double bonds
& The two carbons of aliphatic carbon-carbon double bonds and their four neighbours are within \qty{0.25}{\angstrom} of the closest shared plane.
\\
Internal steric clash
& The interatomic distance between pairs of non-covalently bound atoms is above \num{0.8} of the lower bound determined by distance geometry. 
\\
Energy ratio %
& The calculated energy of the input molecule is no more than \num{100} times the average energy of an ensemble of \num{50} conformations generated for the input molecule. The energy is calculated using the UFF\cite{rappe1992uff} in RDKit and the conformations are generated with ETKDGv3 followed by force field relaxation using the UFF with up to \num{200} iterations.
\\
\addlinespace
\SetCell[c=2]{l} \bfseries Intermolecular validity & \\
\addlinespace 
Minimum protein-ligand distance
& The distance between protein-ligand atom pairs is larger than  \num{0.75} times the sum of the pairs van der Waals radii. 
\\
Minimum distance to organic cofactors
& The distance between ligand and organic cofactor atoms is larger than  \num{0.75} times the sum of the pairs van der Waals radii. 
\\
Minimum distance to inorganic cofactors
& The distance between ligand and inorganic cofactor atoms is larger than  \num{0.75} times the sum of the pairs covalent radii. 
\\
Volume overlap with protein
& The share of ligand volume that intersects with the protein is less than \qty{7.5}{\percent}. The  volumes are defined by the van der Waals radii around the heavy atoms scaled by \num{0.8}. 
\\
Volume overlap with organic cofactors
& The share of ligand volume that intersects with organic cofactors is less than \qty{7.5}{\percent}. The  volumes are defined by the van der Waals radii around the heavy atoms scaled by \num{0.8}. 
\\
Volume overlap with inorganic cofactors
& The share of ligand volume that intersects with inorganic cofactors is less than \qty{7.5}{\percent}. The  volumes are defined by the van der Waals radii around the heavy atoms scaled by \num{0.5}. 
\\
\bottomrule
\end{tblr}
\end{table*}

\subsubsection{Chemical validity and consistency}
The first test in PoseBusters checks whether the ligand passes \added{the} RDKit's sanitisation. \added{The} RDKit's sanitisation processes information on the valency, aromaticity, radicals, conjugation, hydridization, chirality tags, and protonation to check whether a molecule can be represented as an octet-complete Lewis dot structure \cite{landrum2023rdkit}.
Passing \added{the} RDKit's sanitisation is a commonly-used test for chemical validity in cheminformatics, for example in \emph{de novo} molecular generation \cite{brown2019guacamol}.

The next test in PoseBusters checks for docking-relevant chemical consistency between the predicted and the true ligands by generating `standard InChI' strings \cite{heller2015inchi} for the input and output ligands after removing isotopic information and neutralising charges by adding or removing hydrogens where possible. 
InChI is the \emph{de facto} standard for molecular comparison \cite{goodman2021inchi}, 
and the `standard InChI' strings generated include the layers for the molecular formula (/), molecular bonds (/c), hydrogens (/h), net charge (/q),  protons (/p), tetrahedral chirality (/t), and double bond stereochemistry (/b). 
Standardisation of the ligand's protonation and charge state is needed because the stereochemistry layer is dependent on the hydrogen (/h), net charge (/q) and proton (/p) layers. These can unexpectedly change during docking even though most docking software considers the charge distribution and protonation state of a ligand as fixed \cite{trott2009autodock, torres2019key}. 
The normalisation protocol also removes the stereochemistry information of double bonds in primary ketimines which only depends on the hydrogen atom's ambiguous location.

\subsubsection{Intramolecular validity}
The first set of physical plausibility tests in the PoseBusters test suite validates bond lengths, bond angles, and internal distances between non-covalently bound pairs of atoms in the docked ligand against the corresponding limits in the distance bounds matrix obtained from \added{the} RDKit's Distance Geometry module.
To pass the tests, all molecular measurements must lie within the user-specified tolerances. 
The tolerance used throughout this manuscript is \qty{25}{\percent} for bond lengths and bond angles and \qty{30}{\percent} for non-covalently bound pairs of atoms e.g.: if a bond is less than \qty{75}{\percent} of the Distance Geometry bond length lower bound, it is treated as anomalous. This was selected as all but one of the crystal ligands in the Astex Diverse set and all of those in the PoseBuster\added{}{s} Benchmark set pass at this threshold.

The PoseBusters test for flatness checks that groups of atoms lie in a plane by calculating the closest plane to the atoms and checking that all atoms are within a user-defined distance from this plane. 
This test is performed for 5- and 6-membered aromatic rings and non-ring non-aromatic carbon-carbon double bonds. The chosen threshold of \qty{0.25}{\angstrom} admits all Astex Diverse and PoseBusters Benchmark set crystal structures by a wide margin and as with all other thresholds can be adjusted by the user. 

The final test for intramolecular physicochemical plausibility carried out by PoseBusters is an energy calculation to detect unlikely conformations.
Our metric for this is the ratio of the energy of the docked ligand conformation to the mean of the energies of a set of 50 generated unconstrained conformations as in \citeauthor{wills2022use} \cite{wills2022use}. The conformations are generated using \added{the} RDKit's ETKDGv3 conformation generator \cite{riniker2015better} followed by a force field relaxation using the UFF \cite{rappe1992uff} and up to \num{200} iterations. The test suite rejects conformations for which this ratio is larger than a user-specified threshold. 
\citeauthor{wills2022use} set a ratio of 7 based on the value where \qty{95}{\percent} of the crystal ligands in the PDBbind data set are considered plausible 
\cite{wills2022use}. We selected a less strict ratio of \num{100} where only  one structure each from the Astex Diverse and PoseBuster\added{s} Benchmark set is rejected.

\subsubsection{Intermolecular validity}

Intermolecular interactions are evaluated by two sets of tests in the PoseBusters test suite. The first set checks the minimum distance between molecules and the second checks the share of overlapping volume. Both sets of tests report on intermolecular interactions of the ligand with four types of molecules: the protein, organic cofactors, and inorganic cofactors.

For the distance-based intermolecular tests PoseBusters calculates the ratio of the pairwise distance between pairs of heavy atoms of two molecules and the sum of the two atoms' van der Waals radii. If this ratio is smaller than a user-defined threshold then the test fails. The default threshold is \num{0.75} for all pairings. For inorganic cofactor-ligand pairings the covalent radii are used. All crystal structures in the Astex Diverse set and all but one in the PoseBuster\added{s} Benchmark set pass at this threshold.

For the second set of intermolecular checks, PoseBusters calculates the share of the van der Waals volume of the heavy atoms of the ligand that overlaps with the van der Waals volume of the heavy atoms of the protein using \added{the} RDKit's \texttt{ShapeTverskyIndex} function. The tests have a configurable scaling factor for the volume-defining van der Waals radii and a threshold that defines how much overlap constitutes a clash. A threshold is necessary because many crystal structures already contain clashes. For example, \citeauthor{verdonk2003improved} found that \num{81} out of \num{305} selected high-quality protein-ligand complexes from the PDB 
contain steric clashes \cite{verdonk2003improved}. 
The overlap threshold is \qty{7.5}{\percent} for all molecule pairings and the scaling factor is \num{0.8} for protein-ligand and organic cofactor-ligand pairings and \num{0.5} for inorganic cofactor-ligand pairings. %

\subsection{Quality of fit}
PoseBusters calculates the minimum heavy-atom symmetry-aware root-mean-square deviation (RMSD) between the predicted ligand binding mode and the closest crystallographic ligand \added{using the RDKit’s \texttt{GetBestRMS} function}. Coverage, a metric often used for testing docking methods, is the share of predictions that are within a user adjustable threshold which by default is \qty{2}{\angstrom} RMSD. This value is arbitrary but commonly-used and recommended for regular-size ligands \cite{cole2005comparing}.

\subsection{Sequence identity}
In this paper, sequence identity between two amino acid chains is the number of exact residue matches after sequence alignment divided by the number of residues of the query sequence. 
The sequence alignment used is the Smith–Waterman algorithm \cite{smith1981identification} implemented in Biopython \cite{cock2009biopython} using an open gap score of \num{-11} and an extension gap score of \num{-1} and the \texttt{BLOSUM62} substitution matrix. 
Unknown amino acid residues are counted as mismatches. 

\subsection{Molecular mechanics energy minimisation}
Post-docking energy minimisation of the ligand structure in the binding pocket was performed using the AMBER ff14sb force field \cite{maier2015ff14sb} and the Sage small molecule force field 
\cite{boothroyd2023development} in OpenMM \cite{eastman2017openmm}.
The protein files were prepared using PDBfixer \cite{eastman2017openmm} and all protein atom positions were fixed in space only allowing updates to the ligand atoms positions. %
Minimisation was performed until energy convergence within \qty{0.01}{\kilo\joule\per\mole}.

\subsection{Data}
\label{sec:methods_data}

\subsubsection{Astex Diverse set}
The Astex Diverse set \cite{hartshorn2007diverse} published in 2007 is a set of hand-picked, relevant, diverse, and high-quality protein-ligand complexes from the PDB \cite{berman2000protein}. 
The complexes were downloaded from the PDB as MMTF files \cite{bradley2017mmtf} and PyMOL \cite{schrodingerllc2015pymol} was used to remove solvents and all occurrences of the ligand of interest from the complexes before saving the proteins with the cofactors in PDB files and the ligands in SDF files.

\subsubsection{PoseBusters Benchmark set}
The PoseBusters Benchmark set is a new set of carefully-selected publicly-available crystal complexes from the PDB.
It is a diverse set of recent high-quality protein-ligand complexes which contain drug-like molecules. It only contains complexes released since 2021 and therefore does not contain any complexes present in the PDBbind General Set v2020 used to train many of the methods. Table~S2 lists the steps used to select the \replaced{\num{308}}{\num{428}} unique proteins and \replaced{\num{308}}{\num{428}} unique ligands in the PoseBuster\added{s} Benchmark set. 
The complexes were downloaded from the PDB as MMTF files and PyMOL was used to remove solvents and all occurrences of the ligand of interest before saving the proteins with the cofactors in PDB files and the ligands in SDF files. 

\section{Results}

The following section presents the analysis of the PoseBusters test suite on the re-docked ligands of five DL-based docking methods and two standard non-DL-based docking methods on the \num{85} ligands of the Astex Diverse set and the \replaced{\num{308}}{\num{428}} ligands of the PoseBusters Benchmark set into the receptors crystal structures.

\subsection{Results on the Astex Diverse set}

\begin{figure}
\centering
\includegraphics[width=\columnwidth]{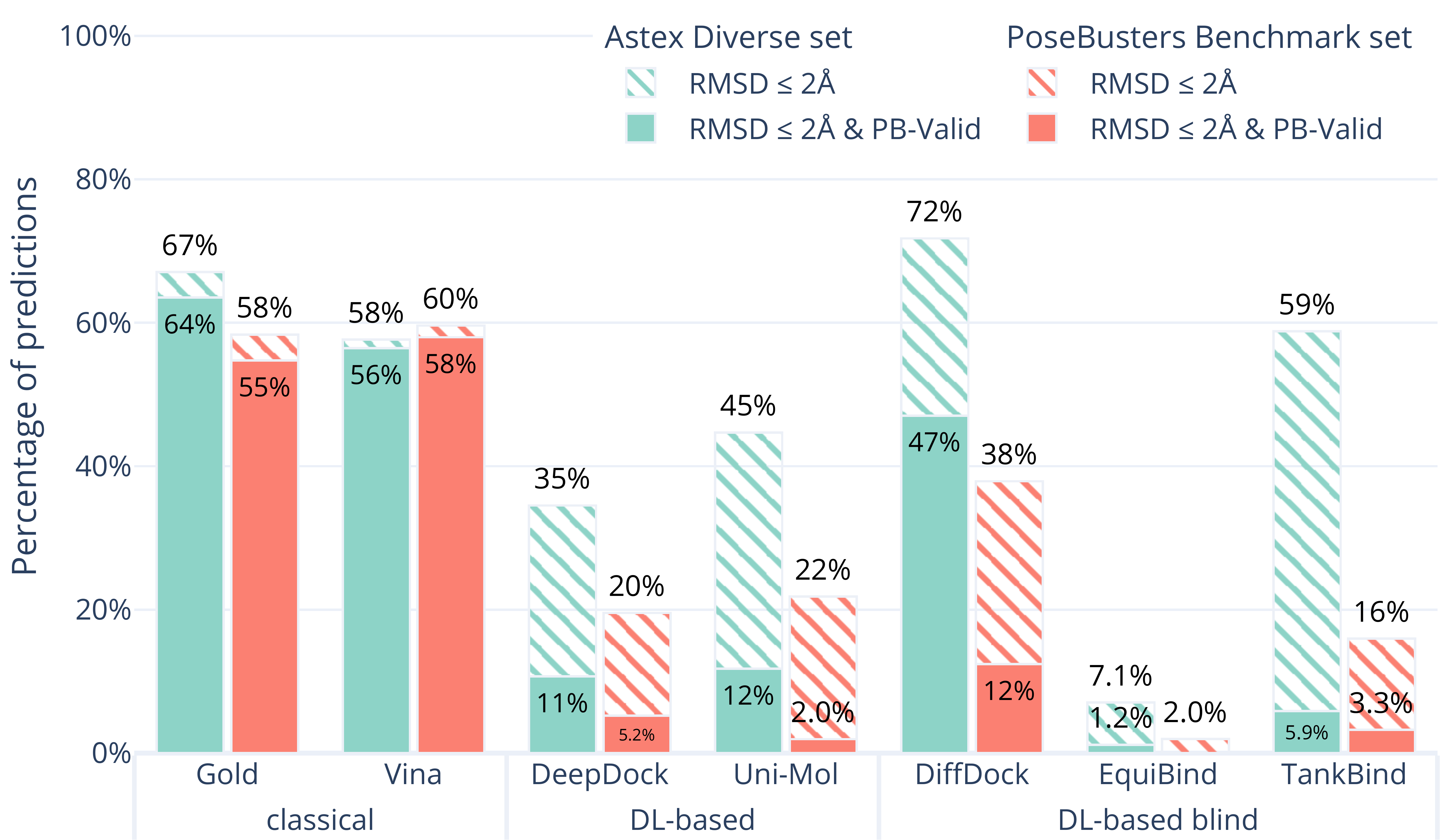}
\caption{
Comparative performance of the docking methods.
The Astex Diverse set (\num{85} cases) was chosen as an easy test set containing many complexes the five DL-based methods were trained on while the PoseBusters Benchmark set (\replaced{\num{308}}{\num{428}} cases) was chosen to be a difficult test set containing complexes none of the methods was trained on. 
The striped bars show the share of predictions of each method that have an RMSD within \qty{2}{\angstrom} and the solid bars show the subset that in addition have valid geometries and energies, i.e., pass all PoseBuster tests and are therefore `PB-Valid'.
DiffDock appears to outperform the classical methods on the Astex Diverse set when only binding mode RMSD is considered (striped teal bars).
However, when physical plausibility is also considered (solid teal bars) \emph{or} when presented with the PoseBusters Benchmark set (coral bars), AutoDock Vina and Gold outperform all DL-based methods. 
}
\label{fig:bars_datasets}
\end{figure}

Figure~\ref{fig:bars_datasets} shows the overall results of the seven (AutoDock Vina \cite{trott2009autodock}, Gold \cite{verdonk2003improved}, DeepDock \cite{mendezlucio2021geometric}, DiffDock \cite{corso2023diffdock}, EquiBind \cite{stark2022equibinda}, TankBind \cite{lu2022tankbinda}, Uni-Mol \cite{zhou2023unimol}) docking methods on the Astex Diverse set in ocean green. The striped bars show the performance only in terms of RMSD coverage ($\text{RMSD} \leqslant \qty{2}{\angstrom}$) and the solid bars show the performance after also considering physical plausibility, i.e., only predictions which in addition pass all tests in PoseBusters and are therefore PB-valid. 

\begin{figure}
\centering
\includegraphics[width=1.0\linewidth]{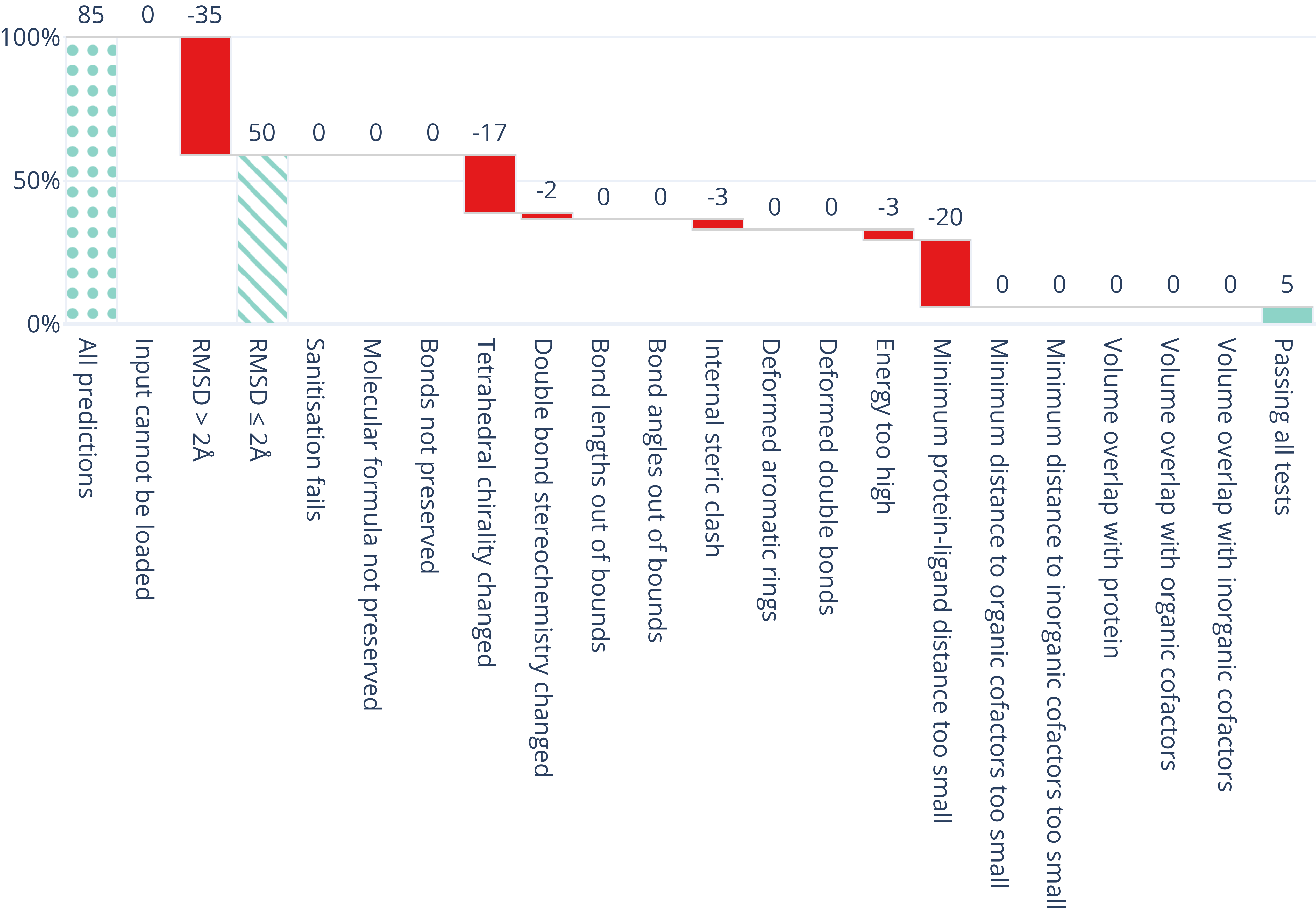}
\caption{Waterfall plot showing the PoseBusters tests as filters for the TankBind predictions on the Astex Diverse data set. The tests in the PoseBuster test suits are described in Table~\ref{tab:methods_tests}. The leftmost (dotted) bar shows the number of complexes in the test set. The red bars show the number of predictions that fail with each additional test going from left to right. The rightmost (solid) bar indicates the number of predictions that pass all tests, i.e. those that are `PB-Valid'. For the \num{85} test cases in the Astex Diverse set \num{50} (\qty{59}{\percent}) predictions have RMSD within \qty{2}{\angstrom} RMSD and \num{5} (\qty{5.9}{\percent}) pass all tests. Figures~S5 and S6 in the Supplementary Information show waterfall plots for all methods and both data sets.}
\label{fig:waterfall_text}
\end{figure}
 
The Astex Diverse set is a well-established and commonly-used benchmark for evaluating docking methods.
Good performance on this set is expected because the five DL-based methods evaluated here have been trained on most of these complexes. \num{47} of the \num{81} complexes in the Astex Diverse set are in the PDBbind 2020 General Set
and \num{67} out of the \num{81} of the Astex Diverse set proteins have more than \qty{95}{\percent} sequence identity with proteins found in PDBbind 2020 General Set. AutoDock Vina may also perform well on this data set because the linear regression model behind the scoring function was trained on an earlier version of PDBbind  \cite{trott2009autodock} which already included most of the Astex Diverse set.

The RMSD criterion alone (striped green bars in Figure~\ref{fig:bars_datasets}) gives the impression that DiffDock (\qty{72}{\percent}) performs better than TankBind (\qty{59}{\percent}), Gold (\qty{67}{\percent}), AutoDock Vina (\qty{58}{\percent}) and Uni-Mol (\qty{45}{\percent}). However, when we look closer, accepting only ligand binding modes that are physically sensible, i.e., those predictions that pass all PoseBusters tests and are therefore PB-valid (solid green bars in Figure~\ref{fig:bars_datasets}), many of the apparently impressive DL predictions are removed. The best three methods when considering RMSD and physical plausibility are Gold (\qty{64}{\percent}), AutoDock Vina (\qty{56}{\percent}), and DiffDock (\qty{47}{\percent}) followed by Uni-Mol (\qty{12}{\percent}), DeepDock (\qty{11}{\percent}) and TankBind (\qty{5.9}{\percent}). DiffDock is therefore the only DL-based method that has comparable performance to the standard methods on the Astex Diverse set when considering physical plausibility of the predicted poses.

\begin{figure}

\subfloat[Double bond stereochemistry not preserved. DiffDock prediction for ligand VDX of protein-ligand complex 7QPP. RMSD \qty{1.9}{\angstrom}.]{
\begin{minipage}[c]{0.98\linewidth}
\hfill
\includegraphics[width=0.49\linewidth,trim={0 18 0 22},clip]{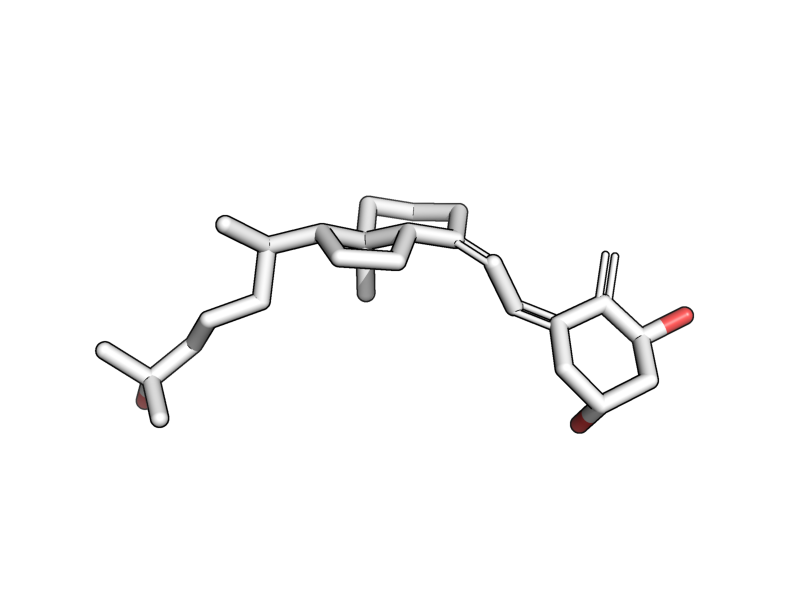}
\hfill
\includegraphics[width=0.49\linewidth,trim={0 18 0 22},clip]{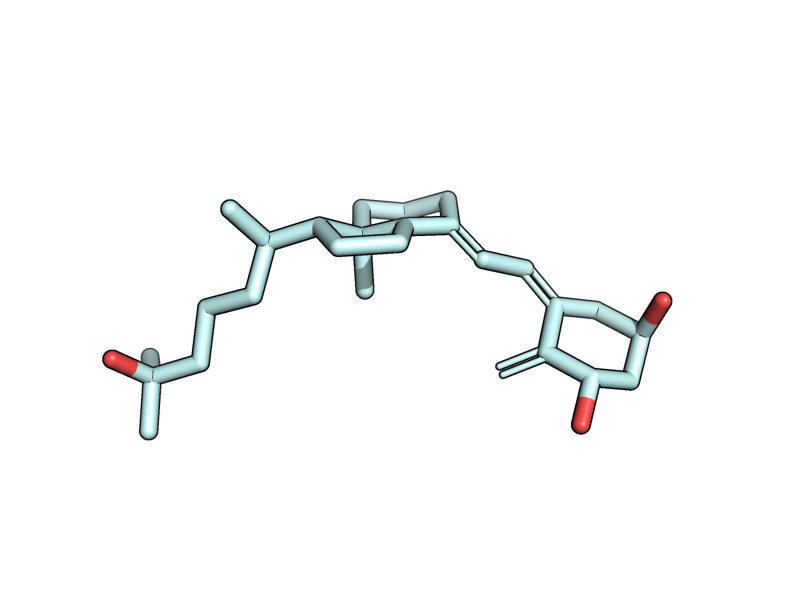}
\hspace*{\fill}
\end{minipage}
}

\subfloat[Bond lengths too long. Uni-Mol prediction for ligand P16 of protein-ligand complex 1OPK. RMSD \qty{1.5}{\angstrom}.]{
\begin{minipage}[c]{0.98\linewidth}
\hfill
\includegraphics[width=0.49\linewidth,trim={5 20 5 20},clip]{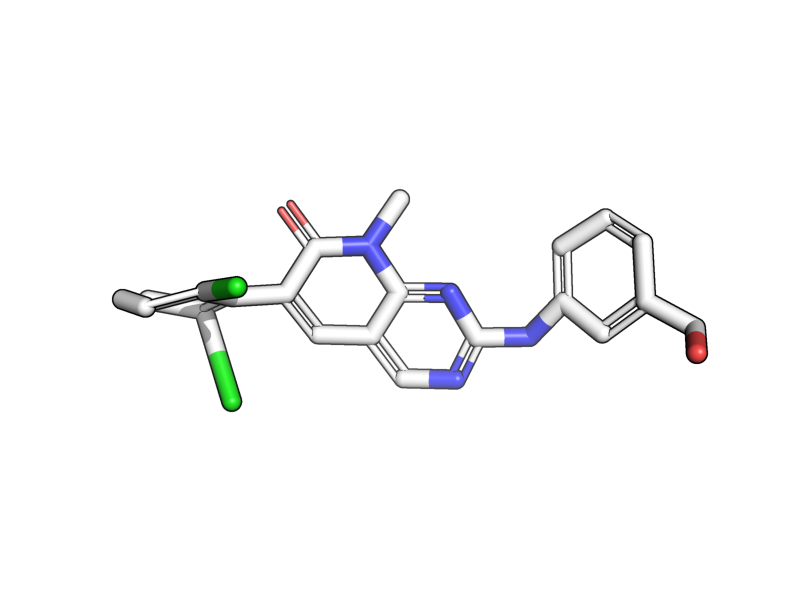}
\hfill
\includegraphics[width=0.49\linewidth,trim={5 20 5 20},clip]{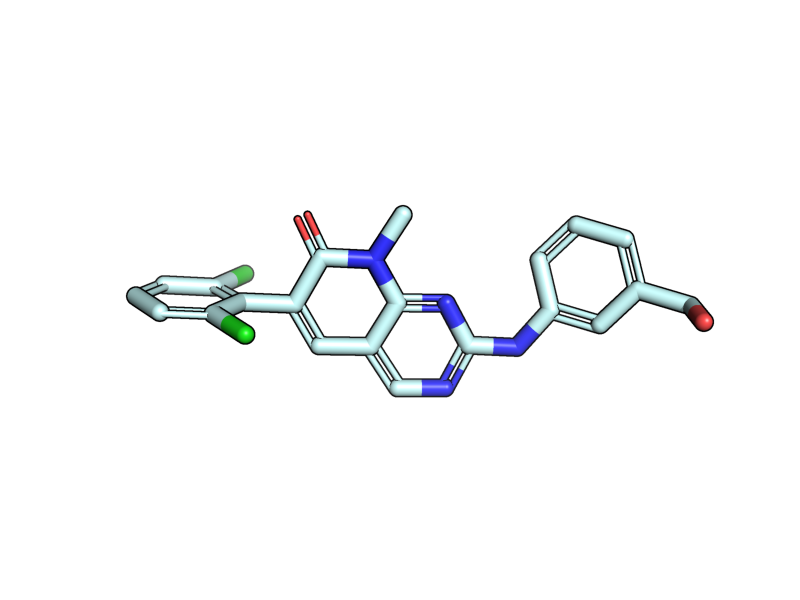}
\hspace*{\fill}
\end{minipage}
}

\subfloat[Bond angles too extreme. Uni-Mol prediction for ligand FR4 of protein-ligand complex 1UML. RMSD \qty{1.4}{\angstrom}.]{
\begin{minipage}[c]{0.98\linewidth}
\hfill
\includegraphics[width=0.49\linewidth,trim={5 13 5 10},clip]{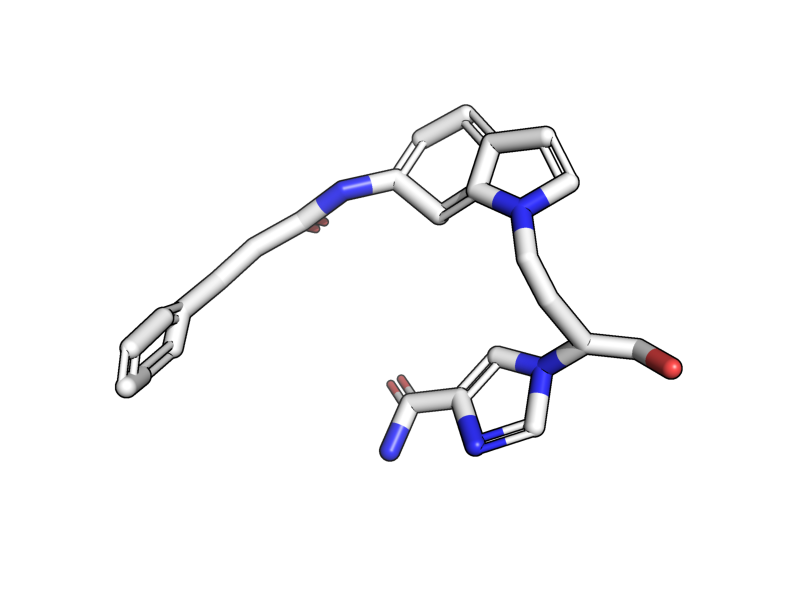}
\hfill
\includegraphics[width=0.49\linewidth,trim={5 13 5 10},clip]{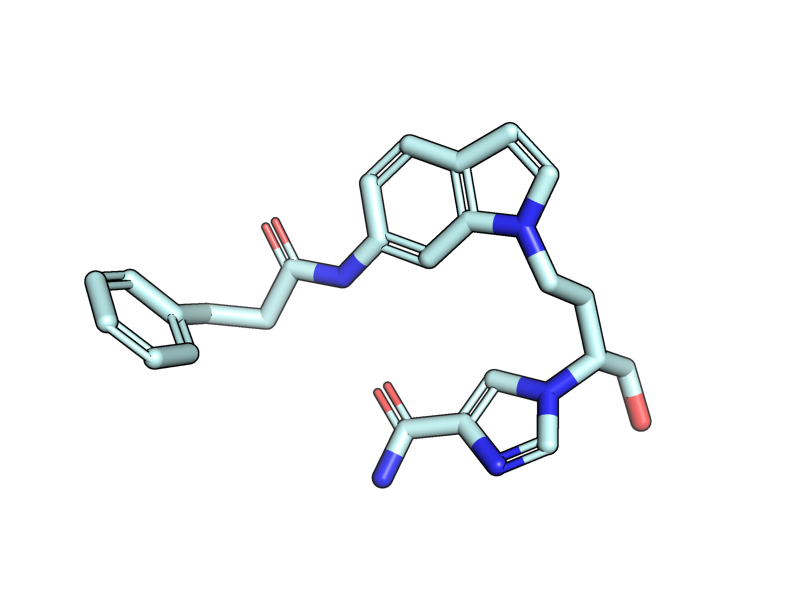}
\hspace*{\fill}
\end{minipage}
}

\subfloat[Internal clash. DeepDock prediction for ligand BDI of protein-ligand complex 1N2V. RMSD \qty{1.6}{\angstrom}.]{
\begin{minipage}[c]{0.98\linewidth}
\hfill
\includegraphics[width=0.49\linewidth,trim={0 20 0 16}, clip]{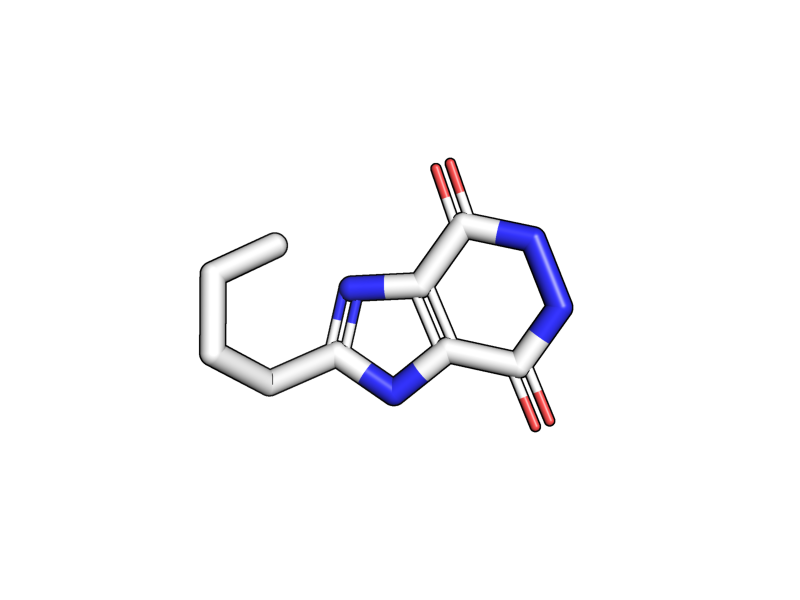}
\hfill
\includegraphics[width=0.49\linewidth,trim={0 20 0 16}, clip]{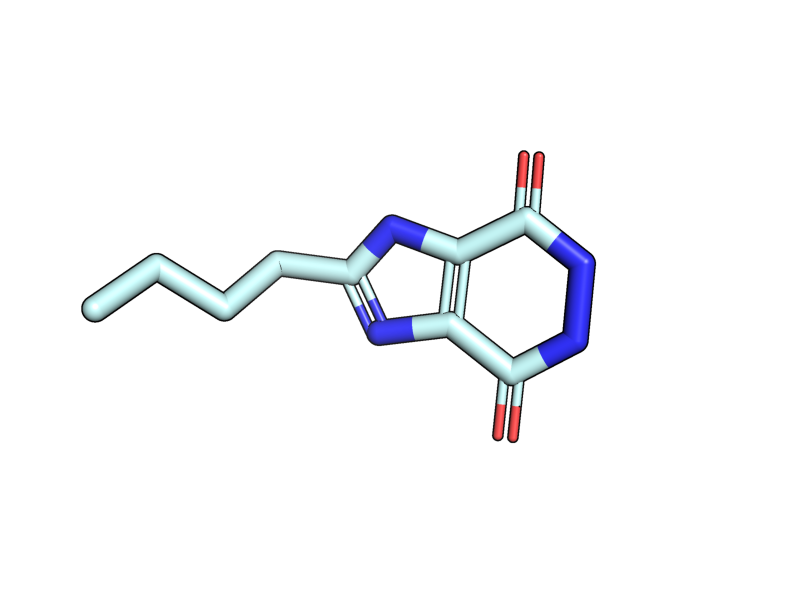}
\hspace*{\fill}
\end{minipage}
}

\subfloat[Aromatic rings not flat. TankBind prediction for ligand CRZ of protein-ligand complex 1TOW. RMSD \qty{2.2}{\angstrom}.]{
\begin{minipage}[c]{0.98\linewidth}
\hfill
\includegraphics[width=0.37\linewidth,trim={0 12 0 3}, clip]{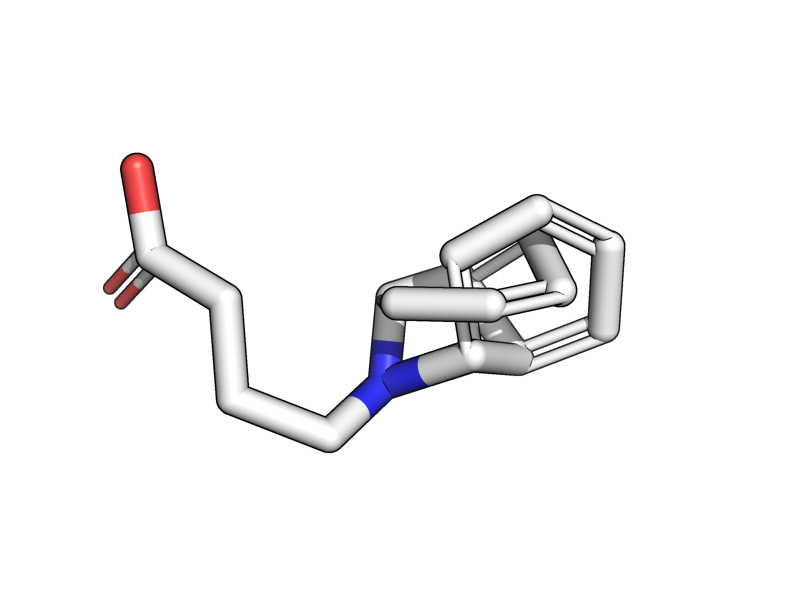}
\hfill
\includegraphics[width=0.37\linewidth,trim={0 12 0 3}, clip]{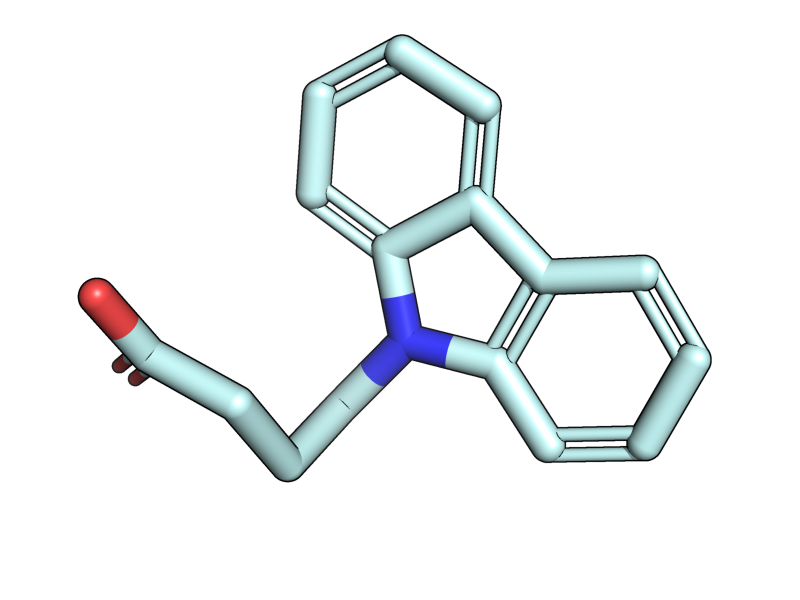}
\hspace*{\fill}
\end{minipage}
}

\subfloat[Double bond not flat. TankBind prediction for ligand DBQ of protein-ligand complex 1U4D. RMSD \qty{1.7}{\angstrom}.]{
\begin{minipage}[c]{0.98\linewidth}
\hfill
\includegraphics[width=0.42\linewidth,trim={0 13 0 15}, clip]{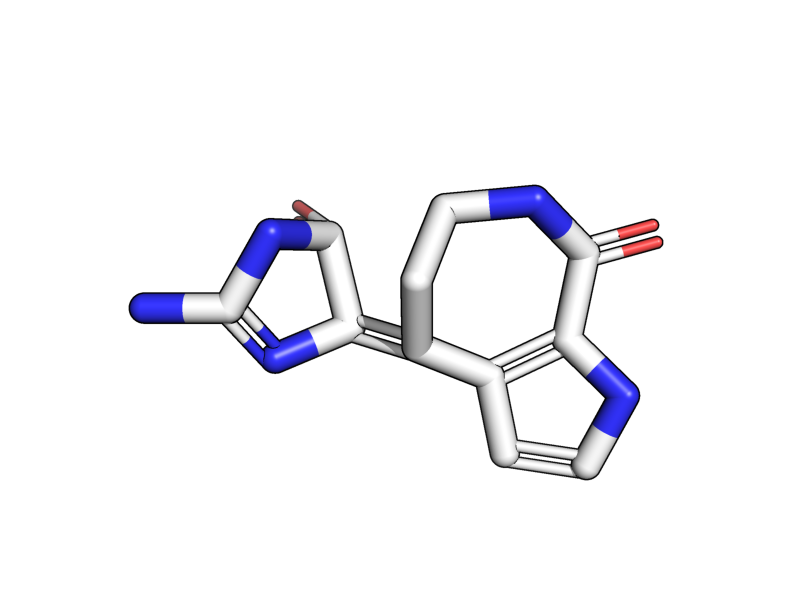}
\hfill
\includegraphics[width=0.42\linewidth,trim={0 13 0 15}, clip]{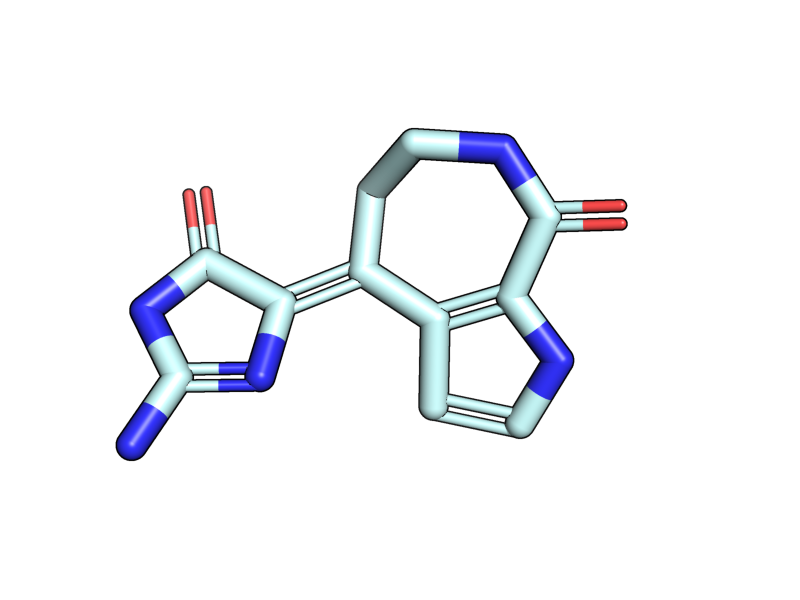}
\hspace*{\fill}
\end{minipage}
}

\subfloat[Energy ratio too high. AutoDock Vina prediction for ligand IFM of protein-ligand complex 7LOU. RMSD \qty{1.9}{\angstrom}.]{
\begin{minipage}[c]{0.98\linewidth}
\hfill
\includegraphics[width=0.23\linewidth,trim={0 8 0 5}, clip]{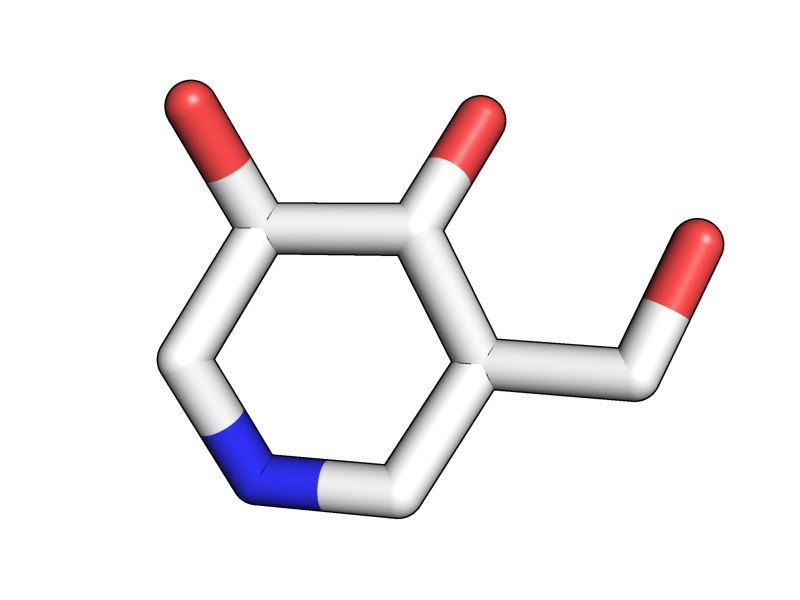}
\hfill
\includegraphics[width=0.23\linewidth,trim={0 8 0 5}, clip]{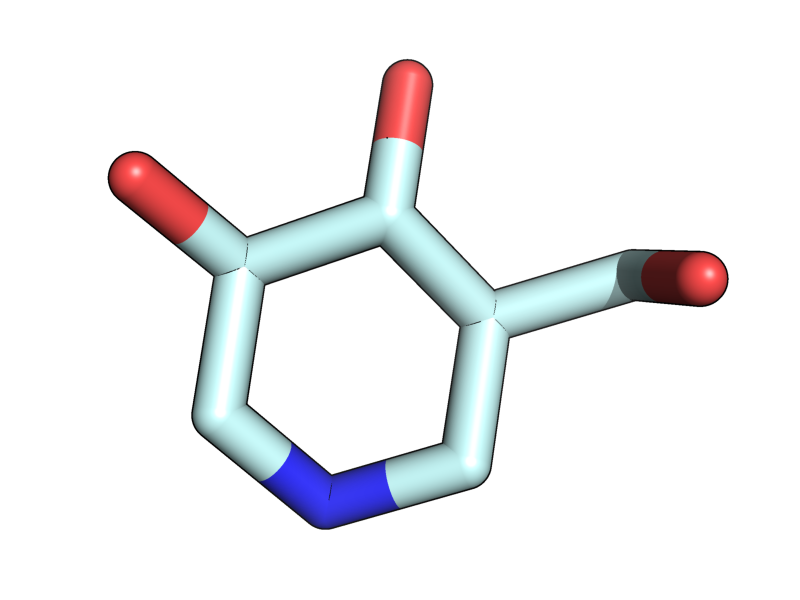}
\hspace*{\fill}
\end{minipage}
}

\subfloat[Clash with protein. DiffDock prediction for ligand XQ1 of protein-ligand complex 7L7C. RMSD \qty{1.6}{\angstrom}.]{
\begin{minipage}[c]{0.98\linewidth}
\includegraphics[width=0.49\linewidth,trim={0 15 0 15}, clip]{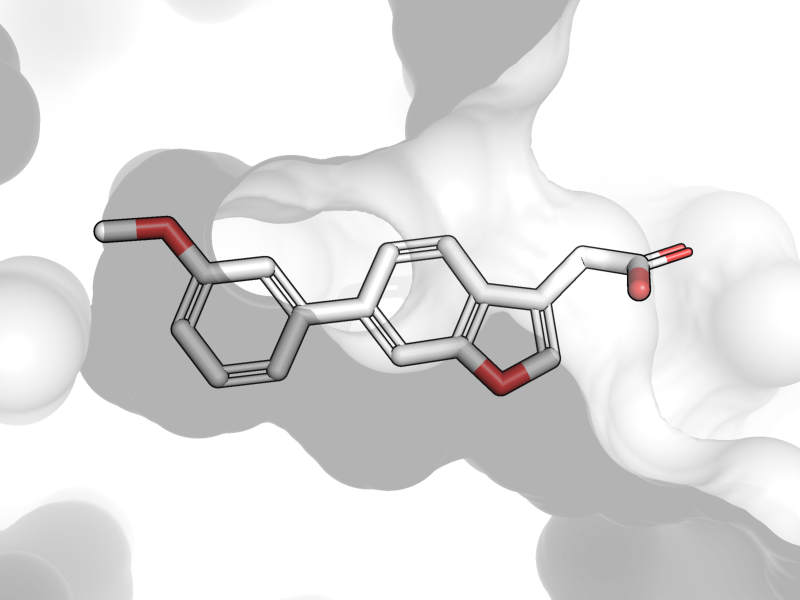}
\hfill
\includegraphics[width=0.49\linewidth,trim={0 15 0 15}, clip]{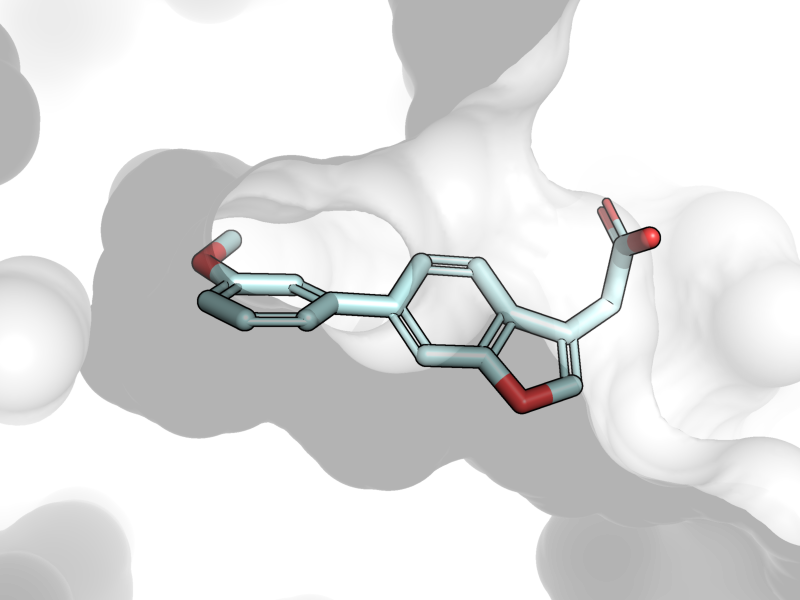}
\end{minipage}
}

\caption{Examples of failure modes that PoseBusters is able to detect. Predictions are shown on the left with white carbons and the crystal structures on the right have cyan carbons. Oxygen atoms are red, nitrogen atoms are dark blue, chlorine atoms are green. Most of the shown predictions have a RMSD within \qty{2}{\angstrom} but all are physically invalid.}
\label{fig:example_fails}
\end{figure}

All five DL-based docking methods struggle with physical plausibility, but even the poses produced by the classical methods Gold and AutoDock Vina do not always pass all the checks.
Figure~\ref{fig:waterfall_text} shows a waterfall plot that indicates how many predicted binding modes fail each test. The waterfall plots for the remaining methods are shown in SI~Figure~S5.
The DL-based methods fail on different tests. TankBind habitually overlooks stereochemistry, Uni-Mol very often fails to predict valid bond lengths, and EquiBind tends to produce protein-ligand clashes.
The classical methods Gold and AutoDock Vina pass most tests but also generated a few protein-ligand clashes. 
Figure~\ref{fig:example_fails} shows examples of poses generated by the methods illustrating various failure modes. 

The results on the Astex Diverse set suggest that despite what the $\text{RMSD} \leqslant \qty{2}{\angstrom}$ criterion would indicate, no DL-based method outperforms classical docking methods when the physical plausibility of the ligand binding mode is taken into account. However, DiffDock in particular is capable of making a large number of useful predictions.

\subsection{Results on the PoseBusters Benchmark set}

The results of the seven (AutoDock Vina, Gold, DeepDock, DiffDock, EquiBind, TankBind, Uni-Mol) docking methods on the PoseBuster\added{s} Benchmark set are shown in coral in Figure~\ref{fig:bars_datasets}. The striped bars show the performance only in terms of coverage (RMSD within 2Å) and the solid bars show the performance 
in terms of PB-validity (passing all of PoseBusters tests). 

The PoseBuster\added{s} Benchmark set was designed to contain no complexes that are in the training data for any of the methods. Performing well on this data set requires a method 
to be able to generalise well. 

All methods perform worse on the PoseBusters Benchmark set than the Astex Diverse set. 
Gold (\replaced{\qty{55}{\percent}}{\qty{48}{\percent}}) and AutoDock Vina (\replaced{\qty{58}{\percent}}{\qty{51}{\percent}}) perform the best out of the seven methods with (solid coral bars) and without (striped coral bars) considering PB-validity. 
On the PoseBusters Benchmark set, the best performing DL method, DiffDock (\replaced{\qty{12}{\percent}}{\qty{14}{\percent}}), does not compete with the two standard docking methods.
Gold and AutoDock Vina again pass the most tests but for a few protein-ligand clashes.

The waterfall plots in SI~Figure~S6 show which tests fail for each method on the PoseBusters Benchmark set.
Again, the methods have different merits and shortcomings. Out of the five DL-based methods, DiffDock still produces the most physically valid poses but few predictions lie within the \qty{2}{\angstrom} RMSD threshold. EquiBind, Uni-Mol and TankBind generate almost no physically valid poses that pass all tests. Uni-Mol has a relatively good RMSD score (\replaced{\qty{22}{\percent}}{\qty{23}{\percent}}) but struggles to predict planar aromatic rings and correct bond lengths.

\begin{figure}
\centering
\includegraphics[width=\columnwidth]{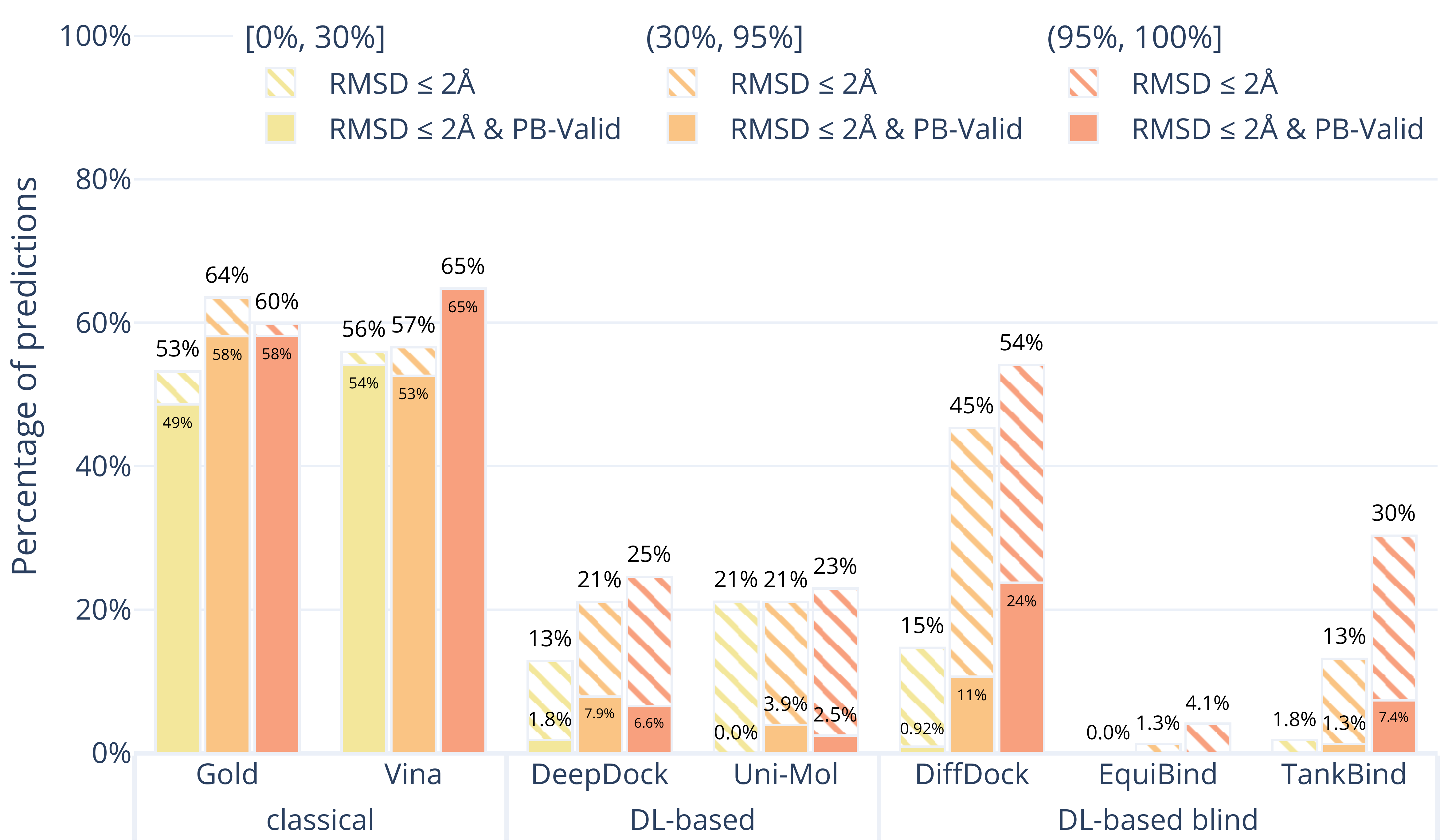}
\caption{
Comparative performance of docking methods on the PoseBusters Benchmark set stratified by sequence identity relative to the PDBBind General Set v2020. 
The sequence identity is the maximum sequence identity between all chains in the PoseBuster test protein and all chains in the PDBBind General Set v2020. 
The striped bars show the share of predictions of each method that have an RMSD within \qty{2}{\angstrom} and the solid bars show those predictions which in addition pass all PoseBuster tests and are therefore PB-valid.
The DL-based methods perform far better on proteins that \deleted{they} are similar to those they were trained on.
}
\label{fig:bars_seq_id}
\end{figure}

Figure~\ref{fig:bars_seq_id} shows the results of the docking methods on the PoseBusters Benchmark set but stratified by the target protein receptor's maximum sequence identity with the proteins in the PDBbind 2020 General Set \cite{liu2017forging}. As the DL-based methods were all trained on subsets of the PDBbind 2020 General Set, this roughly quantifies how different the test set protein targets are from those that the methods were trained on. 
We bin the test cases into three categories 
low [\num{0}, \qty{30}{\percent}], medium (\qty{30}{\percent}, \qty{90}{\percent}], and high (\qty{90}{\percent}, \qty{100}{\percent}] maximum percentage sequence identity. %
Without considering physical plausibility (striped bars), the classical methods appear to perform as well on the three protein similarity bins while the DL-based methods perform worse on the proteins with lower sequence identity. 
This suggests that the DL-based methods are overfitting to the protein targets in their training sets. 

We also compared the performance of the docking methods on the PoseBusters Benchmark set stratified by whether protein-ligand complexes contain cofactors (SI~Figure~S3).
Here, we loosely define cofactors as non-protein non-ligand compounds such as metal ions, iron-sulfur clusters, and organic small molecules in the crystal complex within \qty{4.0}{\angstrom} of any ligand heavy atom. About 45\% of protein-ligand complexes in the PoseBusters Benchmark set have a cofactor (SI Figure S2). The classical methods perform slightly better when a cofactor is present while the DL-based docking methods perform worse on those systems. 

\subsection{Results with pose-docking energy minimisation}

\begin{figure}
\centering
\includegraphics[width=\columnwidth]{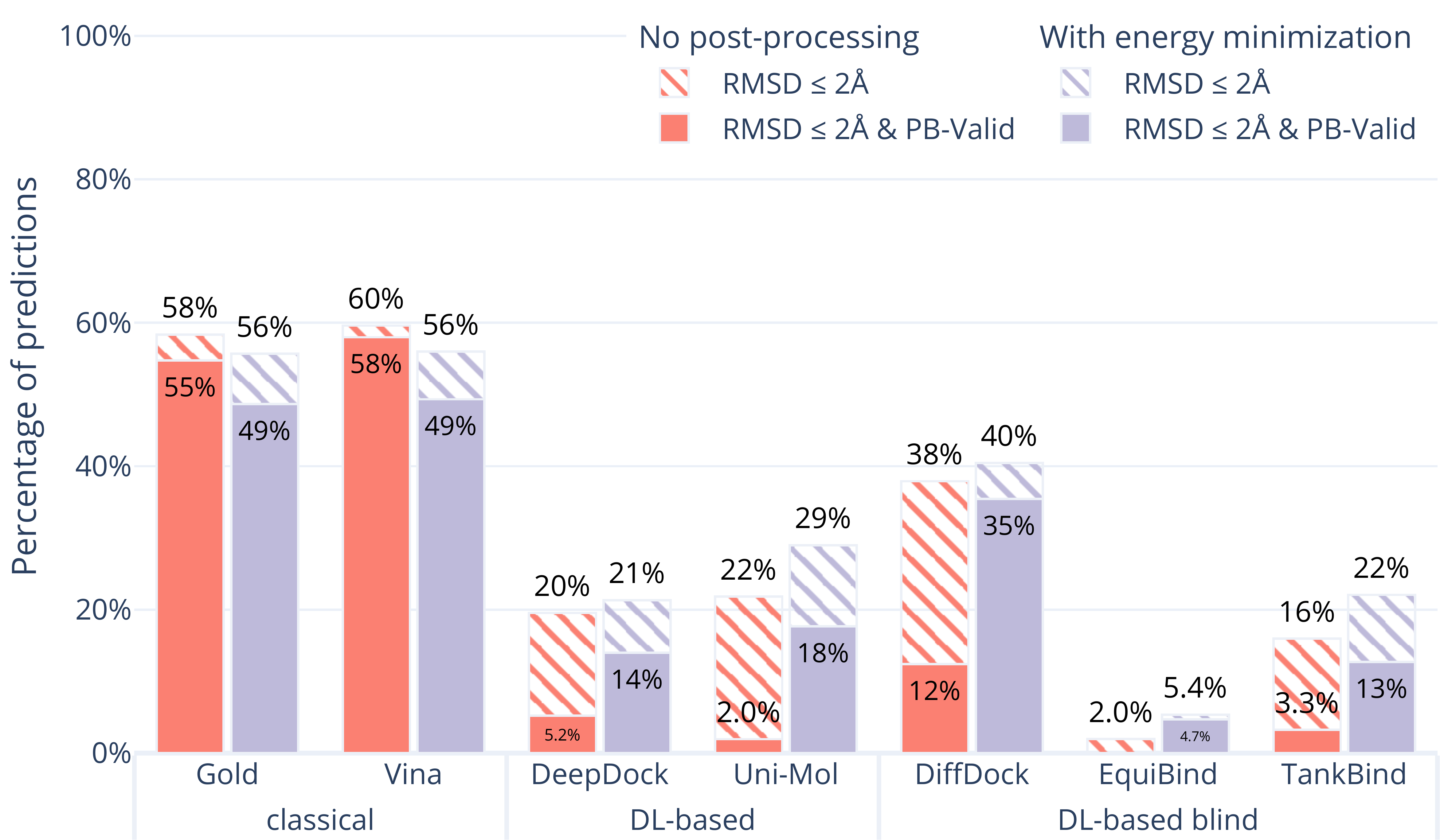}
\caption{
Comparative performance of docking methods with post-\replaced{d}{p}ocking energy minimisation of the ligand (while keeping the protein fixed) on the PoseBusters Benchmark set. 
The striped bars show the share of predictions of each method that have an RMSD within \qty{2}{\angstrom} of the crystal pose and the solid bars show those predictions which in addition pass all PoseBuster tests and are therefore PB-valid.
Post-docking energy minimisation significantly improves the relative physical plausibility of the DL-based methods' predictions.
This indicates that force fields contain docking-relevant physics which is missing from DL-based methods.  
}
\label{fig:bars_post_proc}
\end{figure}

\begin{figure}[H]
\centering
\includegraphics[width=0.49\linewidth,trim={2.5cm 1.6cm 2.95cm 1.5cm},clip]{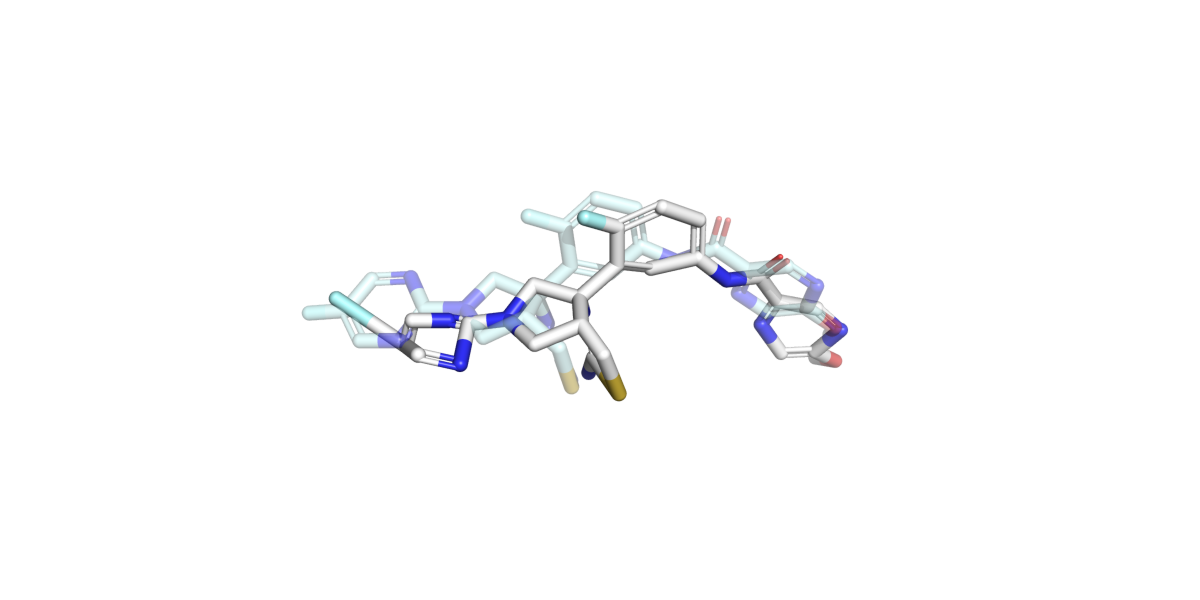}
\hfill
\includegraphics[width=0.49\linewidth,trim={2.5cm 1.6cm 2.95cm 1.5cm},clip]{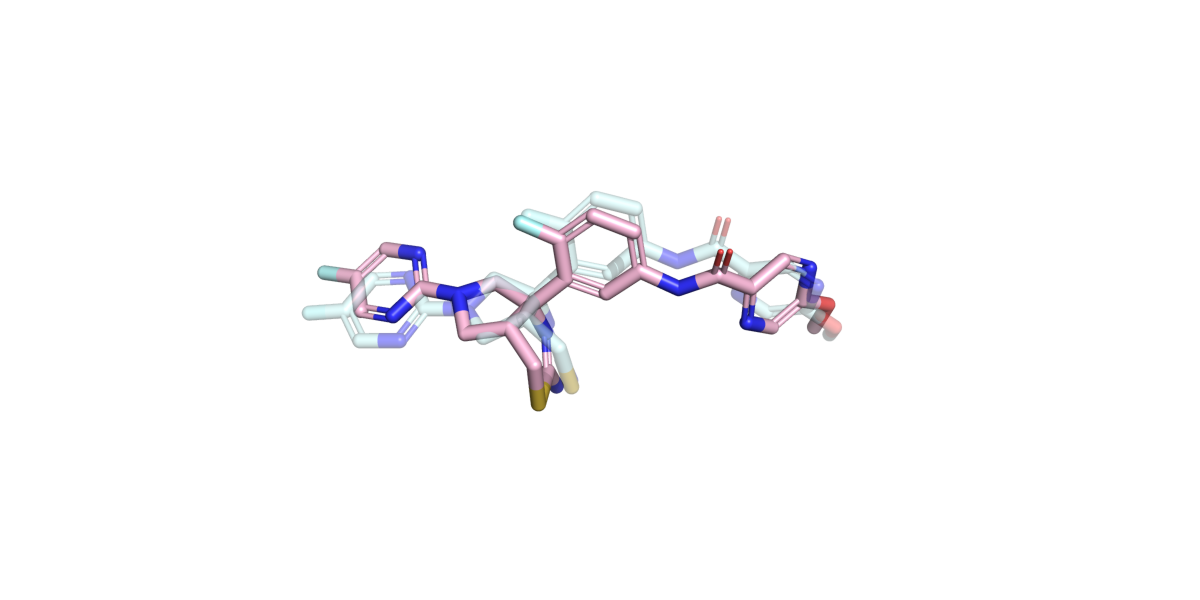}
\caption{\added{Example of a prediction that was fixed by the post-docking energy minimisation. The Uni-Mol prediction (RMSD \qty{2.0}{\angstrom}) is shown in white, the optimised prediction (RMSD \qty{1.1}{\angstrom}) is shown in pink, and the crystal ligand is shown as reference in light blue. Note how the aromatic rings are flattened and the leftmost bond is shortened by the optimisation making the prediction pass all PoseBusters checks.}}
\label{fig:em_example_repair}
\end{figure}

In order to examine whether the outputs of the DL-based methods can be made physically plausible we performed an additional post-docking energy minimisation of the ligand structures in the binding pocket for the PoseBusters Benchmark set (Figure~\ref{fig:bars_post_proc}). Again, striped bars indicate predictions with $\text{RMSD} \leqslant \qty{2}{\angstrom}$ while solid bars indicate which of those are 
also PB-valid
i.e., pass all PoseBusters tests. 
The figure shows that post-docking energy minimisation significantly increases the number of physically plausible structures of the DL-based methods DiffDock, DeepDock, TankBind, and Uni-Mol but does not improve the poses predicted by AutoDock Vina and Gold.  
\added{We also performed energy minimisation on the Uni-Mol results for the minimal \qty{6}{\angstrom} pocket (SI~Figure~S22). The number of poses that pass the tests increase to about the same level as DiffDock.}
The fact that energy minimisation is able to repair many of the DL methods predicted poses and increase coverage
shows that at least some force field physics is missing from DL-based docking methods.
\added{An example of a predicted pose that was fixed is shown in Figure~\ref{fig:em_example_repair}}.
However, even with the energy minimisation step, the best DL-based docking method DiffDock still performs worse than the classical methods Gold and AutoDock Vina.

\section{Discussion}

We present PoseBusters, a test suite designed and built to identify chemically inconsistent and physically implausible ligand poses predicted by protein-ligand docking and molecular generation methods. We show the results of applying the PoseBusters test suite to the output of seven different docking methods, five current DL-based docking methods (DeepDock, DiffDock, EquiBind, TankBind, and Uni-Mol) and two standard methods (AutoDock Vina and Gold). 

We find that no DL-based docking method yet outperforms standard docking methods when \emph{both} physical plausibility and binding mode RMSD is taken into account.
Our work demonstrates the need for physical plausibility to be taken into account when assessing docking tools because it is possible to perform well on an RMSD-based metric while predicting physically implausible ligand poses (Figure~\ref{fig:example_fails}). 
Using the tests in the PoseBusters test suite as an additional criterion when developing DL-based docking methods will help improve methods and the development of more accurate and realistic predictions. 

In addition, the individual tests in the PoseBusters test suite highlight docking-relevant failure modes.
The results show that Uni-Mol for example predicts non-standard bond lengths and TankBind creates internal ligand clashes. The ability to identify such failure modes in predicted ligand poses makes PoseBusters a helpful tool for developers to identify inductive biases that could improve their binding mode prediction methods.

Our results also show that, unlike classical docking methods, DL-based docking methods do not generalise well to novel data.
The performance of the DL-based methods on the PoseBusters Benchmark set overall was poor and the subset of the PoseBusters Benchmark set with low sequence identity to PDBbind 2020 revealed that DL-based methods are prone to overfitting to the proteins they were trained on.
Our analysis of the targets with sequence identity lower than \qty{30}{\percent} to any member of PDBbind General Set v2020 revealed that across all of the DL-based docking methods almost no physically valid poses were generated within the \qty{2}{\angstrom} threshold.

The most commonly-used train-test approach for building DL-based docking models is time-based, e.g., complexes released before a certain date are used for training and complexes released later for testing. Based on our results, we argue that this is insufficient for testing generalisation to novel targets and the sequence identity between the proteins in the training and test must be reported on.

Post-docking energy minimisation of the ligand using force fields can considerably improve the docking poses generated by DL-based methods. However, even with an energy minimisation step, the best DL-based method, DiffDock, does not outperform classical docking methods like Gold and AutoDock Vina. This shows that at least some key aspects of chemistry and physics encoded in force fields are missing from deep learning models.

The PoseBusters test suite provides a new criterion, PB-validity, beyond the traditional ``$\text{RMSD} \leqslant \qty{2}{\angstrom}$'' rule to evaluate the predictions of new DL-based methods, and hopefully will help to identify inductive biases needed for the field to improve docking and molecular generation methods, ultimately resulting in more accurate and realistic predictions.
\deleted{PoseBusters is made available as a pip-installable Python package and as open source code under the BSD-3-Clause license at \url{github.com/maabuu/posebusters}.}
The next generation of DL-based docking methods should aim to outperform standard docking tools on both RMSD criteria and in terms of chemical consistency, physical plausibility, and generalisability. 

\section*{Data availability}
\added{PoseBusters is made available as a pip-installable Python package and as open source code under the BSD-3-Clause license at \url{github.com/maabuu/posebusters}.
Data for this paper, including the Astex Diverse set and PoseBusters Benchmark set, as well as the individual tabulated test results for each docking are available at Zenodo at \url{https://zenodo.org/records/8278563}.}

\section*{Author Contributions}
The manuscript was conceptualised and written through contributions of all authors. M. B. performed the computational experiments, created the PoseBusters software, and wrote the original draft. C. M. D. and G. M. M. supervised the work and reviewed and edited the manuscript.

\section*{Conflicts of interest}
There are no conflicts to declare.

\section*{Acknowledgements}
\added{
We thank Andrew Henry for bringing to our attention the ligands with crystal contacts, enhancing the quality and accuracy of the PoseBusters Benchmark set.
We thank Eric Alcaide for identifying a discrepancy in the preprint's description of the pre-processing for Uni-Mol.
}

\bibliography{references} %
\bibliographystyle{rsc} %

\end{document}


\maketitle

\tableofcontents

\clearpage
\section{Docking protocols}
\label{sec:docking_protocols}
The following protocols detail how the seven docking methods were used to re-dock the ligands into the crystal structures of the Astex Diverse set and the PoseBuster\added{s Benchmark} set.
Methods that require an initial ligand conformation were given identical starting conformations generated with \added{the} RDKit's ETKDGv3 conformer generator \cite{wang2020improving} followed by an energy minimisation using the universal force field (UFF) \cite{tosco2014bringing}. \added{All docking protocols were given receptors prepared without waters as none of the DL-based methods supports docking with waters.}

\setlist[description]{font=\normalfont\itshape}

\subsubsection*{AutoDock Vina}
\begin{description}
\item[Software version] Vina 1.2.3, Meeko 0.4.0, Reduce 4.9.210817, ADFRsuite 1.0, RDKit 2022.09.1 %
\item[Ligand preparation] The initial ligand conformations described above were prepared with Meeko using standard settings.
\item[Protein preparation] Hydrogen atoms were added with reduce and then the PDBQT files were generated with the ADFR \verb|prepare_receptor| script.
\item[Parameters] A bounding box with side-length \qty{25}{\angstrom} was created around the centroid of the crystal ligand. Vina was used to create 40 poses with an exhaustiveness setting of \num{32} and the top-ranked pose was selected. 
\end{description}

\subsubsection*{CCDC Gold}
\begin{description}
\item[Software version] CCDC Python API version 3.0.14
\item[Ligand preparation] The initial ligand conformations described above were prepared with \verb|LigandPreparation| using the default settings which include adding missing hydrogens, removing unknown atoms, and rule-based protonation of the ligand.
\item[Protein preparation] The protein and co-factors were loaded from separate files and all hydrogens were added.
\item[Parameters] A settings file was created for each complex using the \verb|Docker| class default settings. The binding site was defined around the crystal ligand centroid using \verb|BindingSiteFromPoint| with radius \qty{25}{\angstrom}. The settings used are rescore function \verb|plp|, autoscale \replaced{\qty{100}{\percent}}{\qty{10}}, and early termination off. After generating \num{40} poses only the top-ranked pose was saved.
\end{description}

\subsubsection*{DeepDock}
\begin{description}
\item[Software version] DeepDock commit hash \verb|54a2a64| from authors' public code repository {\url{https://github.com/OptiMaL-PSE-Lab/DeepDock}}, MSMS 2.6.1, PDB2PQR 2.1.1, APBS 3.4.1
\item[Ligand preparation] The generated starting ligand conformations were used without further processing. 
\item[Protein preparation] The steps in example notebook \verb|Docking_example.ipynb| were used to generate protein surface meshes. The function \verb|compute_inp_surface| generated binding site surfaces using the crystal ligands and the crystal protein structures with a distance threshold of \qty{10}{\angstrom}.
\item[Parameters] The protocol and settings in notebook \verb|Docking_example.ipynb| in the DeepDock repository were used for docking. 
\end{description}

\subsubsection*{DiffDock}
\begin{description}
\item[Software version] DiffDock commit hash \verb|fff8f0b| from authors' public code repository {\url{https://github.com/gcorso/DiffDock}}
\item[Ligand preparation] The generated starting ligand conformations were used without further processing.
\item[Protein preparation] ESM was used to generate FASTA files. 
\item[Parameters] The protocol in \verb|README.md| was used to generate ESM embeddings and then to do inference. 40 poses were sampled using 20 inference steps with no noise on the final step. The top-ranked pose was selected. 
\end{description}

\subsubsection*{EquiBind}
\begin{description}
\item[Software version] EquiBind commit hash \verb|41bd00f| from authors' public code  repository {\url{https://github.com/HannesStark/EquiBind}}, Reduce 3.3.160602, Open Babel 3.1.0, RDKit 2022.09.1
\item[Ligand preparation] The generated starting ligand conformations were processed with Open Babel and then with \added{the} RDKit to add missing hydrogens. 
\item[Protein preparation] The receptors were processed with Open Babel. Then reduce was used to correct receptor residues and to add hydrogens. Then the protein chains which have at least one residue within \qty{10}{\angstrom} of the crystal ligand were selected.
\item[Parameters] The configuration file \verb|configs_clean/inference.yml| in the repository was used.
\end{description}

\subsubsection*{TankBind} 
\begin{description}
\item[Software version] TANKBind commit hash \verb|804e9fc| from authors' public code  repository {\url{https://github.com/luwei0917/TankBind}}, p2rank 2.3
\item[Ligand preparation] The notebook \verb|prediction_example_using_PDB_6hd6.ipynb| was used to renumber the ligand atoms and generate features from the ligands. 
\item[Protein preparation] The notebook \verb|prediction_example_using_PDB_6hd6.ipynb| was used to generate features from the crystal protein structures.
\item[Parameters] The steps in the notebook \verb|prediction_example_using_PDB_6hd6.ipynb| were used for inference. The steps are running p2rank to generate a list of binding pockets and then docking using the TankBind model.
\end{description}

\subsubsection*{Uni-Mol}
\begin{description}
\item[Software version] Uni-Mol commit hash \verb|b962451| from authors' public code repository {\url{https://github.com/dptech-corp/Uni-Mol}}
\item[Ligand preparation] The ligands were generated according to the protocol described in the \verb|README.md| file in the top folder of the Uni-Mol repository.
\item[Protein preparation] The binding pockets residues are those within \replaced{\qty{8}{\angstrom}}{\qty{6}{\angstrom}} of any crystal ligand heavy atom.
\item[Parameters] The default arguments (\verb|recycling=3|,  \verb|batch_size=8|,  \verb|dist_threshold=8.0|) were used. 
\end{description}

\clearpage
\section{Search space illustrations}
\begin{figure}[H]
\centering

\subfloat[Gold: Sphere of radius \qty{25}{\angstrom} centered on the geometric centre of the crystal ligand heavy atoms.]{\includegraphics[width=0.49\textwidth]{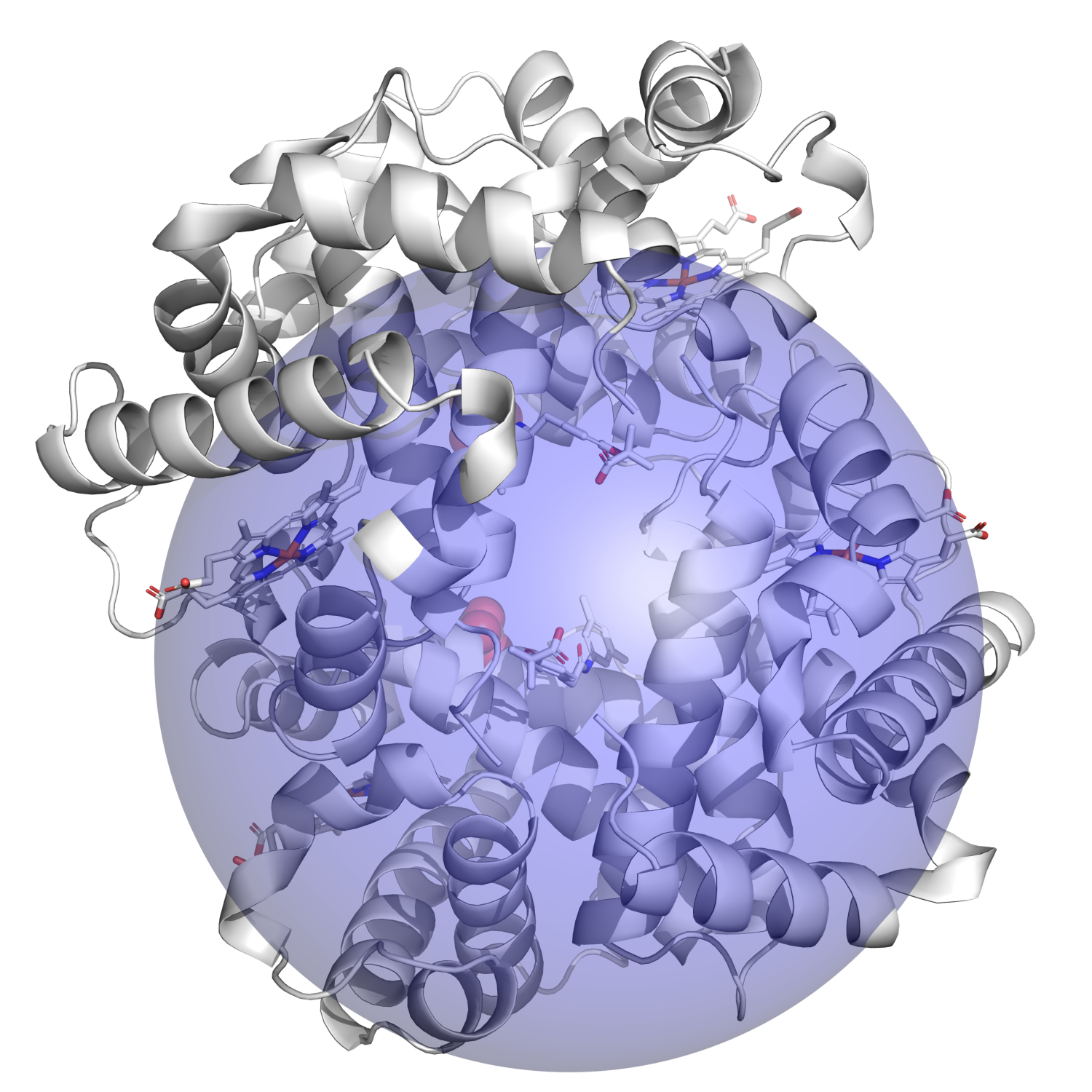}}
\hfill
\subfloat[AutoDock Vina: Cube with side length \qty{25}{\angstrom} centered on the geometric centre of crystal ligand heavy atoms.]{\includegraphics[width=0.49\textwidth]{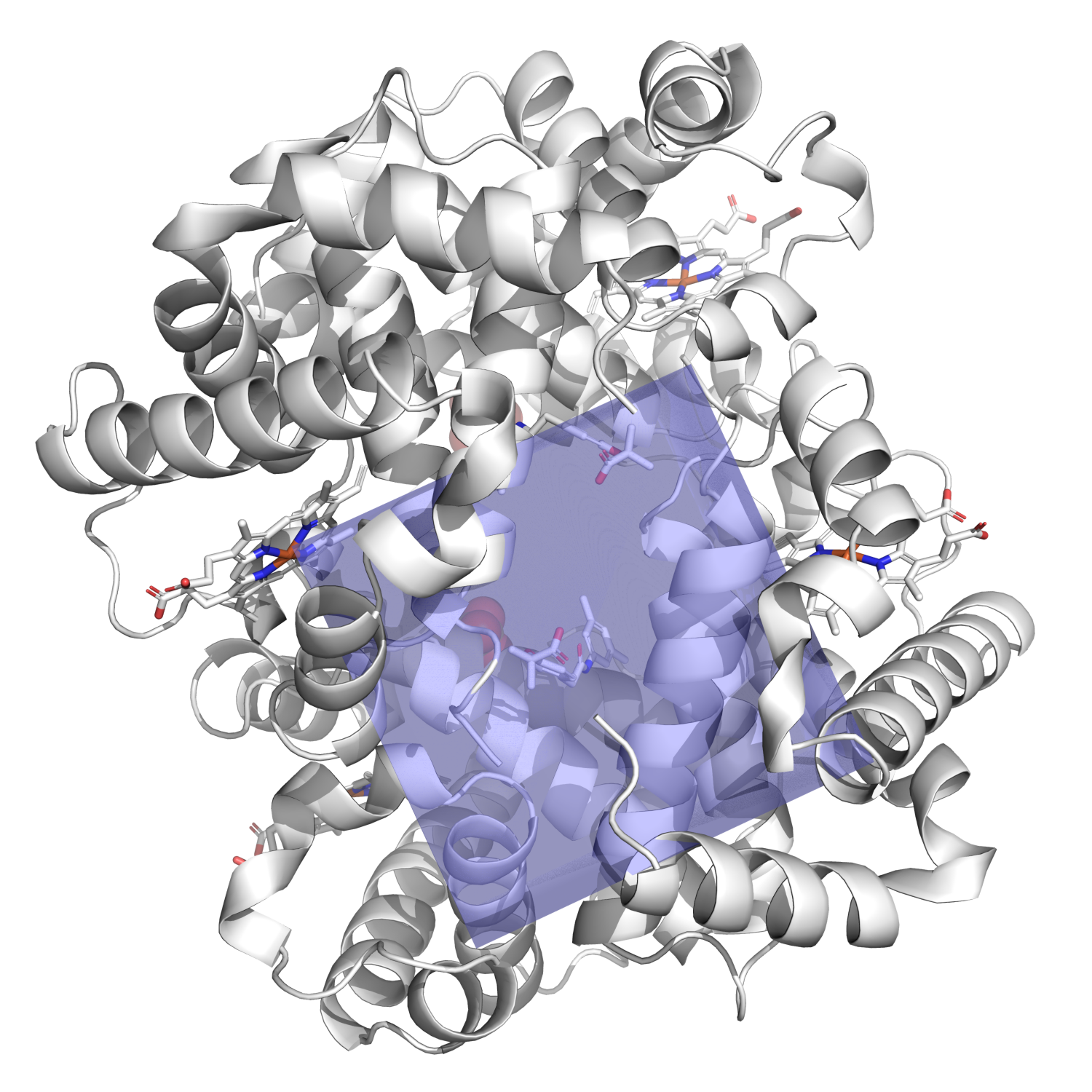}}

\subfloat[DeepDock: Protein surface mesh nodes within \qty{10}{\angstrom} of any crystal ligand atom.]{\includegraphics[width=0.49\textwidth]{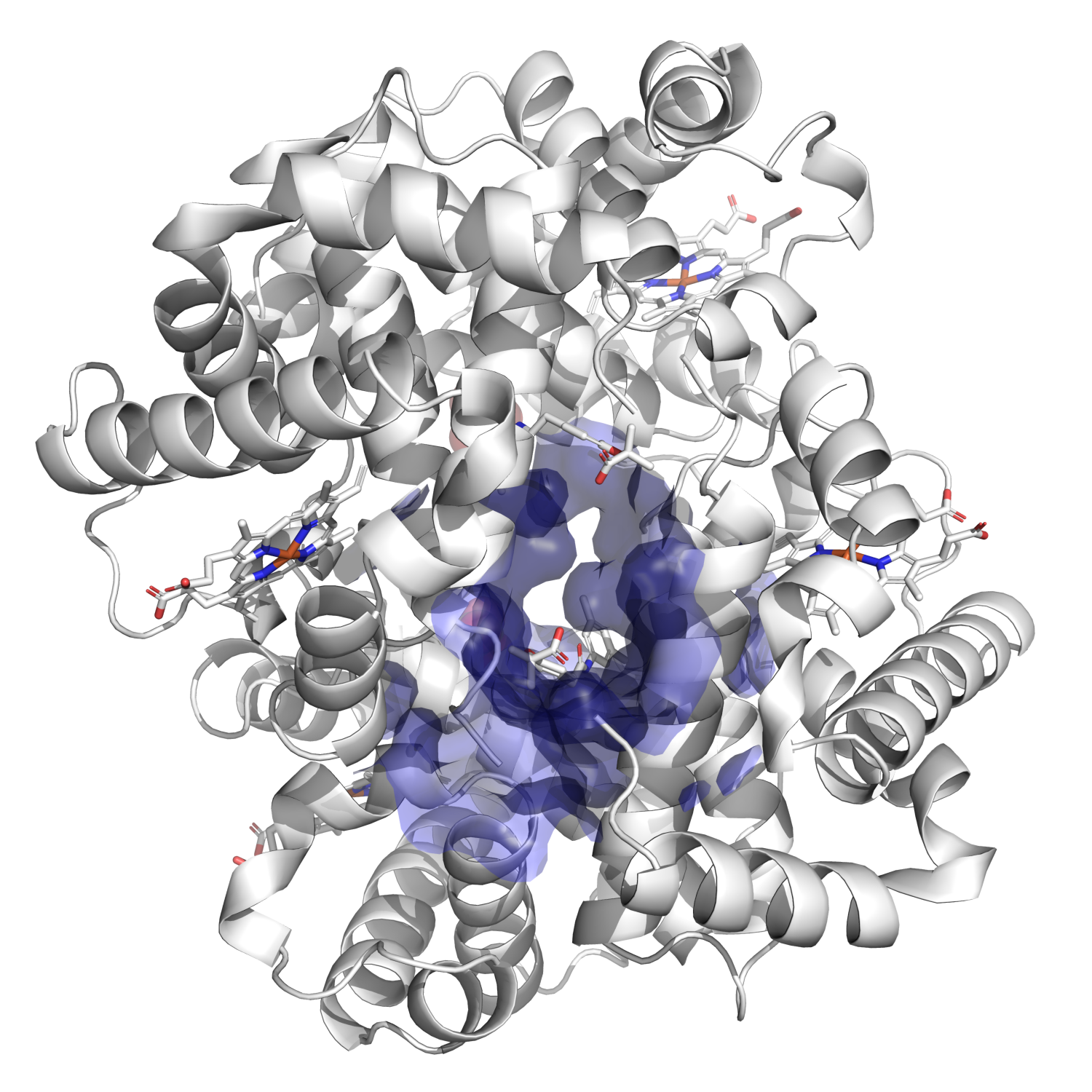}}
\hfill
\subfloat[Uni-Mol: Protein residues within \replaced{\qty{8}{\angstrom}}{\qty{6}{\angstrom}} of any crystal ligand heavy atom.]{\includegraphics[width=0.49\textwidth]{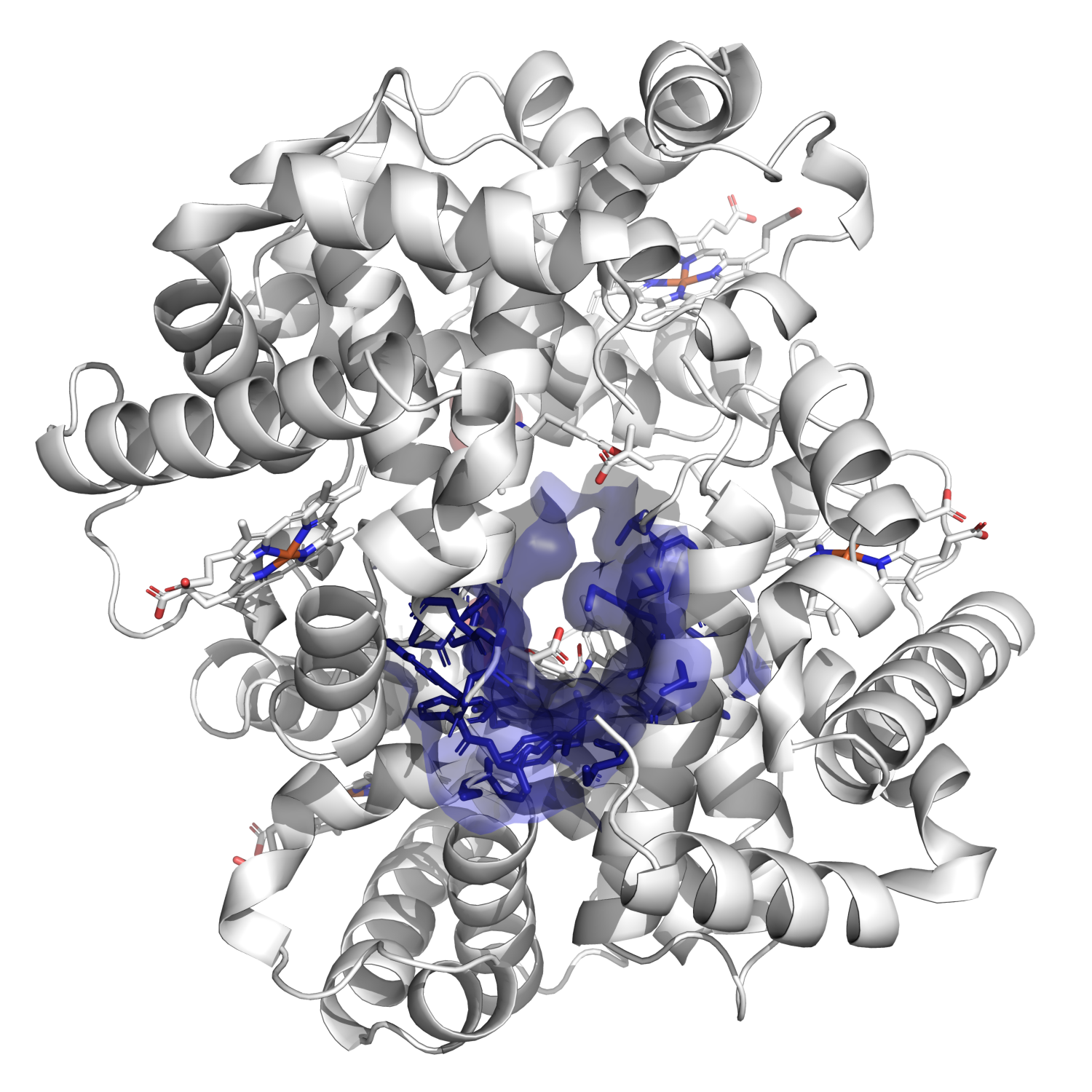}}
 
\caption{Search spaces of the docking methods illustrated on PDB entry 1G9V for ligand RQ3. The search spaces for the blind docking methods DiffDock, EquiBind, and TankBind are the entire protein crystal structure. For more information refer to Table~3 in the main text.}
\label{fig:search_spaces}
\end{figure}

\clearpage
\section{PoseBusters Benchmark set procurement}

\begin{table}[H]
\caption{Selection process of the PDB entries and ligands for the PoseBuster\added{s Benchmark} \deleted{data} set. The filters are based on the PDB meta data, the PDB quality reports, and the PDB structure data. The final PoseBuster\added{s Benchmark} \deleted{data} set consists of \replaced{308}{428} unique PDB entries containing \replaced{308}{428} unique ligands.}
\begin{threeparttable}[b]
\begin{tblr}{width=\linewidth, colspec={X[1,l]Q[r,f]Q[r,f]}}
\toprule
Selection step & {Number of proteins\\ (unique PDB IDs)} & {Number of ligands\\ (unique CCD IDs)} \\
\midrule
PDB entries with a protein and `ligand of interest' released from 1 January 2021 to 30 May 2023 & 10537 & 6635 \\
Ligands weighing from \qtyrange{100}{900}{\dalton} & 10537 & 6424 \\
Ligands with at least 3 heavy atoms & 10537 & 6374 \\
Ligands containing only H, C, O, N, P, S, F, Cl atoms & 10537 & 6271 \\
Ligands that are not covalently bound to protein & 7247 & 4891 \\
Structures with no unknown atoms (e.g. element X) & 7218 & 4881 \\
X-ray structure high resolution limit at most \qty{2}{\angstrom} & 4686 & 3314 \\
Ligand real space R-factor is at most \qty{0.2}{} & 3800 & 2572 \\
Ligand real space correlation coefficient is at least \qty{0.95}{} & 1849 & 1054 \\
Ligand model completeness is 100\% & 1820 & 1039 \\
Ligand starting conformation could be generated with ETKDGv3\cite{riniker2015better} & 1733 & 1019 \\
All ligand SDF files can be loaded with RDKit \cite{landrum2023rdkit} and pass its sanitization & 1706 & 994 \\
PDB ligand report does not list stereochemical errors & 1706 & 994 \\
PDB ligand report does not list any atomic clashes & 1256 & 844 \\
Select single protein-ligand conformation\tnote{1} & 1256 & 844 \\
Intermolecular distance between the ligand(s) of interest and the protein  is at least \qty{0.2}{\angstrom} & 1237 & 834 \\
Intermolecular distance between ligand(s) of interest and other small organic molecules is at least \qty{0.2}{\angstrom} & 1237 & 834 \\
Intermolecular distance between the ligand(s) of interest and ion metals in complex is at least \qty{0.2}{\angstrom} & 1232 & 832 \\
Blocklist for PDB entries\tnote{2} & 1227 & 827 \\
Blocklist for CCD entries\tnote{3} & 1223 & 823 \\
Randomly select PDB entries to get a set with unique ligands & 809 & 823 \\
Randomly select ligands to get a set with unique PDB entries & 809 & 809 \\
Select representative PDB entries by clustering protein sequences\tnote{4} & 428 & 428 \\
\added{Remove ligands which are within \qty{5.0}{\angstrom} of any protein symmetry mate} & \added{308} & \added{308} \\
\bottomrule
\end{tblr}
\begin{tablenotes}
\small
\item [1] The first conformation containing the ligand of interest was chosen when multiple conformations containing the ligand were available in the PDB entry.
\item [2] The blocklist for the PDB entries (by PDB identifier) contains entries removed due to bad ligand conformations (7X48, 7UYC), ligands forming polymers (7WJD, 7DB4), racemeic mixtures of ligands where the stereoisomer has a different CCD identifier (6ZYU, 7W2W), and structures containing elements Te and Yb which AutoDock Vina does not support by default (7ZSQ, 8AVA).
\item [3] The blocklist for the ligands (by CCD identifier) contains the four entries I8P, 5A3, U71, and UEV. These four are omitted because they are highly symmetric and the substructure search yields many possible atom-atom mappings between conformations negatively affecting the RMSD calculation time.
\item [4] Clustering with Diamond\cite{buchfink2021sensitive} is done with an identity cutoff for the clustering of 0\% and a minimum coverage of the cluster member sequences by the representative sequences of 100\% and otherwise default values which includes the \verb|BLOSUM62| substitution matrix. 
\end{tablenotes}
\end{threeparttable}
\end{table}

\clearpage
\section{Data sets description}

\begin{figure}[H]
\centering

\includegraphics[width=0.75\textwidth]{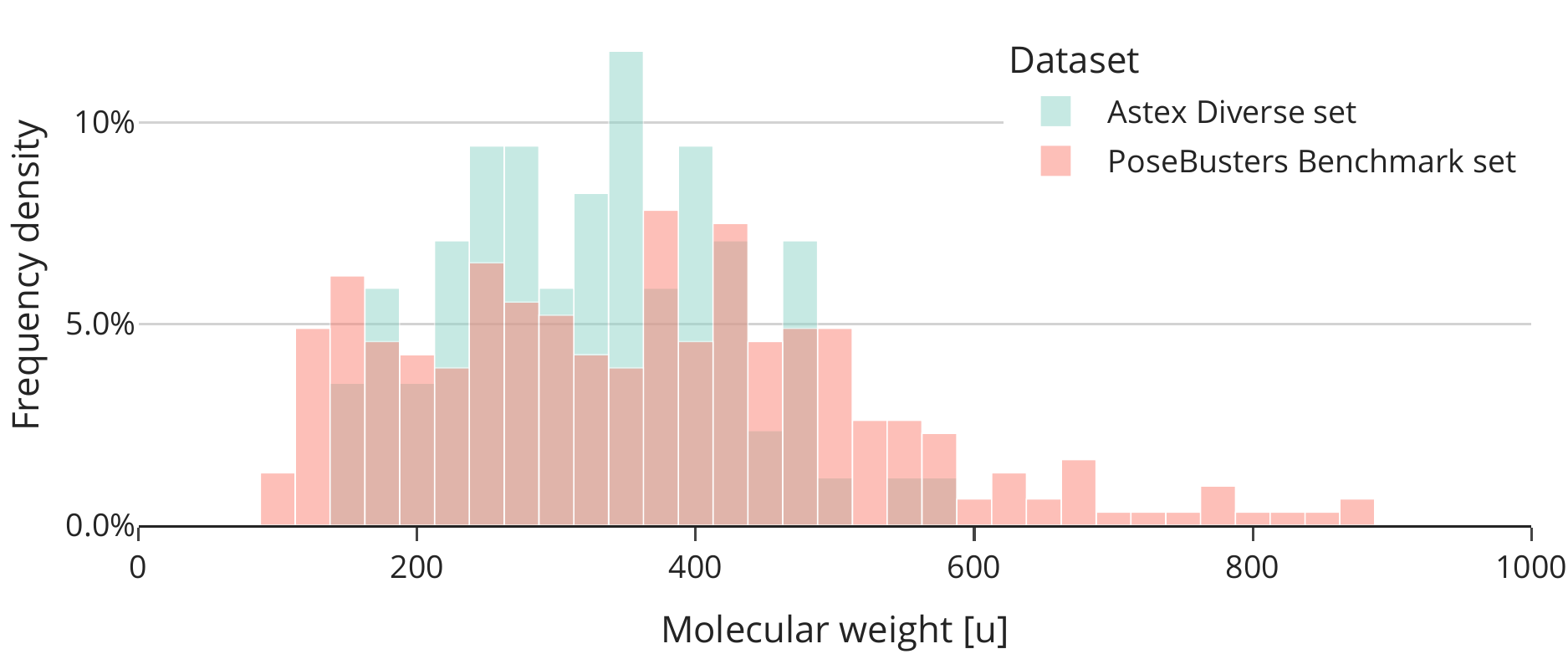}

\includegraphics[width=0.75\textwidth]{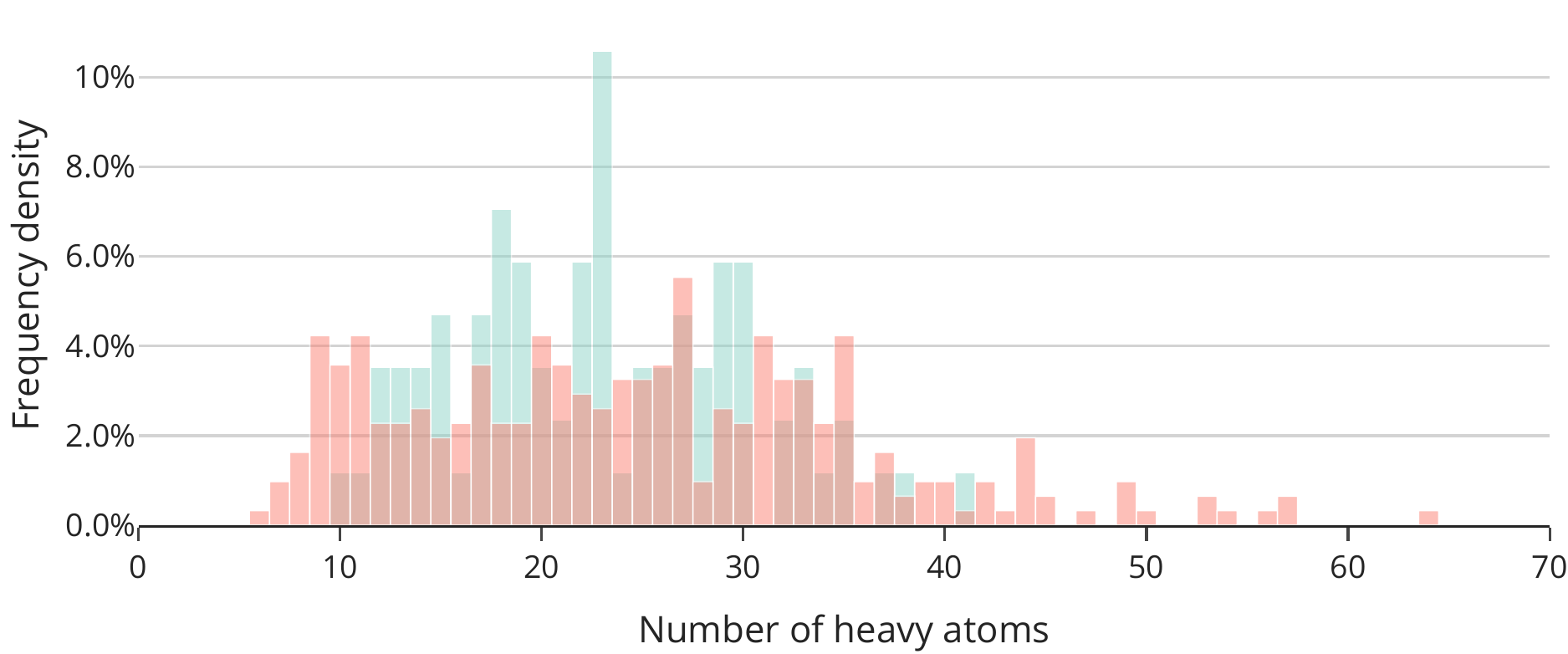}

\includegraphics[width=0.75\textwidth]{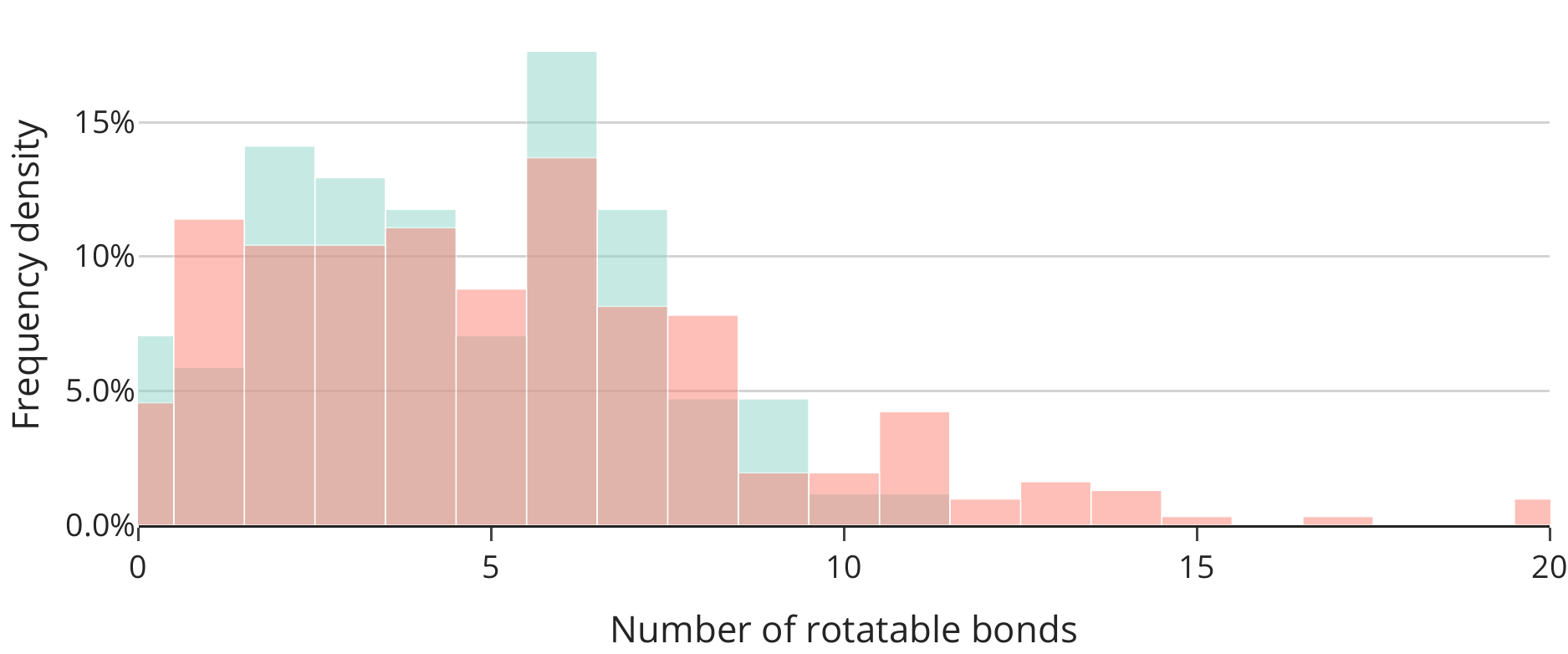}

\includegraphics[width=0.75\textwidth]{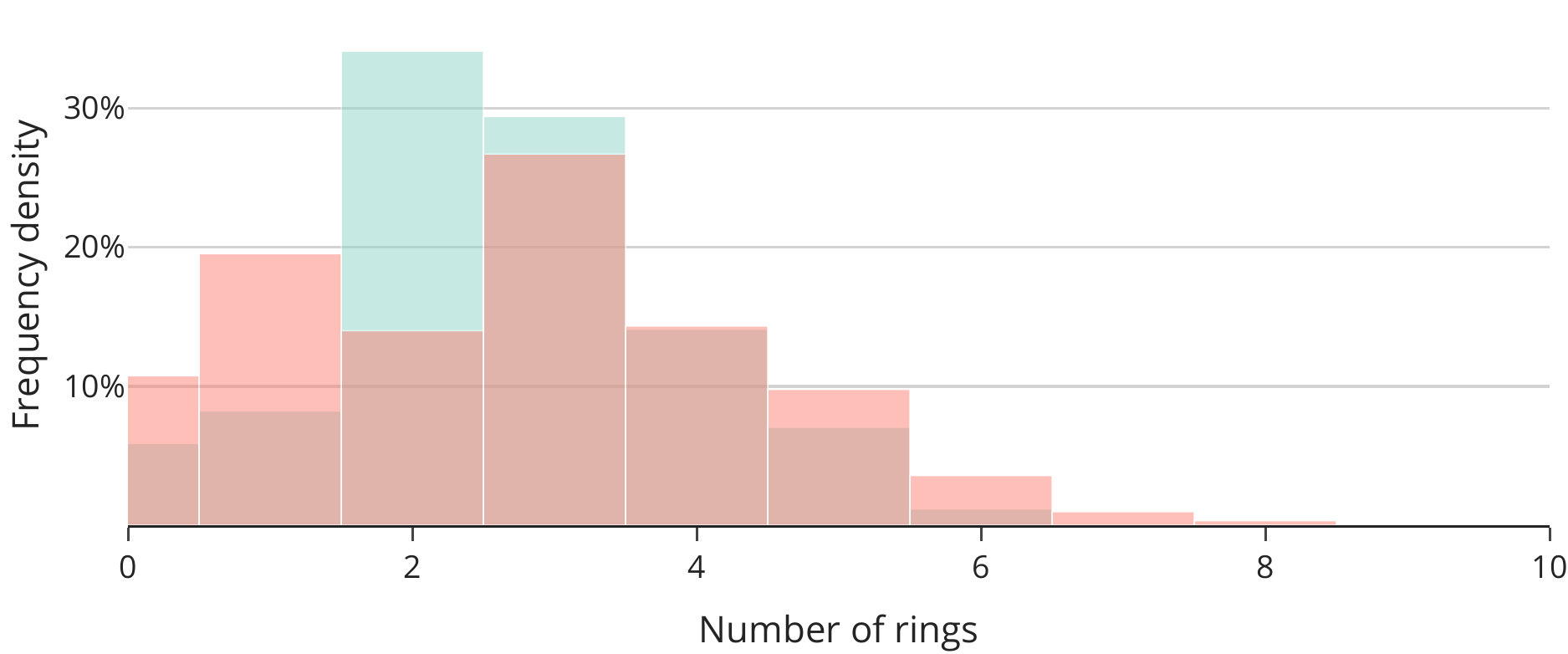}

\caption{Comparison of the 85 ligands in the Astex Diverse \added{set} and the \replaced{308}{428} ligands in the PoseBuster\added{s Benchmark} \deleted{data} set in terms of molecular weight, number of heavy atoms, number of rotatable bonds, and number of rings.}
\label{fig:ligand_comparison}
\end{figure}

\clearpage
\section{Data sets}

The following sections list the protein database\cite{berman2000protein} (PDB) codes and chemical component dictionary\cite{westbrook2015chemical} (CCD) codes for the protein-ligand complexes and the corresponding ligands of interest for the two data sets used.

\subsection*{Astex Diverse set}
1G9V~RQ3, 1GKC~NFH, 1GM8~SOX, 1GPK~HUP, 1HNN~SKF, 1HP0~AD3, 1HQ2~PH2, 1HVY~D16, 1HWI~115, 1HWW~SWA, 1IA1~TQ3, 1IG3~VIB, 1J3J~CP6, 1JD0~AZM, 1JJE~BYS, 1JLA~TNK, 1K3U~IAD, 1KE5~LS1, 1KZK~JE2, 1L2S~STC, 1L7F~BCZ, 1LPZ~CMB, 1LRH~NLA, 1M2Z~DEX, 1MEH~MOA, 1MMV~3AR, 1MZC~BNE, 1N1M~A3M, 1N2J~PAF, 1N2V~BDI, 1N46~PFA, 1NAV~IH5, 1OF1~SCT, 1OF6~DTY, 1OPK~P16, 1OQ5~CEL, 1OWE~675, 1OYT~FSN, 1P2Y~NCT, 1P62~GEO, 1PMN~984, 1Q1G~MTI, 1Q41~IXM, 1Q4G~BFL, 1R1H~BIR, 1R55~097, 1R58~AO5, 1R9O~FLP, 1S19~MC9, 1S3V~TQD, 1SG0~STL, 1SJ0~E4D, 1SQ5~PAU, 1SQN~NDR, 1T40~ID5, 1T46~STI, 1T9B~1CS, 1TOW~CRZ, 1TT1~KAI, 1TZ8~DES, 1U1C~BAU, 1U4D~DBQ, 1UML~FR4, 1UNL~RRC, 1UOU~CMU, 1V0P~PVB, 1V48~HA1, 1V4S~MRK, 1VCJ~IBA, 1W1P~GIO, 1W2G~THM, 1X8X~TYR, 1XM6~5RM, 1XOQ~ROF, 1XOZ~CIA, 1Y6B~AAX, 1YGC~905, 1YQY~915, 1YV3~BIT, 1YVF~PH7, 1YWR~LI9, 1Z95~198, 2BM2~PM2, 2BR1~PFP, 2BSM~BSM

\subsection*{PoseBusters Benchmark set}
\deleted{5S8I~2LY,} 5SAK~ZRY, 5SB2~1K2, 5SD5~HWI, 5SIS~JSM, 6M2B~EZO, 6M73~FNR, 6T88~MWQ, 6TW5~9M2, 6TW7~NZB, \deleted{6VS3~R6V,} 6VTA~AKN, \deleted{6W59~SZD,} 6WTN~RXT, \deleted{6X8D~ARA,} \deleted{6XAF~GDP,} 6XBO~5MC, 6XCT~478, 6XG5~TOP, 6XHT~V2V, 6XM9~V55, \deleted{6XUM~30L,} \deleted{6Y7L~QMG,} \deleted{6YDY~K73,} 6YJA~2BA, 6YMS~OZH, 6YQV~8K2, 6YQW~82I, 6YR2~T1C, 6YRV~PJ8, 6YSP~PAL, 6YT6~PKE, 6YYO~Q1K, 6Z0R~Q4H, 6Z14~Q4Z, 6Z1C~7EY, 6Z2C~Q5E, 6Z4N~Q7B, \deleted{6Z5Z~BDF,} 6ZAE~ACV, 6ZC3~JOR, 6ZCY~QF8, 6ZK5~IMH, 6ZPB~3D1, \deleted{6ZR8~QOZ,} \deleted{6ZT2~QPK,} \deleted{6ZX3~QRZ,} \deleted{6ZXQ~IMO,} 7A1P~QW2, 7A9E~R4W, 7A9H~TPP, \deleted{7AA0~R6B,} 7AFX~R9K, 7AKL~RK5, \deleted{7AMC~73B,} 7AN5~RDH, \deleted{7AS1~21G,} \deleted{7AVI~S2Q,} \deleted{7B0E~C2E,} 7B2C~TP7, 7B94~ANP, \deleted{7BA0~T5H,} 7BCP~GCO, \deleted{7BHX~TO5,} \deleted{7BJ6~TVK,} 7BJJ~TVW, 7BKA~4JC, \deleted{7BLA~WCS,} \deleted{7BLG~GAL,} 7BMI~U4B, 7BNH~BEZ, 7BTT~F8R, 7C0U~FGO, 7C3U~AZG, \deleted{7C6P~SQH,} 7C8Q~DSG, 7CD9~FVR, 7CIJ~G0C, 7CL8~TES, 7CNQ~G8X, 7CNS~PMV, 7CTM~BDP, 7CUO~PHB, \deleted{7D0P~1VU,} 7D5C~GV6, 7D6O~MTE, \deleted{7D8Q~GZF,} \deleted{7D9L~GSF,} \deleted{7DIN~MPO,} 7DKT~GLF, 7DQL~4CL, 7DUA~HJ0, \deleted{7E2S~BLA,} 7E4L~MDN, 7EBG~J0L, 7ECR~SIN, 7ED2~A3P, 7ELT~TYM, \deleted{7EN7~J79,} 7EPV~FDA, 7ES1~UDP, 7F51~BA7, 7F5D~EUO, 7F8T~FAD, 7FB7~8NF, 7FHA~ADX, 7FRX~O88, 7FT9~4MB, 7JG0~GAR, \deleted{7JGW~V9S,} 7JHQ~VAJ, 7JMV~4NC, \deleted{7JNB~A2G,} \deleted{7JR8~VH7,} \deleted{7JUD~MMA,} 7JXX~VP7, 7JY3~VUD, 7K0V~VQP, \deleted{7K41~VUA,} 7KB1~WBJ, 7KC5~BJZ, \deleted{7KFO~IAC,} \deleted{7KLX~WOV,} 7KM8~WPD, \deleted{7KP6~WTP,} 7KQU~YOF, 7KRU~ATP, 7KZ9~XN7, 7L00~XCJ, 7L03~F9F, 7L5F~XNG, \deleted{7L6D~BMF,} 7L7C~XQ1, \deleted{7L81~UD4,} \deleted{7LB3~XXS,} 7LCU~XTA, 7LEV~0JO, 7LJN~GTP, 7LMO~NYO, 7LOE~Y84, 7LOU~IFM, 7LT0~ONJ, 7LZD~YHY, \deleted{7LZQ~YJV,} 7M31~TDR, 7M3H~YPV, \deleted{7M41~YQG,} 7M6K~YRJ, \deleted{7MAE~XUS,} \deleted{7MEU~MGP,} 7MFP~Z7P, 7MGT~ZD4, 7MGY~ZD1, 7MMH~ZJY, 7MOI~HPS, \deleted{7MRH~ZMJ,} \deleted{7MS7~ZQ1,} 7MSR~DCA, 7MWN~WI5, 7MWU~ZPM, 7MY1~IPE, 7MYU~ZR7, \deleted{7MZS~GLA,} 7N03~ZRP, 7N4N~0BK, 7N4W~P4V, 7N6F~0I1, 7N7B~T3F, 7N7H~CTP, \deleted{7NA4~1I9,} \deleted{7NB4~U6Q,} 7NF0~BYN, 7NF3~4LU, 7NFB~GEN, 7NGW~UAW, \deleted{7NLK~UHK,} 7NLV~UJE, \deleted{7NML~I7B,} 7NP6~UK8, 7NPL~UKZ, \deleted{7NR6~UO8,} 7NR8~UOE, 7NSW~HC4, \deleted{7NTG~F6R,} 7NU0~DCL, 7NUT~GLP, 7NXO~UU8, 7O0N~CDP, 7O1T~5X8, \deleted{7OCB~V88,} \deleted{7ODX~DGP,} 7ODY~DGI, 7OEO~V9Z, 7OFF~VCB, 7OFK~VCH, \deleted{7OKC~VFE,} \deleted{7OKF~VH5,} 7OLI~8HG, \deleted{7OLT~58J,} \deleted{7OMJ~GCP,} 7OMX~CNA, 7OP9~06K, 7OPG~06N, \deleted{7ORW~7WA,} 7OSO~0V1, \deleted{7OU8~1XI,} 7OZ9~NGK, 7OZC~G6S, 7P1F~KFN, 7P1M~4IU, 7P2I~MFU, \deleted{7P2W~4QR,} 7P4C~5OV, \deleted{7P4J~5JK,} \deleted{7P4V~DAT,} 7P5T~5YG, \deleted{7P85~5ZG,} \deleted{7PA4~C,} 7PGX~FMN, 7PIH~7QW, 7PJQ~OWH, 7PK0~BYC, 7PL1~SFG, 7POM~7VZ, 7PRI~7TI, 7PRM~81I, 7PT3~3KK, 7PUV~84Z, \deleted{7Q19~DSM,} 7Q25~8J9, 7Q27~8KC, 7Q2B~M6H, 7Q5I~I0F, 7QE4~NGA, 7QF4~RBF, 7QFM~AY3, 7QGP~DJ8, 7QHG~T3B, 7QHL~D5P, \deleted{7QK0~EBL,} 7QPP~VDX, \deleted{7QSW~CAP,} 7QTA~URI, 7R3D~APR, 7R59~I5F, 7R6J~2I7, 7R7R~AWJ, 7R9N~F97, 7RC3~SAH, \deleted{7REE~4LY,} 7RH3~59O, \deleted{7RH8~UTP,} 7RKW~5TV, 7RNI~60I, 7ROR~69X, 7ROU~66I, \deleted{7RPZ~6IC,} 7RSV~7IQ, \deleted{7RUI~7QZ,} \deleted{7RWO~7WN,} 7RWS~4UR, 7RZL~NPO, \deleted{7S45~ACO,} \deleted{7S9H~7PP,} 7SCW~GSP, 7SDD~4IP, \deleted{7SED~8VD,} 7SFO~98L, \deleted{7SGV~L30,} 7SIU~9ID, \deleted{7SNE~9XR,} \deleted{7SSM~B7L,} 7SUC~COM, 7SZA~DUI, 7T0D~FPP, \deleted{7T0U~E3I,} 7T1D~E7K, \deleted{7T2I~E9F,} 7T3E~SLB, \deleted{7T3F~EM0,} \deleted{7T9O~GEI,} 7TB0~UD1, 7TBU~S3P, 7TE8~P0T, 7TH4~FFO, 7THI~PGA, 7TM6~GPJ, 7TOM~5AD, 7TS6~KMI, 7TSF~H4B, 7TUO~KL9, \deleted{7TWC~CXS,} 7TXK~LW8, \deleted{7TXP~0FX,} 7TYP~KUR, 7U0U~FK5, 7U3J~L6U, 7UAS~MBU, 7UAW~MF6, \deleted{7UEY~N0R,} \deleted{7UF2~5SP,} 7UJ4~OQ4, 7UJ5~DGL, 7UJF~R3V, 7ULC~56B, \deleted{7UMV~NUU,} 7UMW~NAD, \deleted{7UP3~NZ0,} 7UQ3~O2U, 7USH~82V, 7UTW~NAI, 7UXS~OJC, 7UY4~SMI, 7UYB~OK0, 7V14~ORU, 7V3N~AKG, 7V3S~5I9, 7V43~C4O, \deleted{7V8Z~5YH,} 7VB8~STL, 7VBU~6I4, 7VC5~9SF, \deleted{7VJT~7IJ,} 7VKZ~NOJ, 7VQ9~ISY, 7VWF~K55, 7VYJ~CA0, 7W05~GMP, 7W06~ITN, \deleted{7W6F~8I6,} 7WCF~ACP, 7WDT~NGS, 7WJB~BGC, 7WKL~CAQ, 7WL4~JFU, \deleted{7WN5~JGL,} 7WPW~F15, 7WQQ~5Z6, 7WUX~6OI, 7WUY~76N, 7WY1~D0L, 7X5N~5M5, 7X9K~8OG, 7XBV~APC, \deleted{7XEK~9YX,} 7XFA~D9J, 7XG5~PLP, 7XI7~4RI, \deleted{7XIJ~EJ3,} 7XJN~NSD, 7XPO~UPG, 7XQZ~FPF, 7XRL~FWK, 7YZU~DO7, 7Z1Q~NIO, 7Z2O~IAJ, 7Z7F~IF3, 7ZCC~OGA, 7ZDY~6MJ, 7ZF0~DHR, 7ZHP~IQY, 7ZL5~IWE, 7ZOC~T8E, 7ZTL~BCN, 7ZU2~DHT, 7ZXV~45D, \deleted{7ZXZ~K9R,} \deleted{7ZYS~KNR,} \deleted{7ZZB~KGX,} 7ZZW~KKW, 8A1H~DLZ, 8A2D~KXY, 8AAU~LH0, \deleted{8ACL~LQL,} 8AEM~LVF, \deleted{8AEU~M0L,} 8AIE~M7L, \deleted{8AIJ~M9I,} \deleted{8AJX~FUM,} 8AP0~PRP, 8AQL~PLG, 8AUH~L9I, 8AY3~OE3, 8B8H~OJQ, \deleted{8BN6~R53,} 8BOM~QU6, \deleted{8BPL~CP,} \deleted{8BRO~R7E,} 8BTI~RFO, 8C3N~ADP, \deleted{8C5D~GTB,} 8C5M~MTA, \deleted{8C7Y~TXV,} \deleted{8CGC~LMR,} \deleted{8CI0~8EL,} 8CNH~V6U, 8CSD~C5P, 8D19~GSH, 8D39~QDB, 8D5D~5DK, 8DHG~T78, 8DKO~TFB, 8DP2~UMA, 8DSC~NCA, \deleted{8DW5~FQ7,} \deleted{8DZT~G4P,} \deleted{8E77~ULP,} 8EAB~VN2, \deleted{8EAD~UY0,} \deleted{8ERS~WQO,} 8EX2~Q2Q, 8EXL~799, 8EYE~X4I, 8F4J~PHO, 8F8E~XJI, 8FAV~4Y5, \deleted{8FLN~Y7W,} 8FLV~ZB9, 8FO5~Y4U, \deleted{8FV9~80J,} 8G0V~YHT, \deleted{8G43~ZU6,} 8G6P~API, 8GFD~ZHR, \deleted{8H0M~2EH,} 8HFN~XGC, 8HO0~3ZI, 8SLG~G5A

\section{Energy minimisation example}
\begin{figure}[H]
\centering

\includegraphics[width=0.70\textwidth,trim={2cm 0.7cm 2cm 1.5cm},clip]{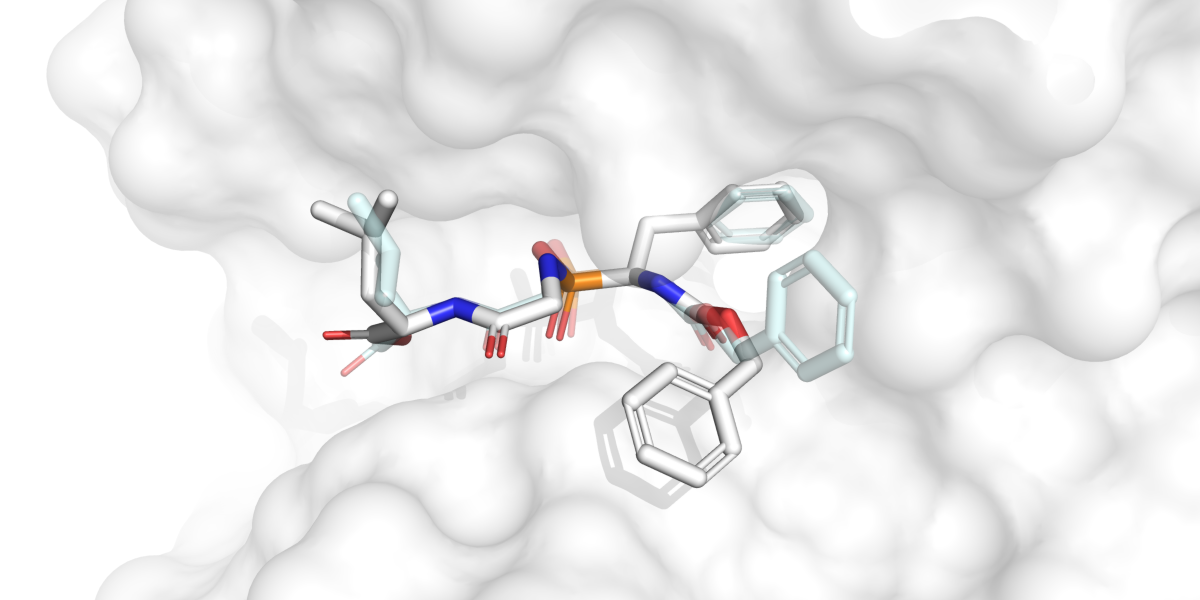}

\includegraphics[width=0.70\textwidth,trim={2cm 0.7cm 2cm 1.5cm},clip]{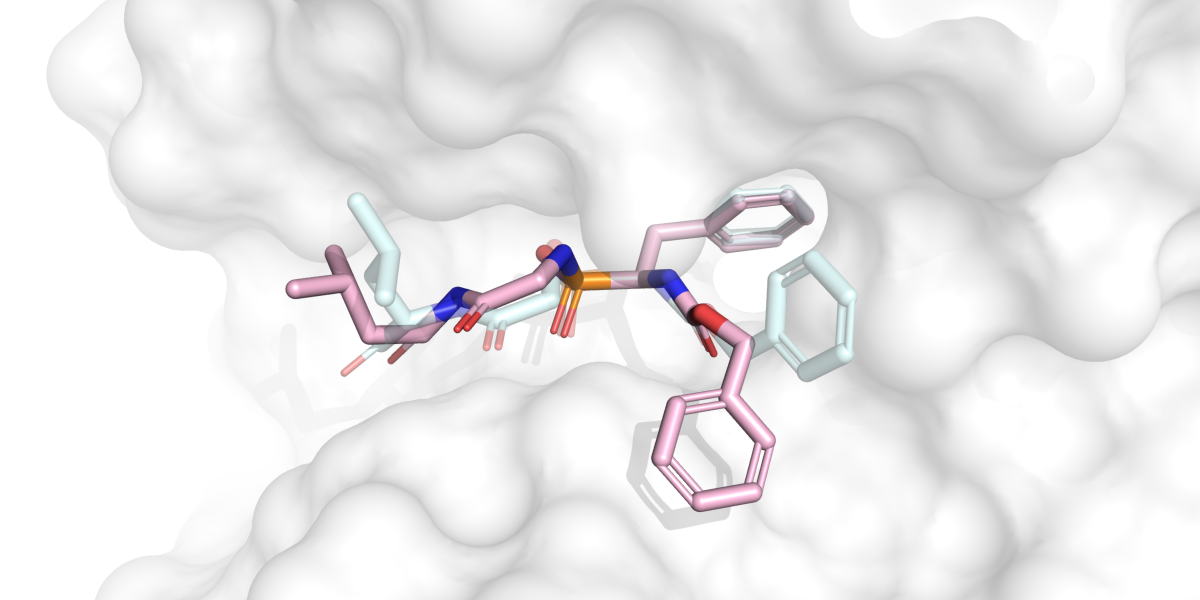}

\caption{\added{Example of a prediction that was ‘destroyed’ by the energy minimisation. The prediction by Autodock Vina passes all PoseBusters checks and has a RMSD of \qty{1.9}{\angstrom} and is shown in white, the optimised predicted ligand has a RMSD of \qty{2.2}{\angstrom} and is shown in pink, and the crystal ligand is shown in light blue.
}}
\label{fig:em_example_destroyed}
\end{figure}

\section{Cofactor analysis}
\begin{figure}[H]
\centering
\includegraphics[width=0.7\columnwidth]{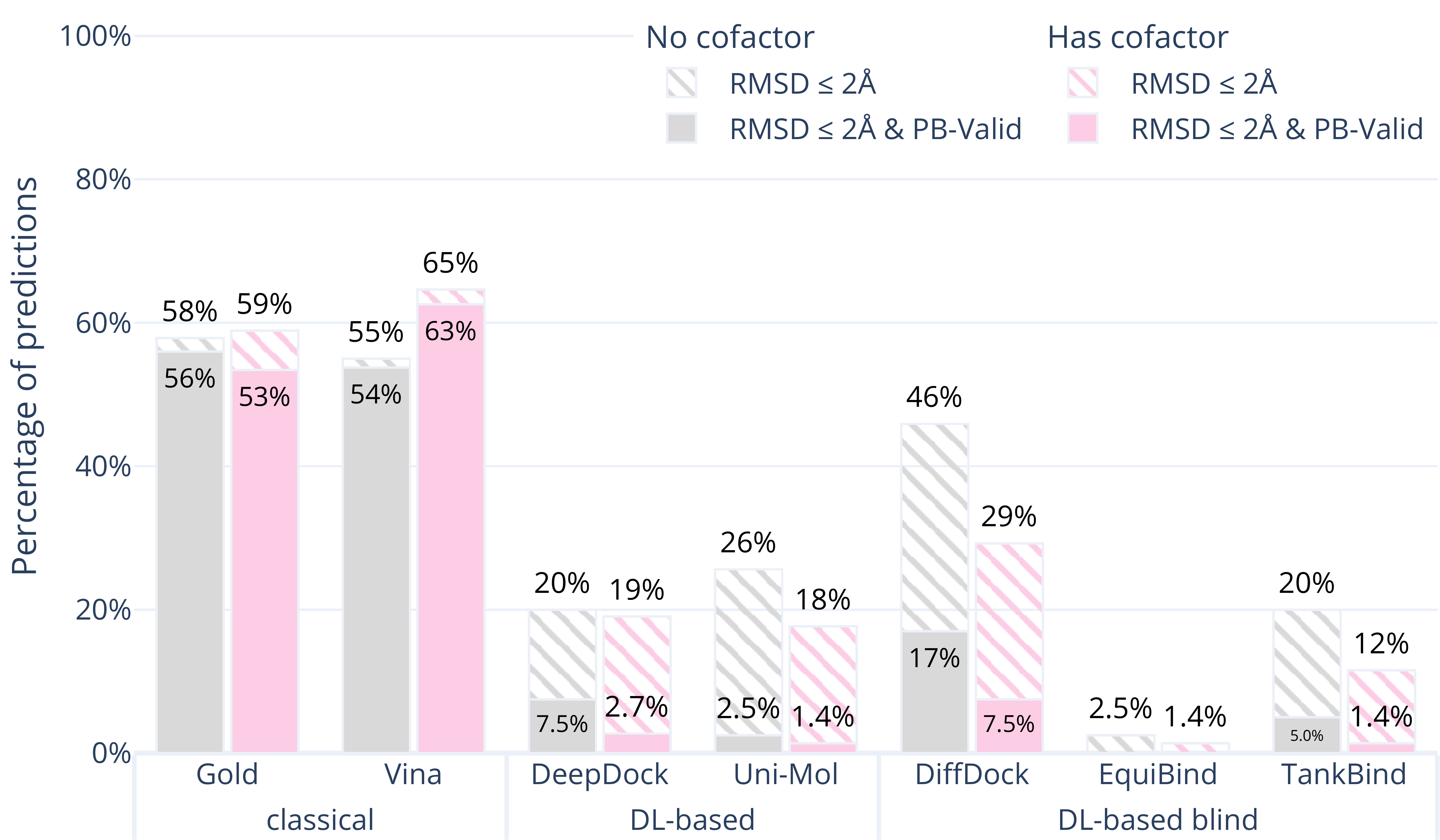}
\caption{
Comparative performance of docking methods on the PoseBusters Benchmark set stratified by the presence of cofactors. 
Cofactors are loosely defined as non-protein non-ligand compounds such as metal ions, iron-sulfur clusters, and organic small molecules that are within \qty{4.0}{\angstrom} of any ligand heavy atom. 
The striped bars show the share of predictions of each method that have an RMSD within \qty{2}{\angstrom} and the solid bars show those predictions which in addition pass all PoseBuster tests and are therefore PB-valid.
The classical docking methods perform better on those systems with cofactors present while the DL-based methods perform worse on those systems.
}
\label{fig:bars_cofactor}
\end{figure}
\begin{figure}[H]
\centering
\includegraphics[width=0.95\textwidth]{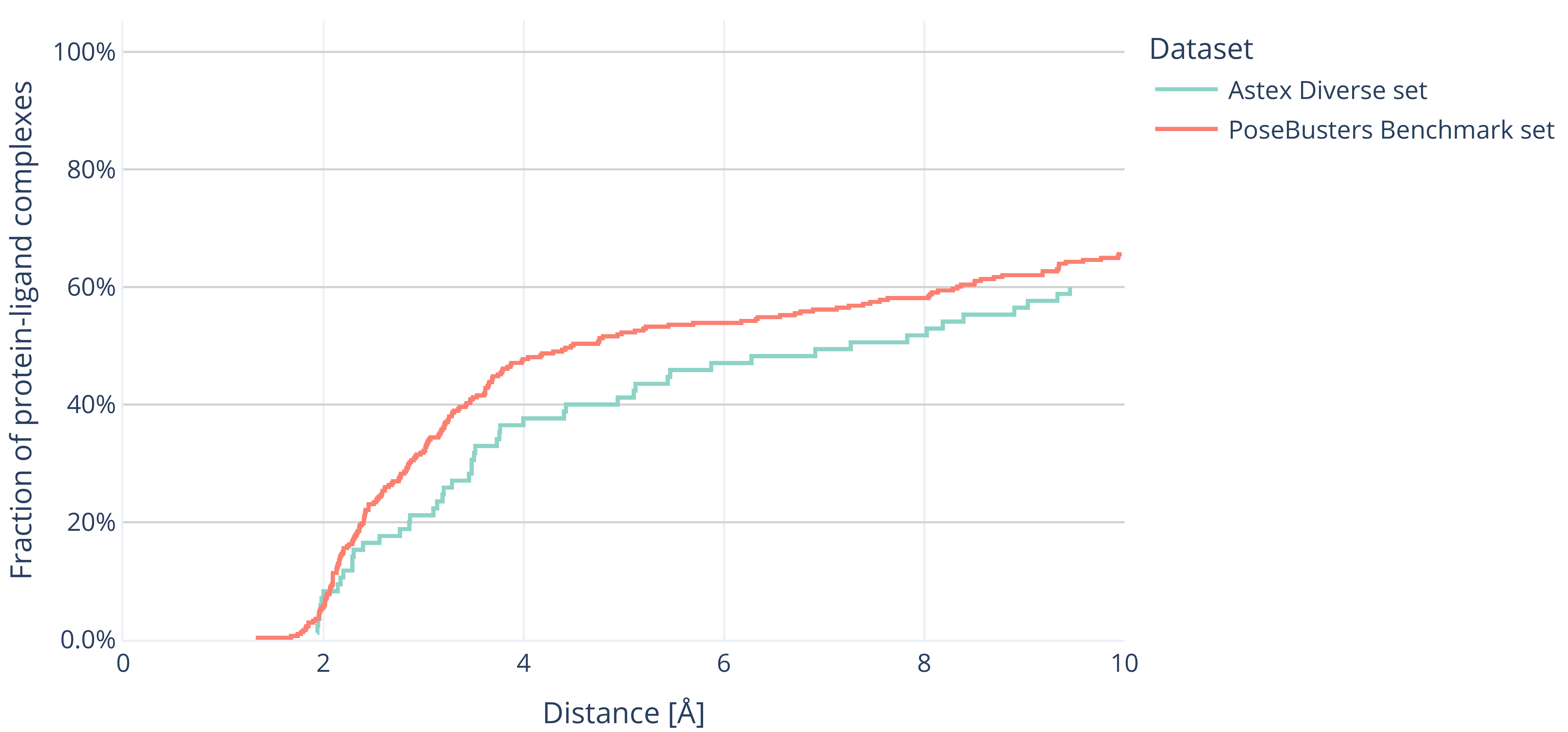}
\caption{
Fraction of protein-ligand complexes with a cofactor close to the ligand as a function of the distance threshold used. 
Here a cofactor is any other compound present in the crystal structure besides the ligand of interest, the protein and solvent. This includes metal ions, iron–sulfur clusters, and small organic molecules. 
46\% of the protein-ligand complexes in the PoseBuster\added{s Benchmark} set have a cofactor within \qty{4}{\angstrom} of the ligand. 
}
\label{fig:ecdf_cofactors}
\end{figure}

\section{Detailed results}

\newlength{\waterfallwidth}
\setlength{\waterfallwidth}{0.6\textwidth}
\begin{figure}[H]
\subfloat[Crystal Structures]{\includegraphics[width=0.49\textwidth]{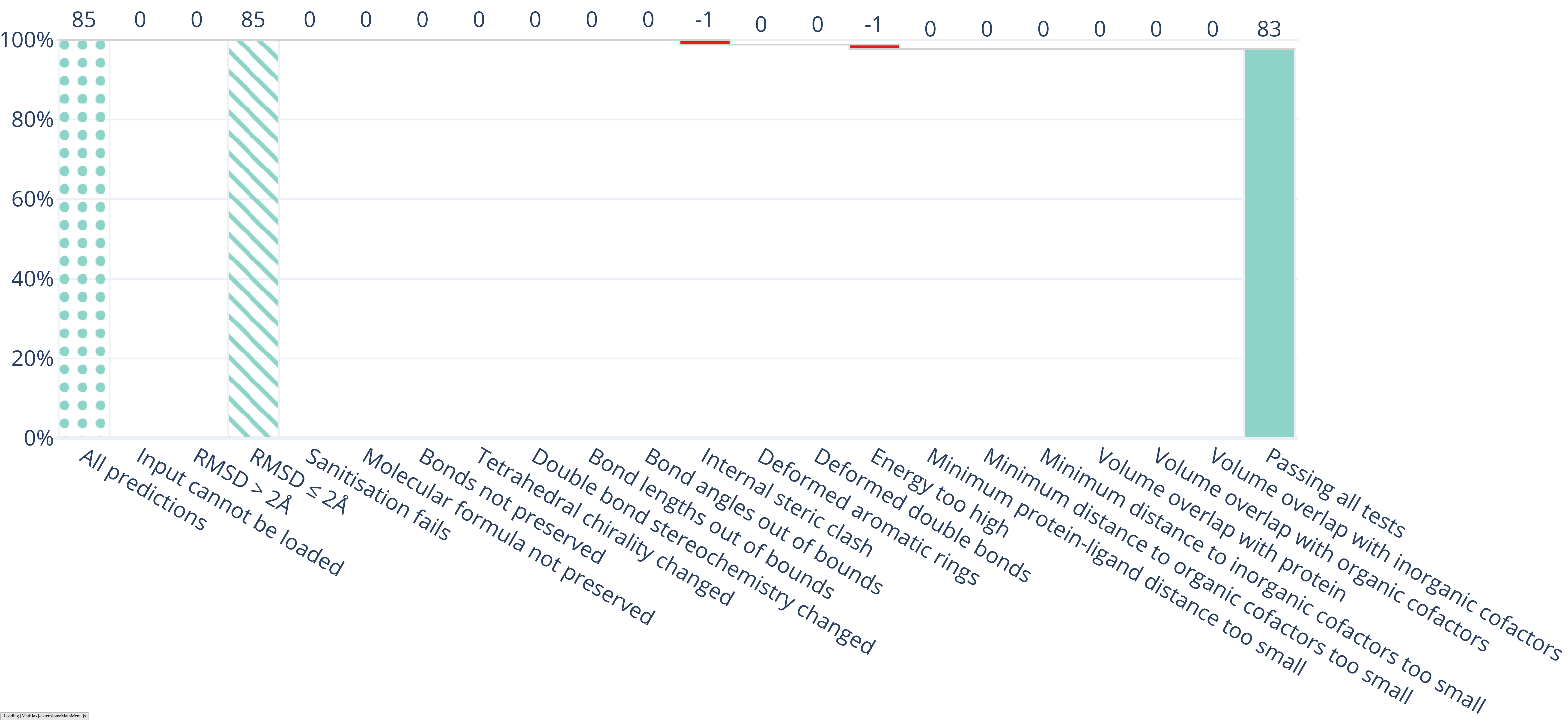}}
\hfill
\subfloat[AutoDock Vina]{\includegraphics[width=0.49\textwidth]{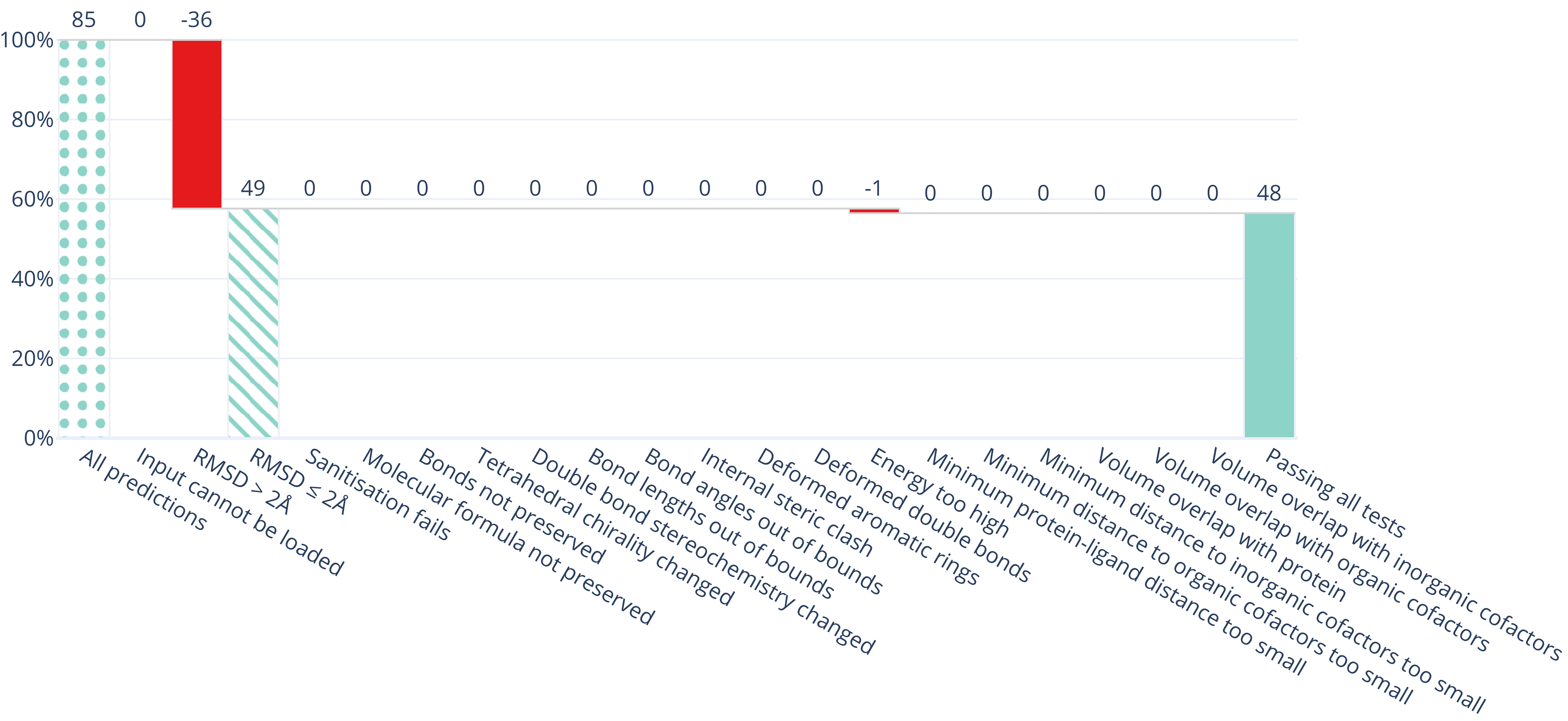}}
\vspace{\intextsep}
\subfloat[CCDC Gold]{\includegraphics[width=0.49\textwidth]{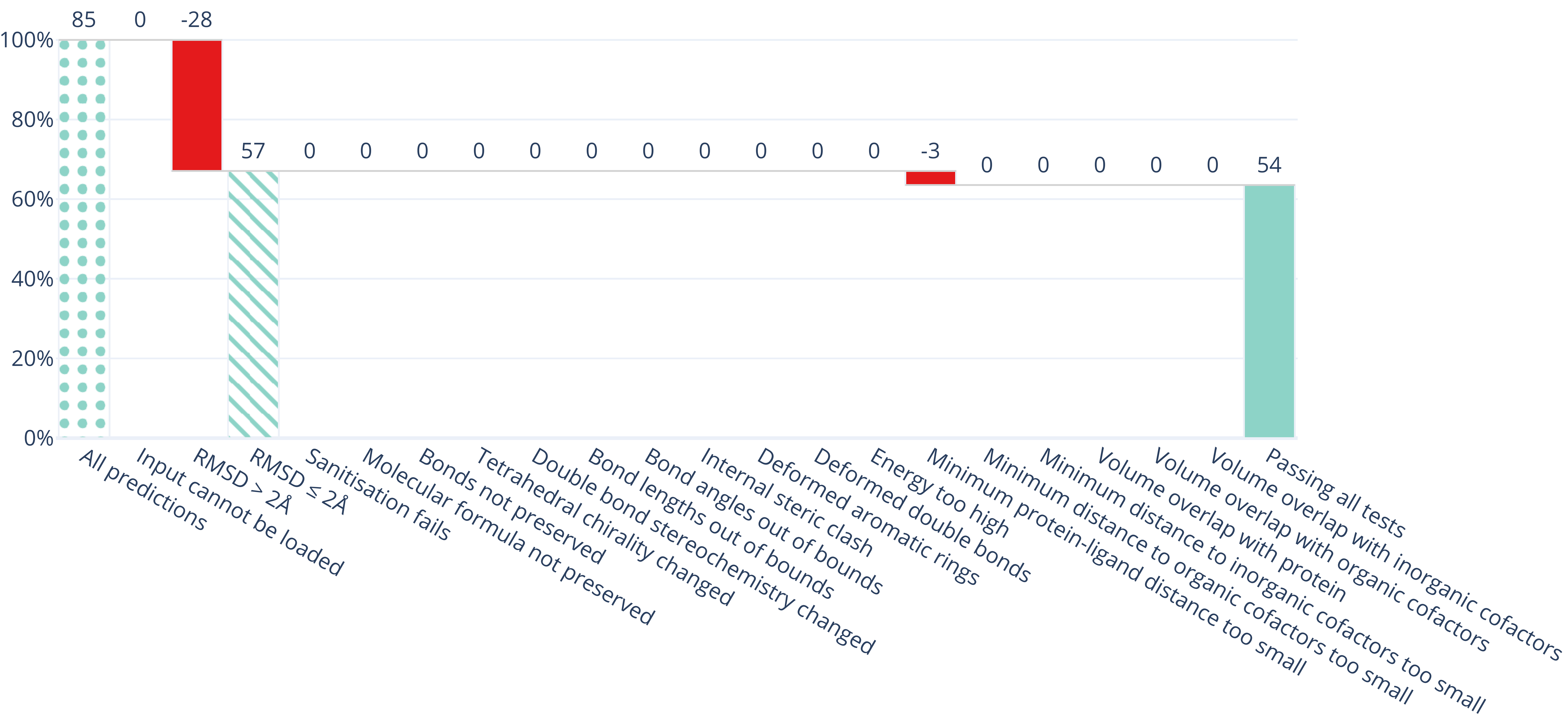}}
\hfill
\subfloat[DeepDock]{\includegraphics[width=0.49\textwidth]{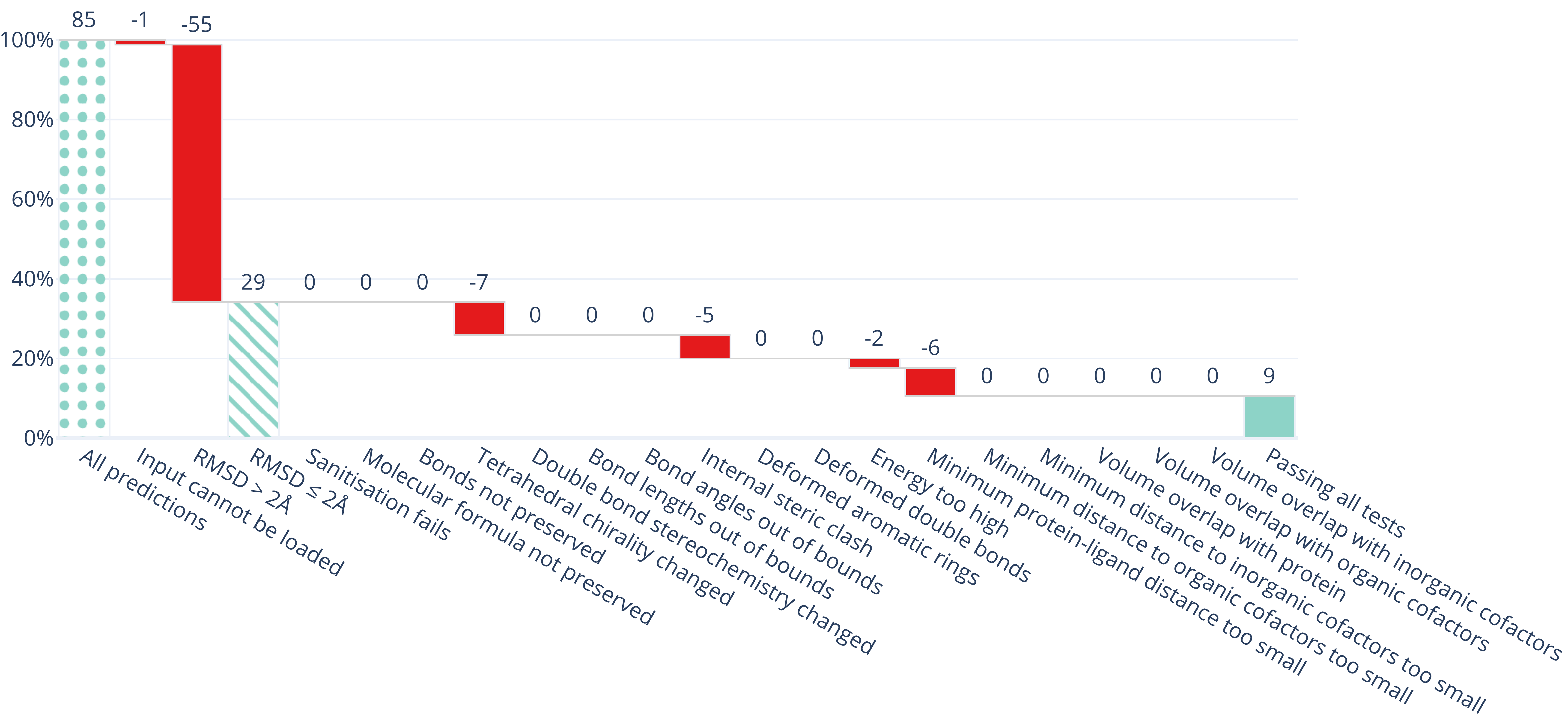}}
\vspace{\intextsep}
\subfloat[DiffDock]{\includegraphics[width=0.49\textwidth]{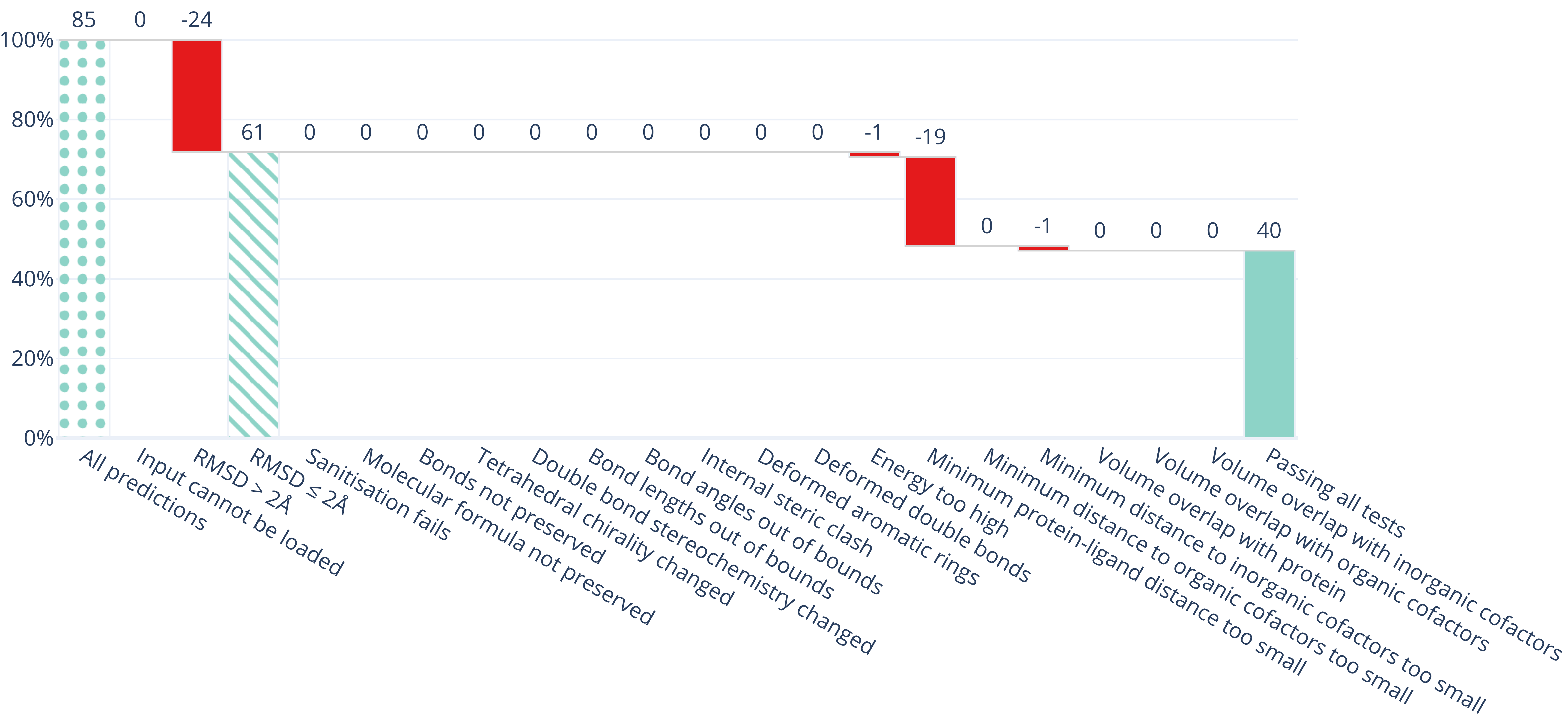}}
\hfill
\subfloat[EquiBind]{\includegraphics[width=0.49\textwidth]{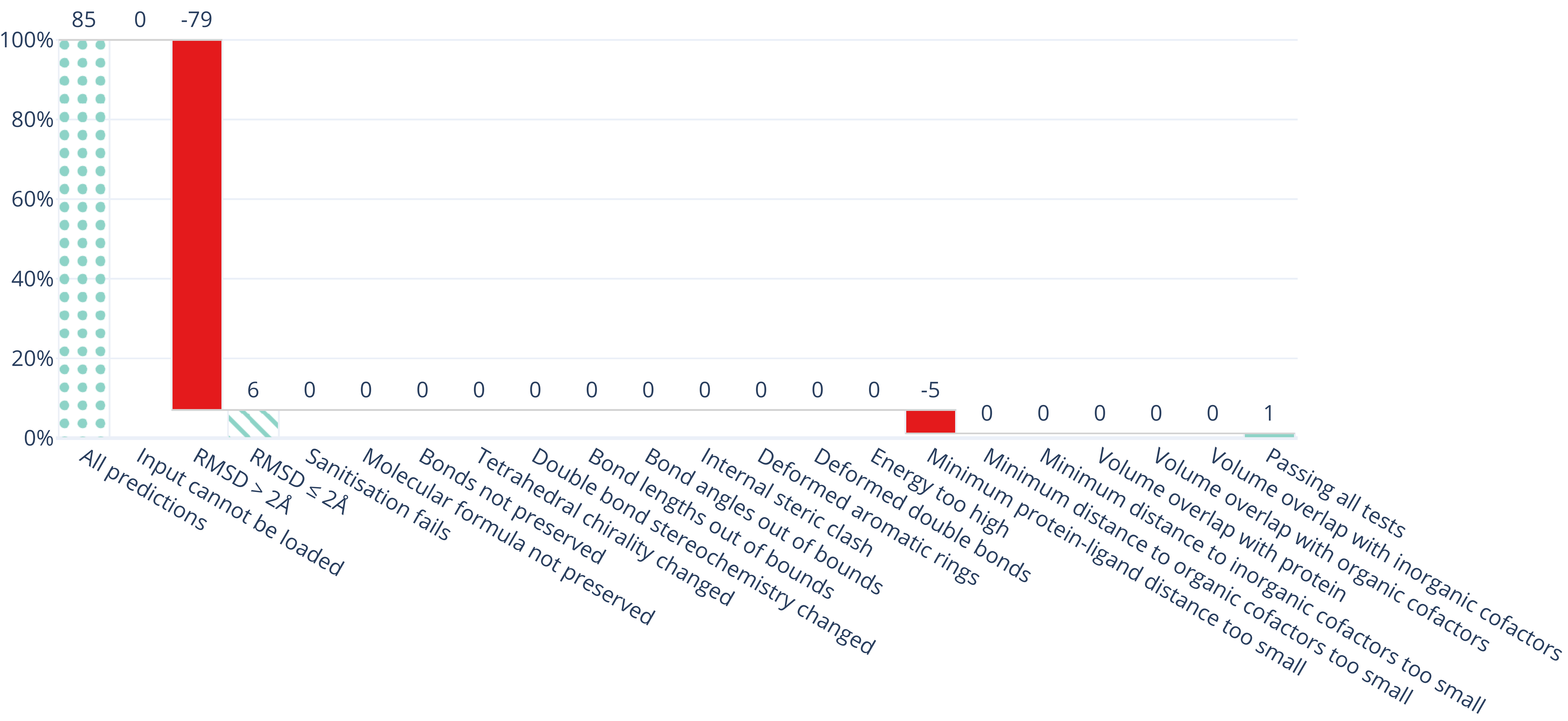}}
\vspace{\intextsep}
\subfloat[TankBind]{\includegraphics[width=0.49\textwidth]{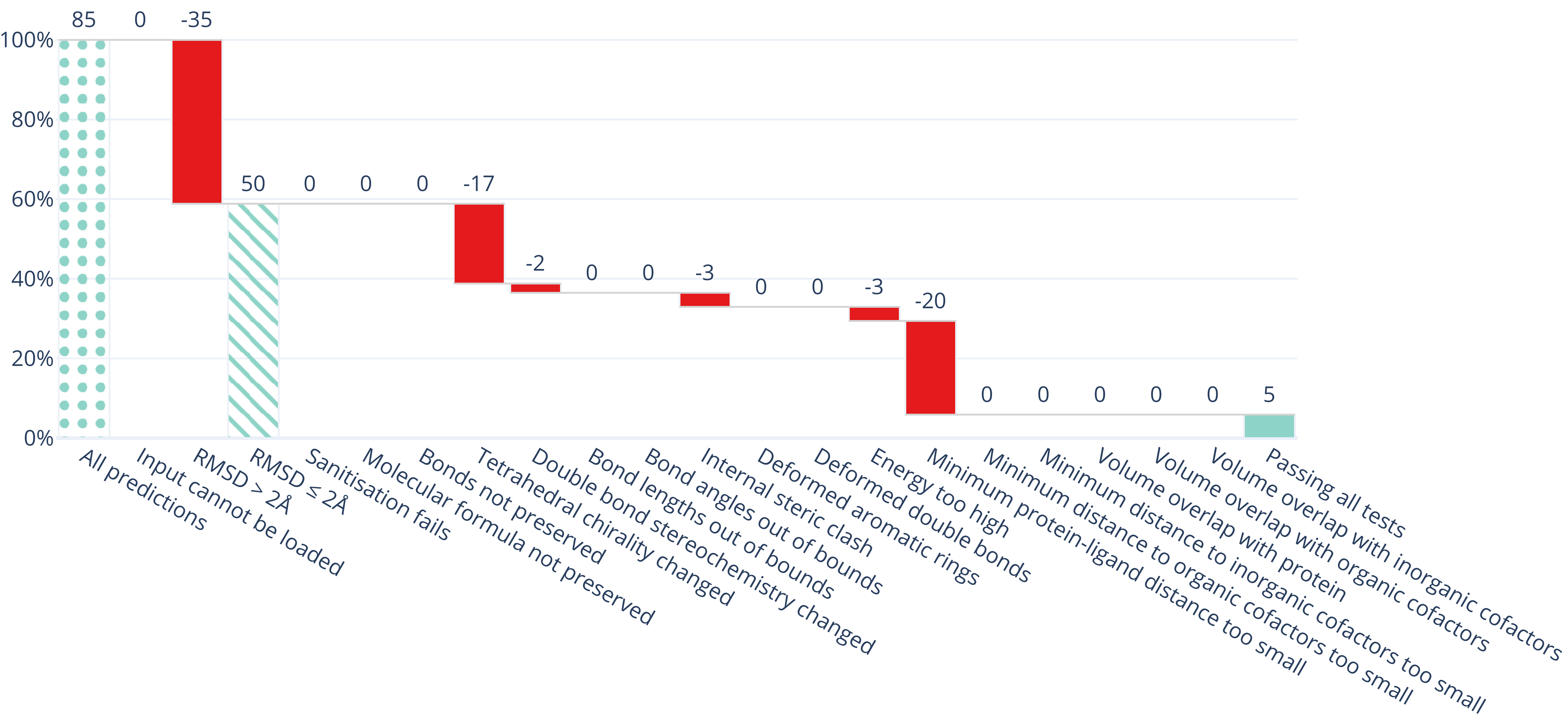}}
\hfill
\subfloat[Uni-Mol]{\includegraphics[width=0.49\textwidth]{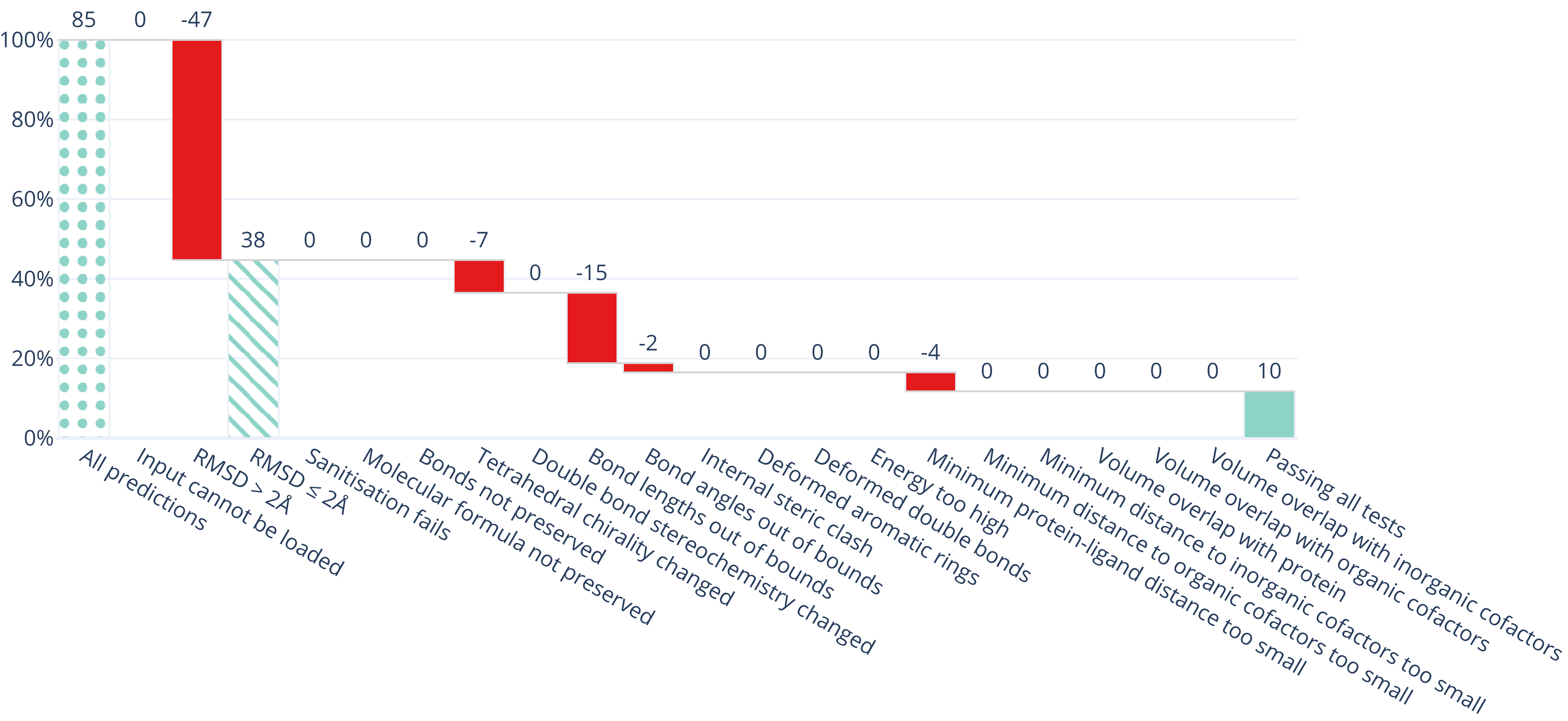}}
\caption{
Waterfall plots showing test results for the Astex Diverse \deleted{data}set. The leftmost (dotted) bars show the number of complexes in the test set. The red bars show the number of predictions that fail with each additional test going from left to right. The right most (solid) bar indicates the number of predictions that pass all tests. 
As a reading example, panel (a) shows that out of AutoDock Vina's 85 predictions 37 are not within \qty{2}{angstrom} RMSD and one additional prediction fails the energy ratio check so that overall 47 ligands have a low RMSD and pass all tests. AutoDock Vina and CCDC Gold pass the most tests.
}
\label{fig:waterfalls_astex}
\end{figure}

\begin{figure}[H]
\subfloat[Crystal Structures]{\includegraphics[width=0.49\textwidth]{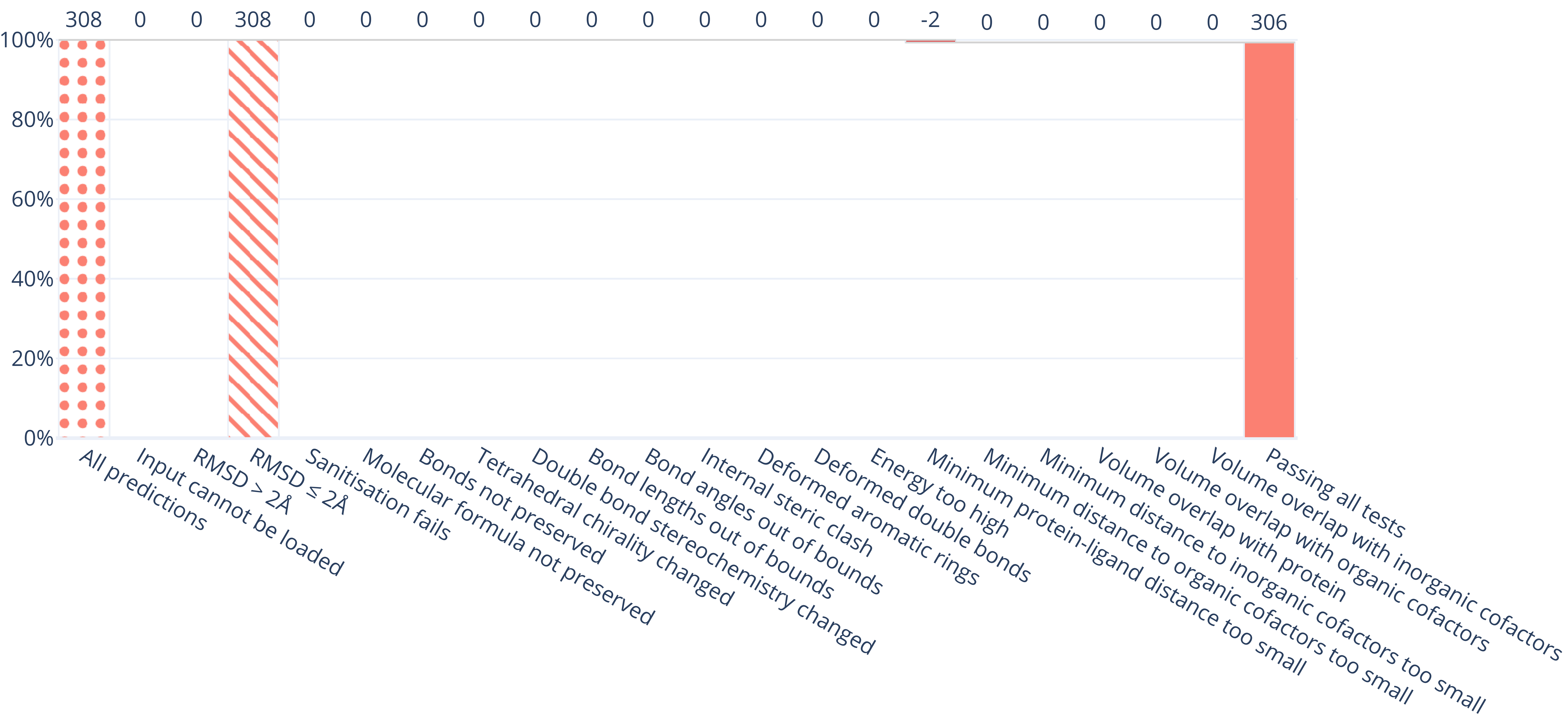}}
\hfill
\subfloat[AutoDock Vina]{\includegraphics[width=0.49\textwidth]{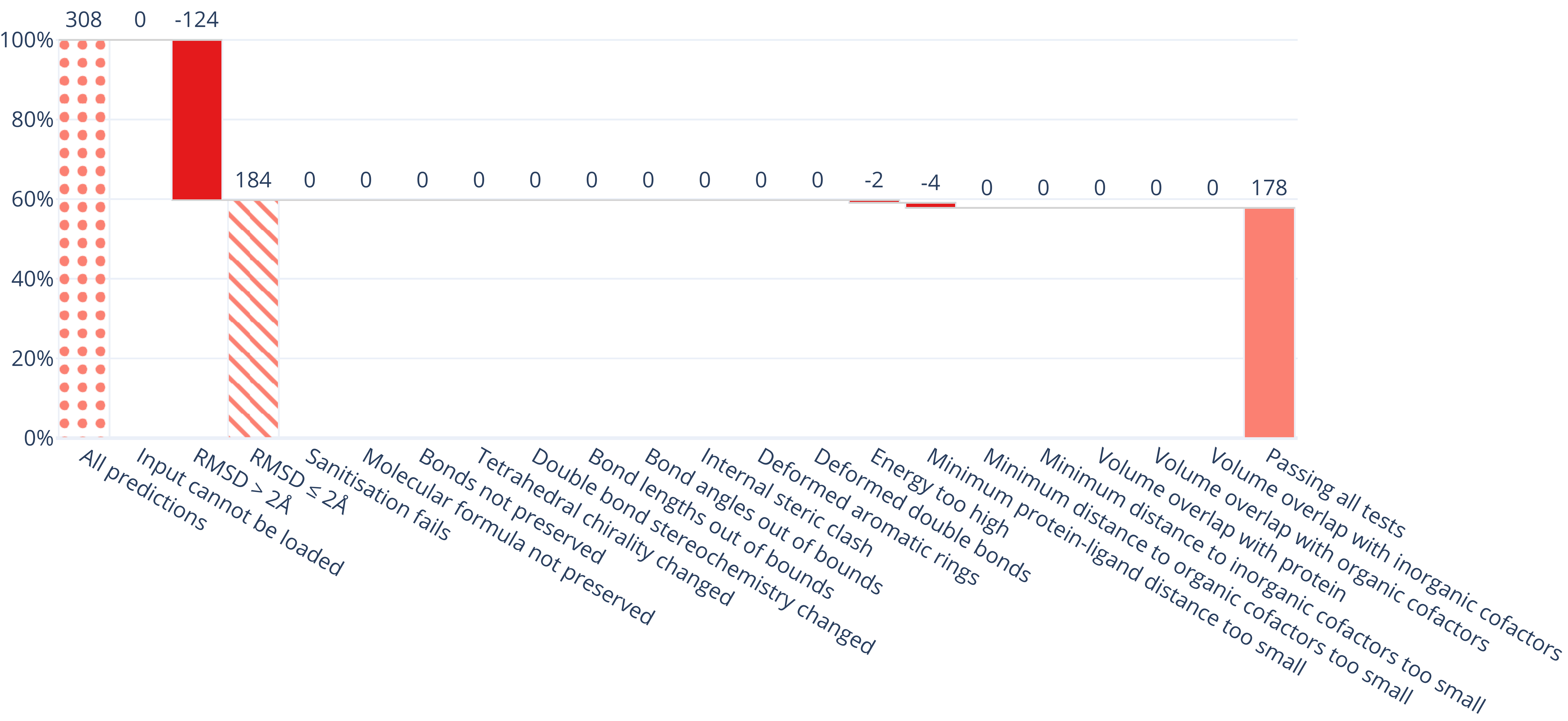}}
\vspace{\intextsep}
\subfloat[CCDC Gold]{\includegraphics[width=0.49\textwidth]{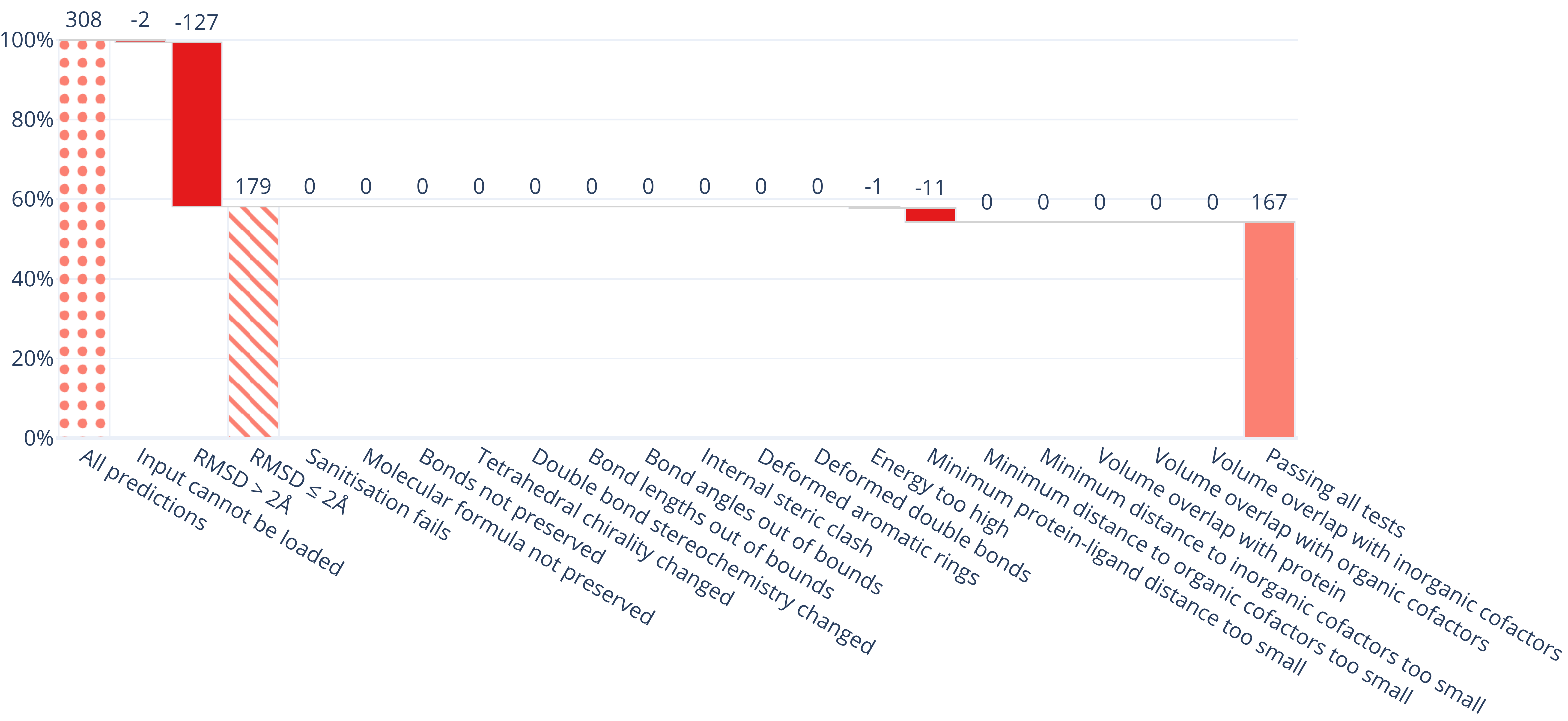}}
\hfill
\subfloat[DeepDock]{\includegraphics[width=0.49\textwidth]{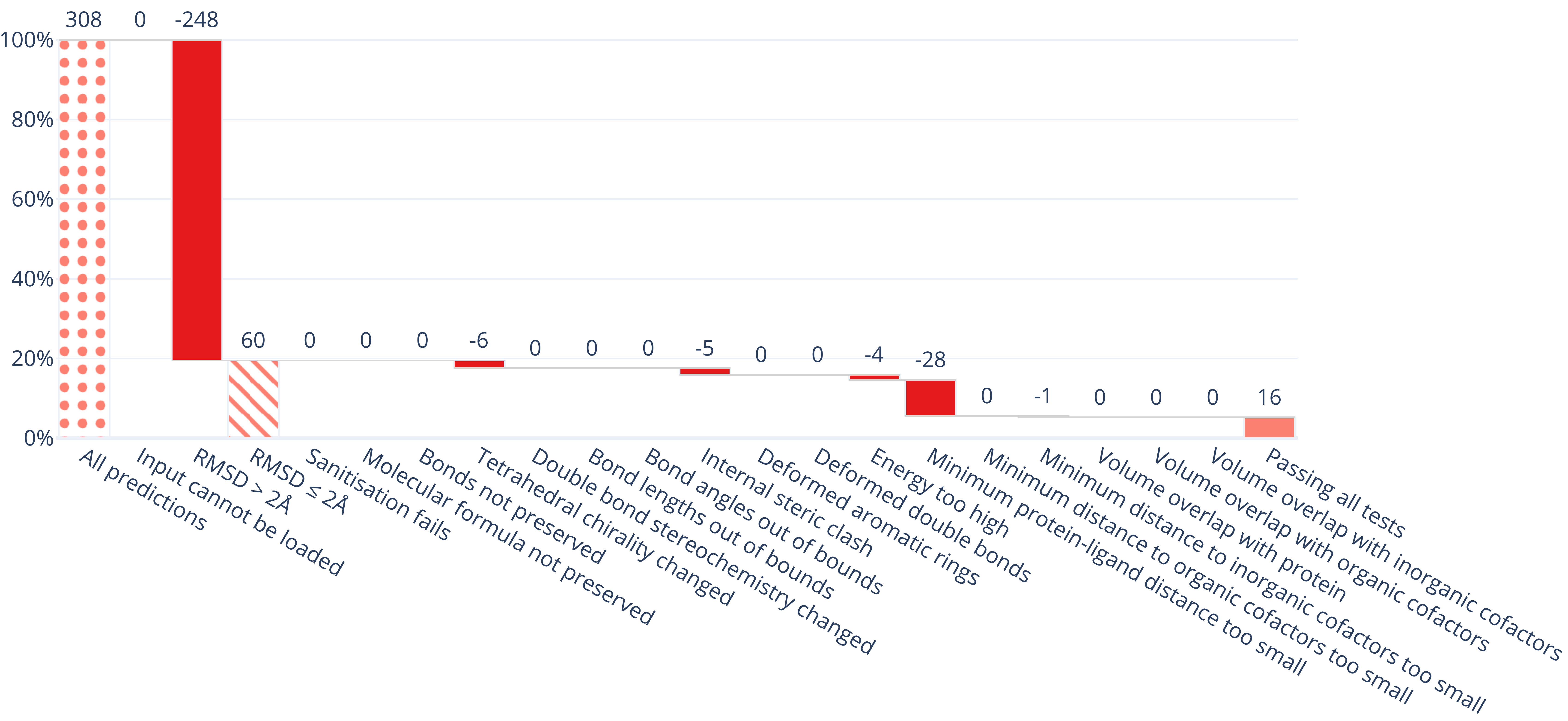}}
\vspace{\intextsep}
\subfloat[DiffDock]{\includegraphics[width=0.49\textwidth]{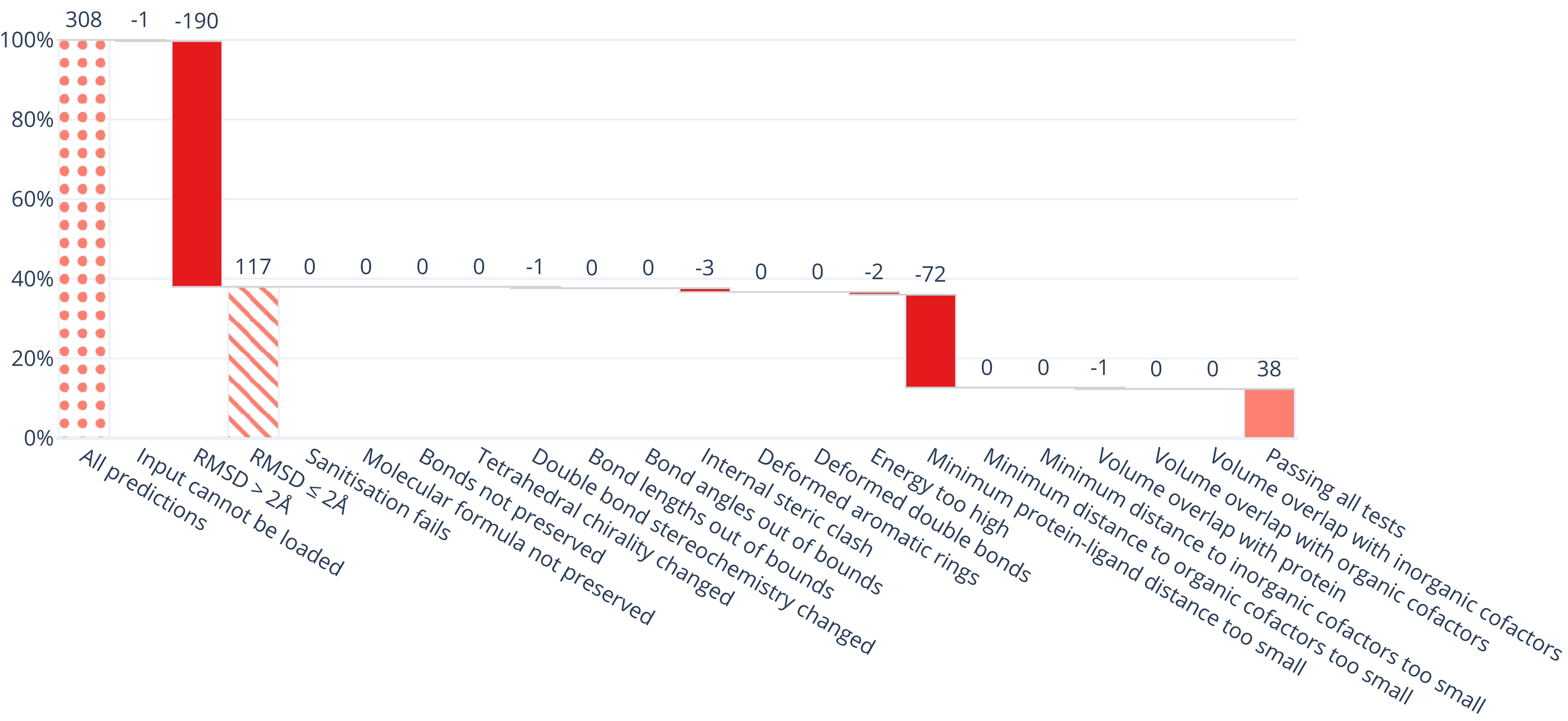}}
\hfill
\subfloat[EquiBind]{\includegraphics[width=0.49\textwidth]{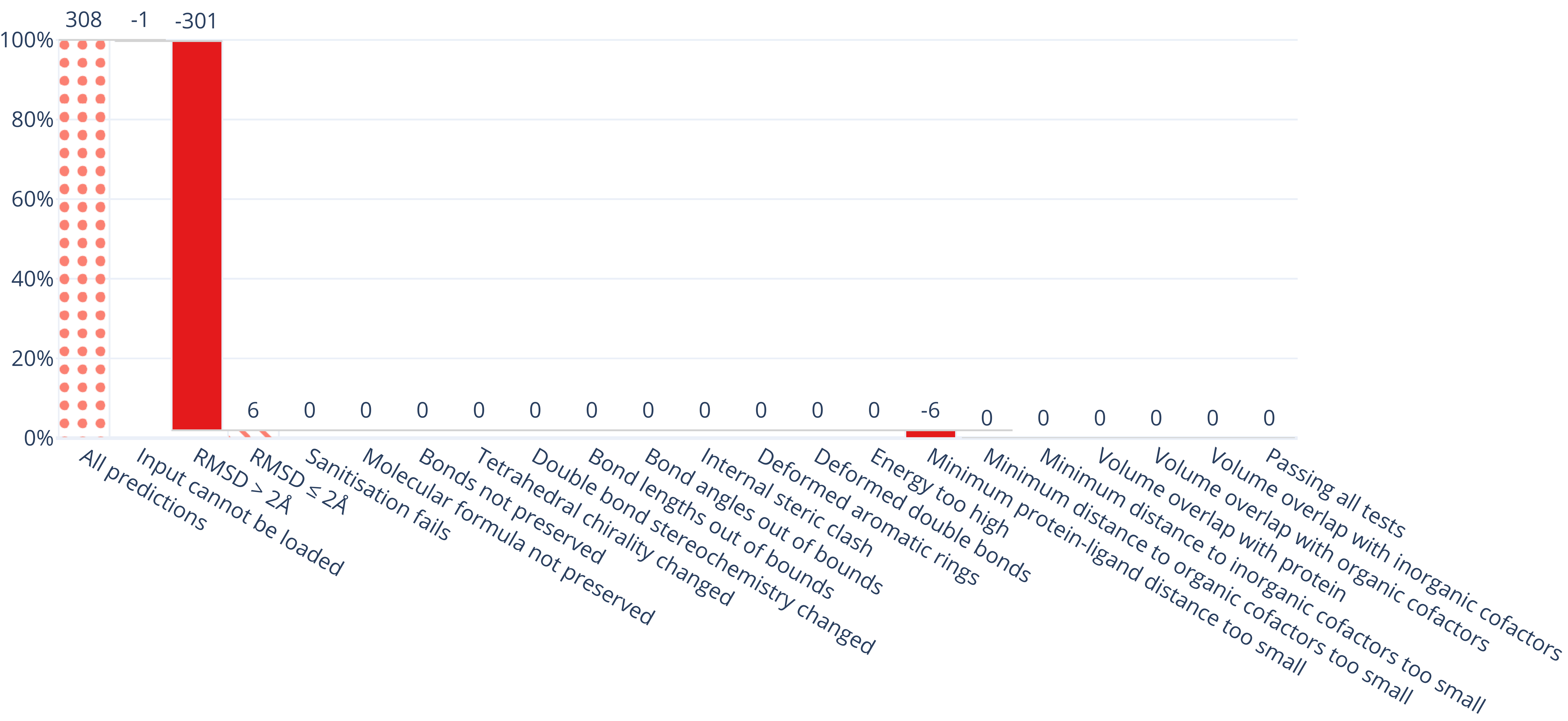}}
\vspace{\intextsep}
\subfloat[TankBind]{\includegraphics[width=0.49\textwidth]{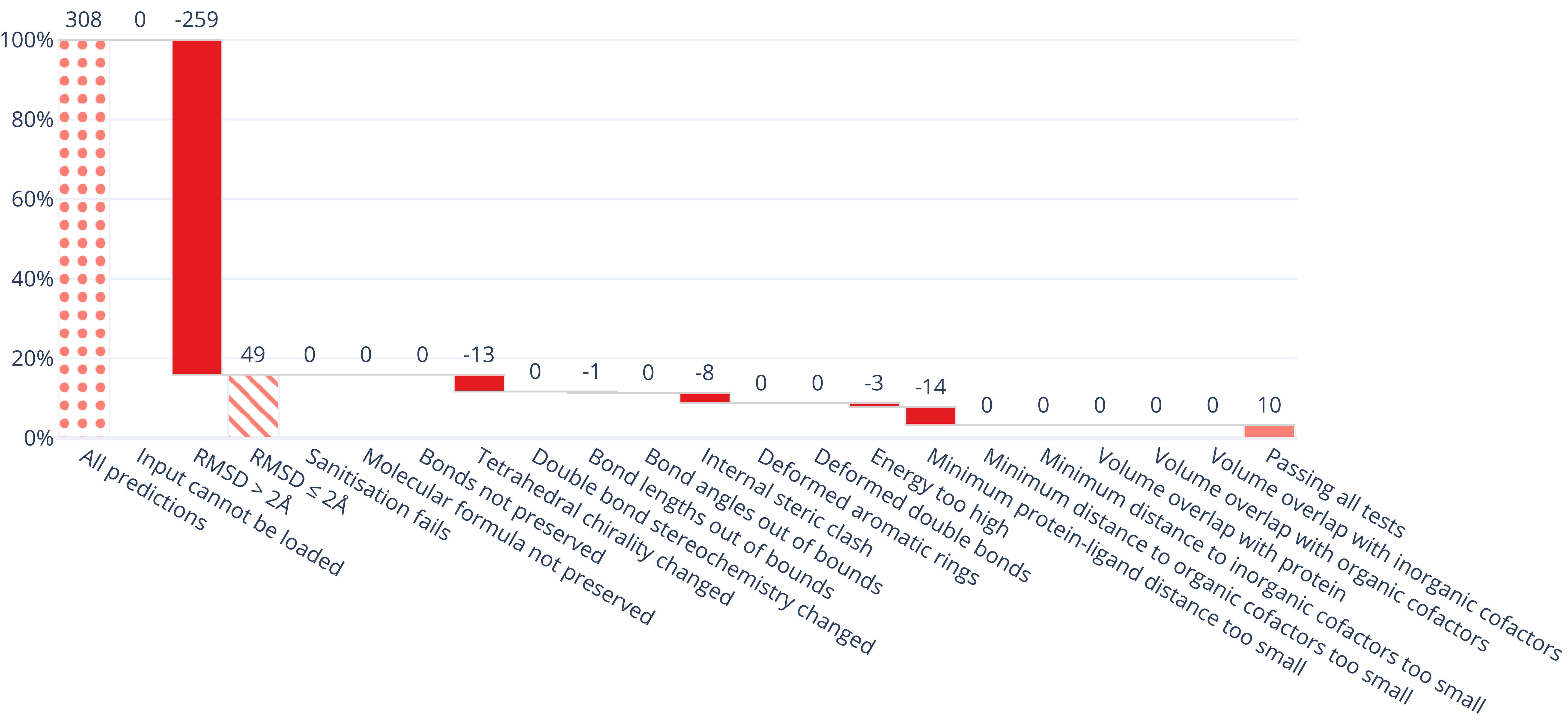}}
\hfill
\subfloat[Uni-Mol]{\includegraphics[width=0.49\textwidth]{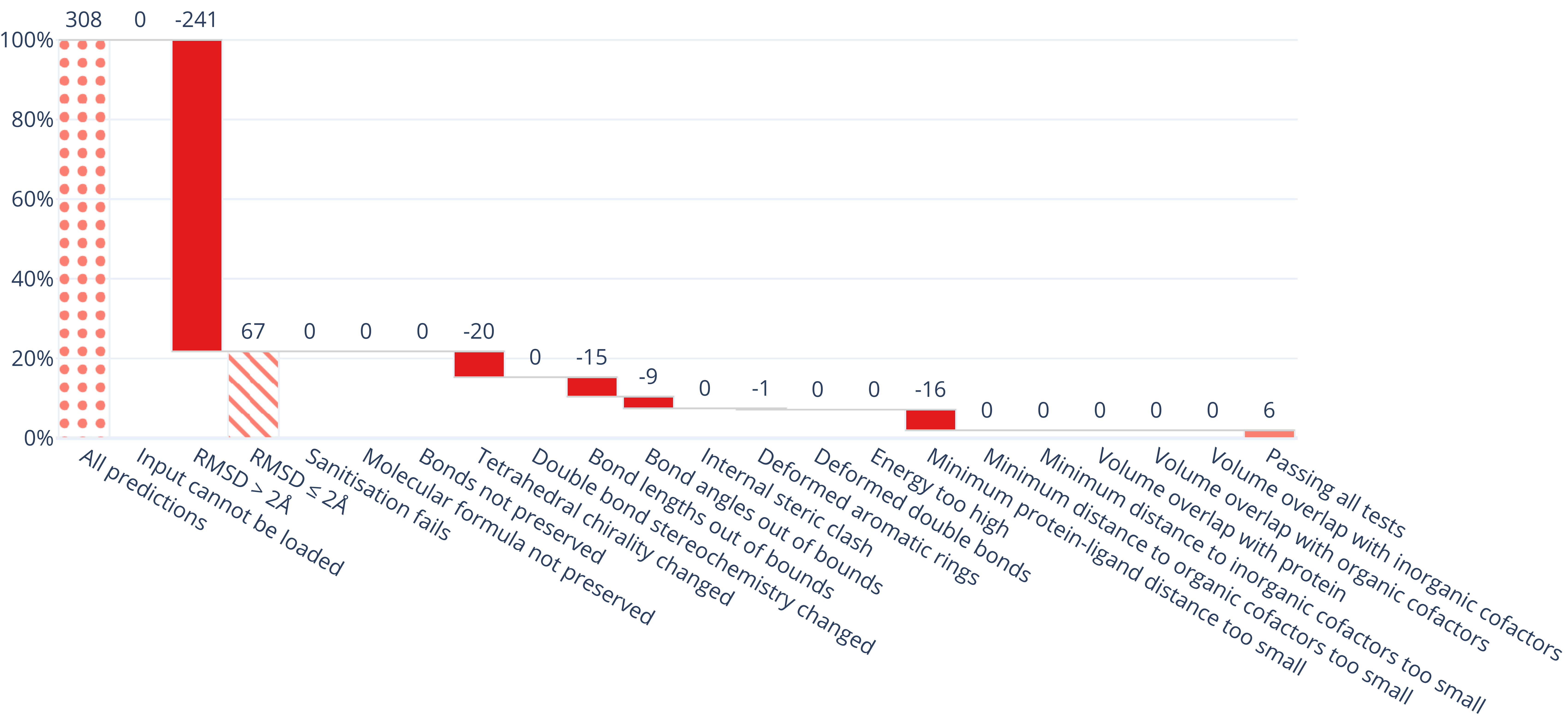}}
\caption{Waterfall plots showing test results for the PoseBuster\added{s Benchmark} \deleted{data} set. The leftmost (dotted) bar shows the number of complexes in the test set. The red bars show the number of predictions that fail with each additional test going from left to right. The right most (solid) bar indicates the number of predictions that pass all tests. Refer to the main article for a description of each test. As a reading example, panel (a) shows that out of AutoDock Vina's \replaced{308}{428} predictions, 200 are not within \qty{2}{\angstrom} RMSD, three clash with the protein and 1 clashes with an organic cofactor leaving 224 prediction with a low RMSD passing all tests. AutoDock Vina and CCDC Gold pass the most tests.}
\label{fig:waterfalls_posebusters}
\end{figure}

\begin{figure}[H]
\centering
\includegraphics[width=\textwidth]{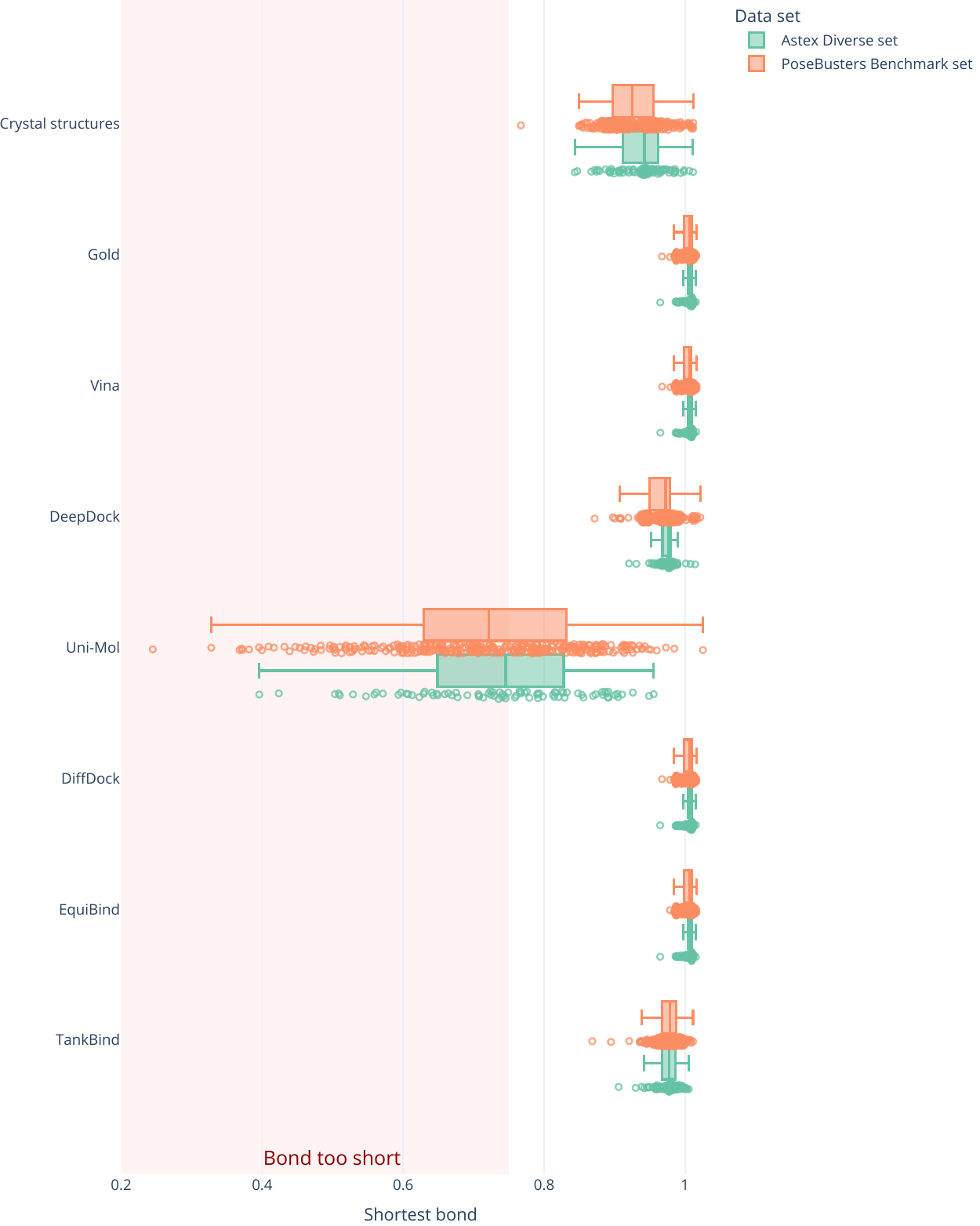}
\caption{
Distribution of shortest bond lengths. Shown are the relatively shortest bonds of each predicted ligand for each method and data set. The bond length is normalized by the lower bound for bond length obtained from Distance Geometry (DG). The lower bound correspond to one. A dot to the left of \qty{0.75} indicates that the relatively shortest bond was more than 25\% shorter than the DG lower bound. All methods except TankBind and Uni-Mol take the bond lengths from the provided ligand starting conformation. Uni-Mol and TankBind generate the bond lengths.
}
\label{fig:bond_length_short}
\end{figure}
\begin{figure}[H]
\centering
\includegraphics[width=\textwidth]{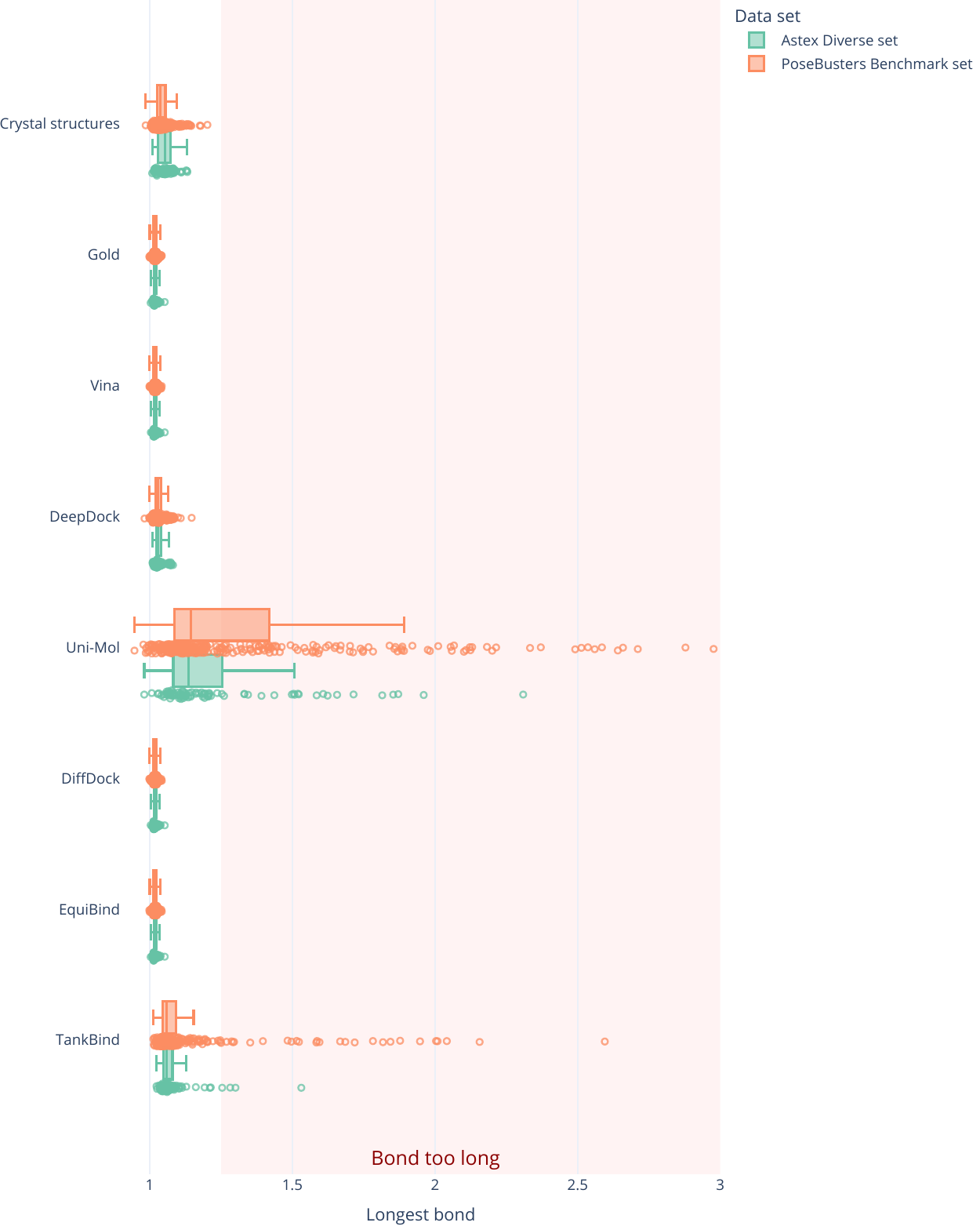}
\caption{
Distribution of longest bond lengths. Shown are the relatively longest bonds of each predicted ligand for each method and data set. The bond length is normalized by the upper bound for bond length obtained from Distance Geometry (DG). The upper bound correspond to one. A dot to the right of \qty{1.25} indicates that the relatively shortest bond was more than \qty{25}\% longer than the DG upper bound. All methods except TankBind and Uni-Mol take the bond lengths from the provided ligand starting conformation. Uni-Mol and TankBind generate the bond lengths.
}
\label{fig:bond_length_long}
\end{figure}
\begin{figure}[H]
\centering
\includegraphics[width=\textwidth]{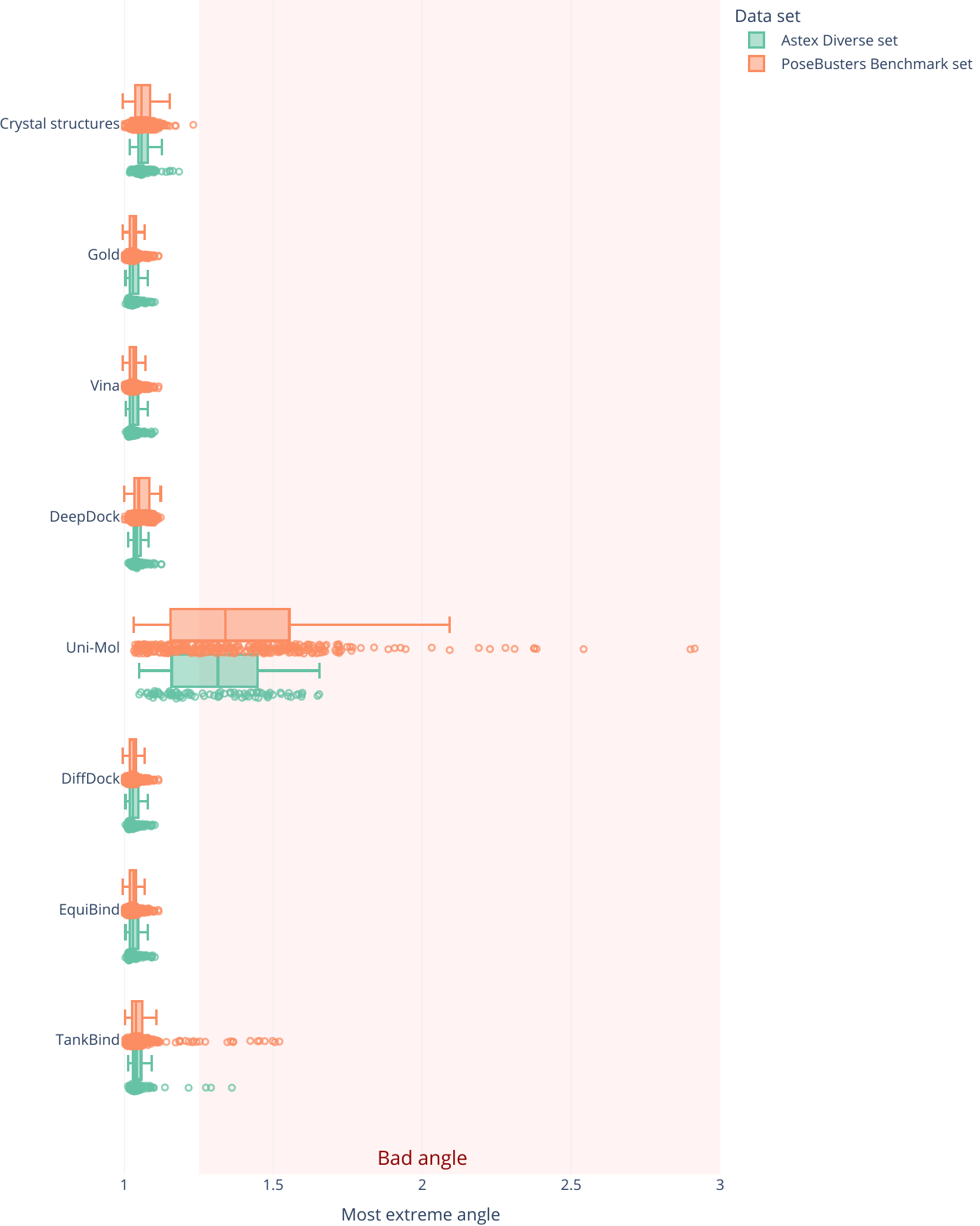}
\caption{Distribution of relative bond angles. Shown are the most extreme angles of each predicted ligand for each method and data set. Each bond angle is normalized by the corresponding bond length bounds obtained from Distance Geometry (DG). The upper bound corresponds to one. A dot to the right of \qty{1.25} indicates that an angle is more than \qty{25}\% larger or shorter than the DG bounds permit. All methods except TankBind and Uni-Mol take the bond angles from the provided ligand starting conformation. Uni-Mol and TankBind generate the angles.}
\label{fig:bond_angle}
\end{figure}

\begin{figure}[H]
\centering
\includegraphics[width=\textwidth]{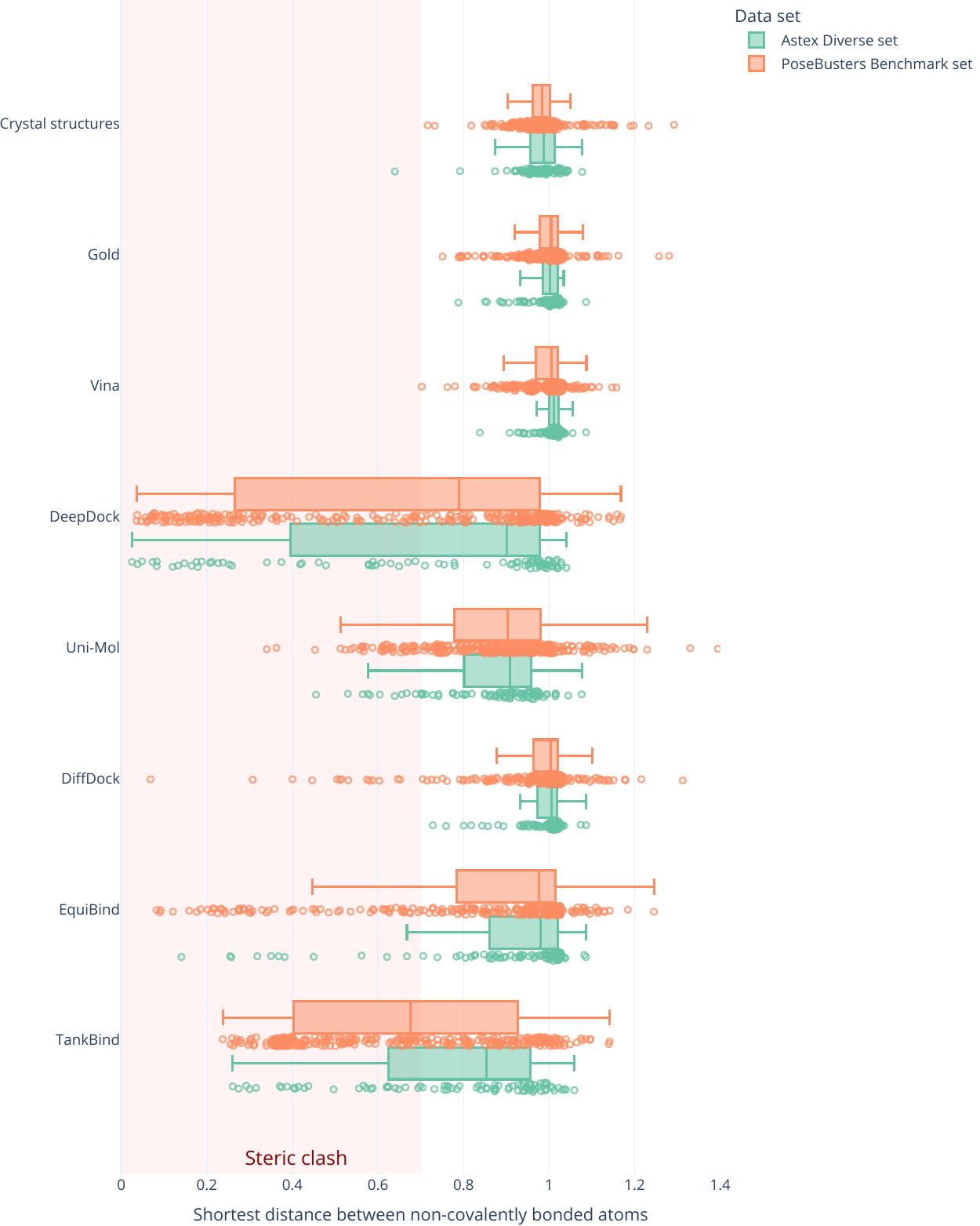}
\caption{Distribution of distances between unbounded atoms. The distribution shows the closest pair of unbounded atoms in each predicted ligand for each method and data set. Each distance is normalized to the Distance Geometry bounds. The lower bound corresponds to one. A dot to the left of 0.7 indicates that a distance was more than 30\% shorter than the lower bound and was counted as a clash.}
\label{fig:internal_clash}
\end{figure}

\begin{figure}[H]
\centering
\includegraphics[width=\textwidth]{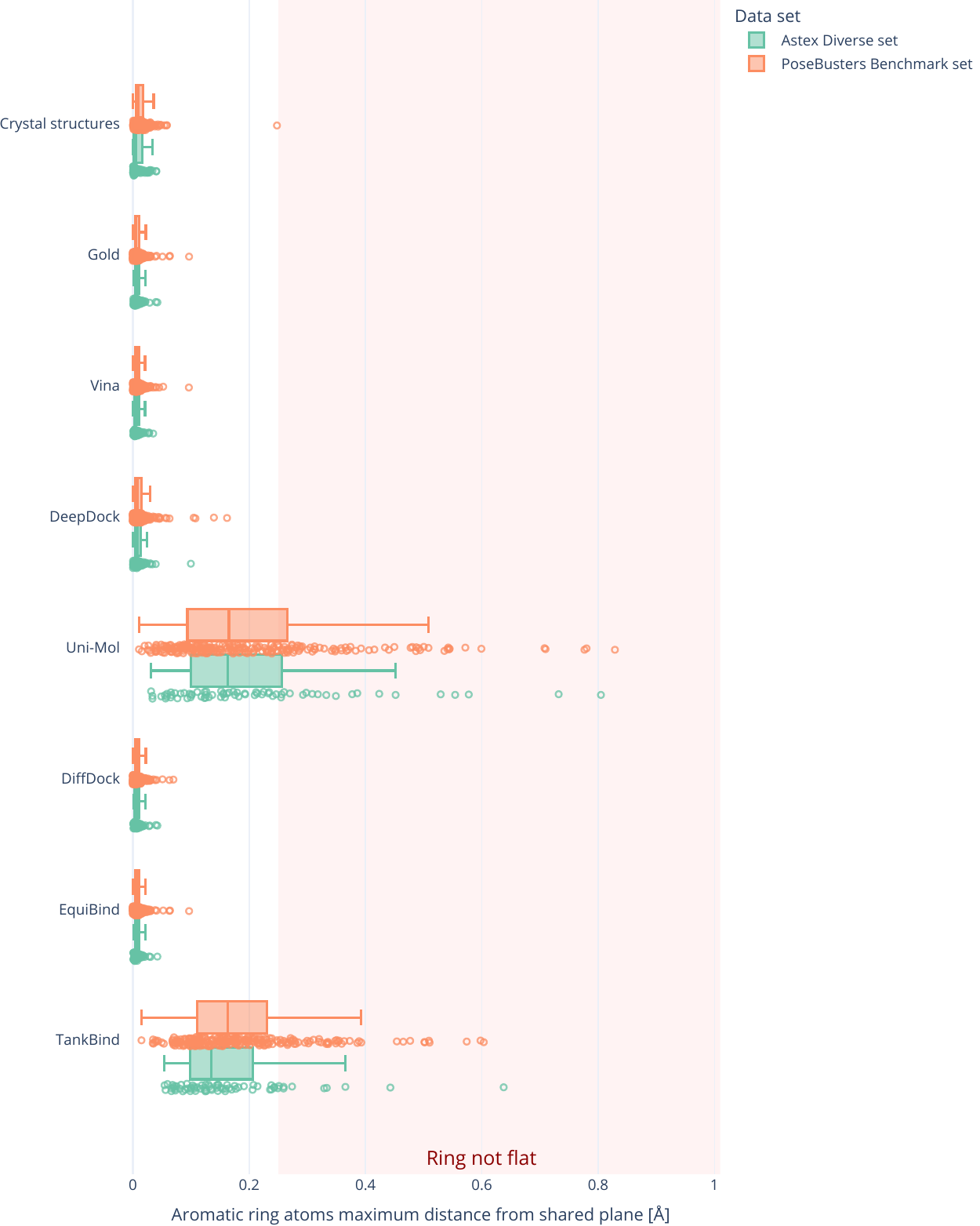}
\caption{
Distance from shared plane of atoms in 5- or 6-membered aromatic rings. The largest distance in Angstrom from the shared planed is shown for each protein ligand complex. If a ligand has no rings it is not shown. TankBind and Uni-Mol generate non-flat rings.
}
\label{fig:flat_rings}
\end{figure}
\begin{figure}[H]
\centering
\includegraphics[width=\textwidth]{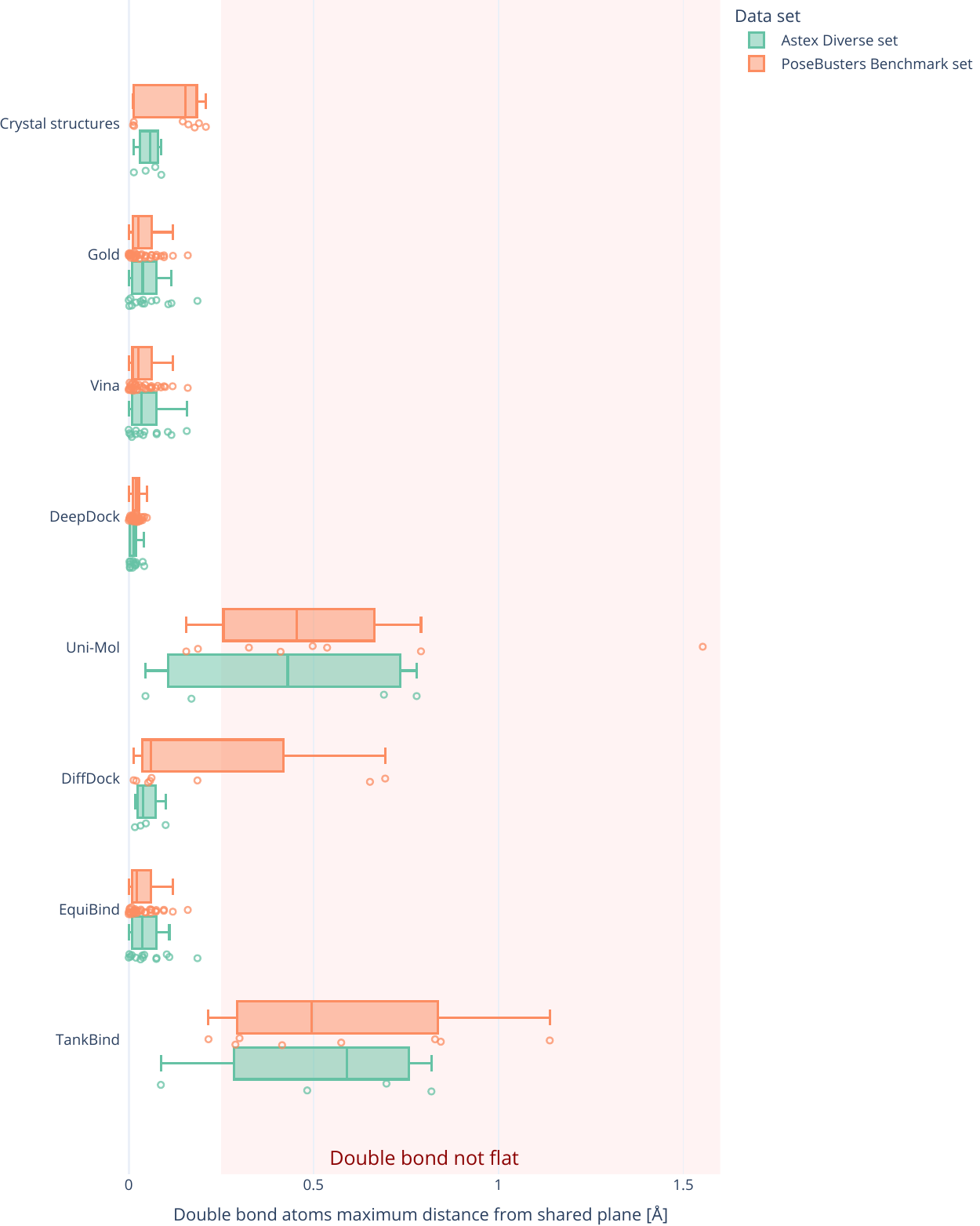}
\caption{
Distance from shared plane of atoms around aliphatic carbon-carbon double bonds. The largest distance in Angstrom from the shared planed of the two carbons and their four neighbours is shown for each protein ligand complex. If a ligand has no aliphatic carbon-carbon double bonds it is not shown. TankBind and Uni-Mol generate non-flat double bonds.
}
\label{fig:flat_bonds}
\end{figure}

\begin{figure}[H]
\centering
\includegraphics[width=\textwidth]{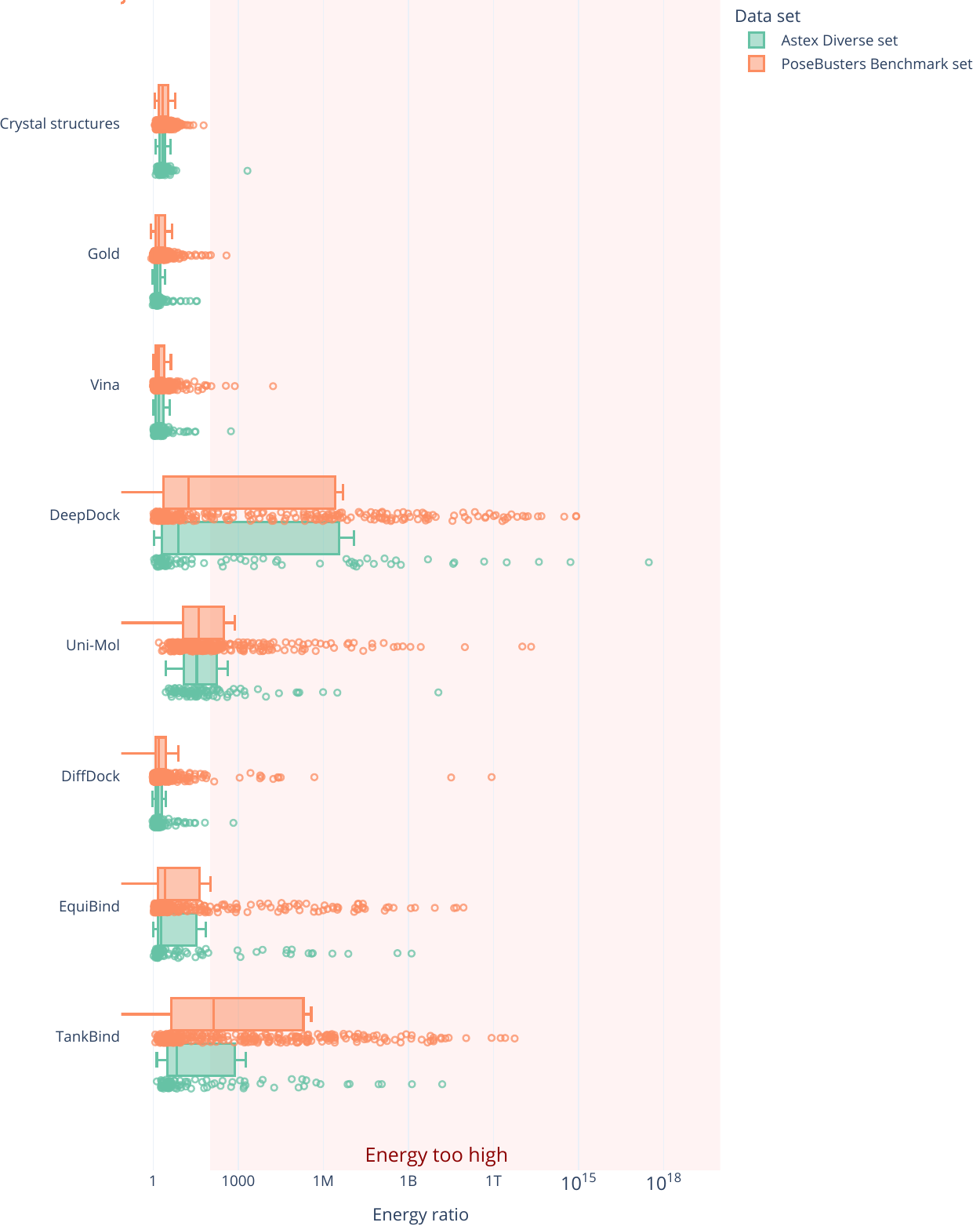}
\caption{
Energy ratio distributions. The ratio is the energy of the predicted ligand conformation over the average energy of an ensemble of 50 conformations generated with ETKDGv3. The UFF implemented in RDKit was used. The dashed red line shows the cutoff value of 100. There is only one crystal ligand in each data set with a higher energy ratio than the cutoff but all docking methods generate multiple high energy conformations above the cutoff.
}
\label{fig:energy_ratio}
\end{figure}

\begin{figure}[H]
\centering
\includegraphics[width=\textwidth]{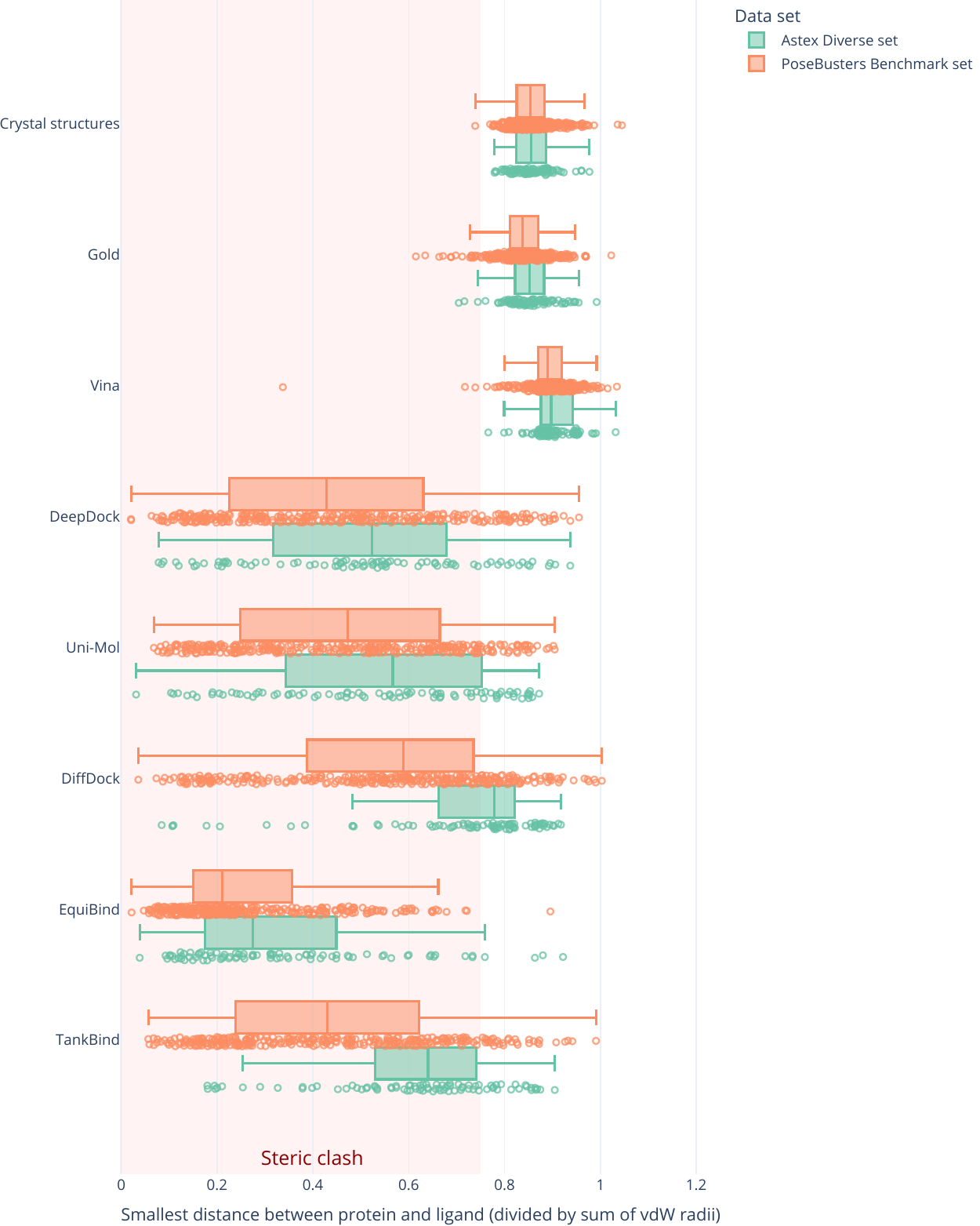}
\caption{
Minimum distances between protein and ligand. Distance is the smallest pairwise distance of heavy atoms of the ligand and protein normalized by their sum of van der Waals radii. The red area highlights the rejection zone below the cutoff of \qty{0.75}.
}
\label{fig:distance_protein}
\end{figure}

\begin{figure}[H]
\centering
\includegraphics[width=\textwidth]{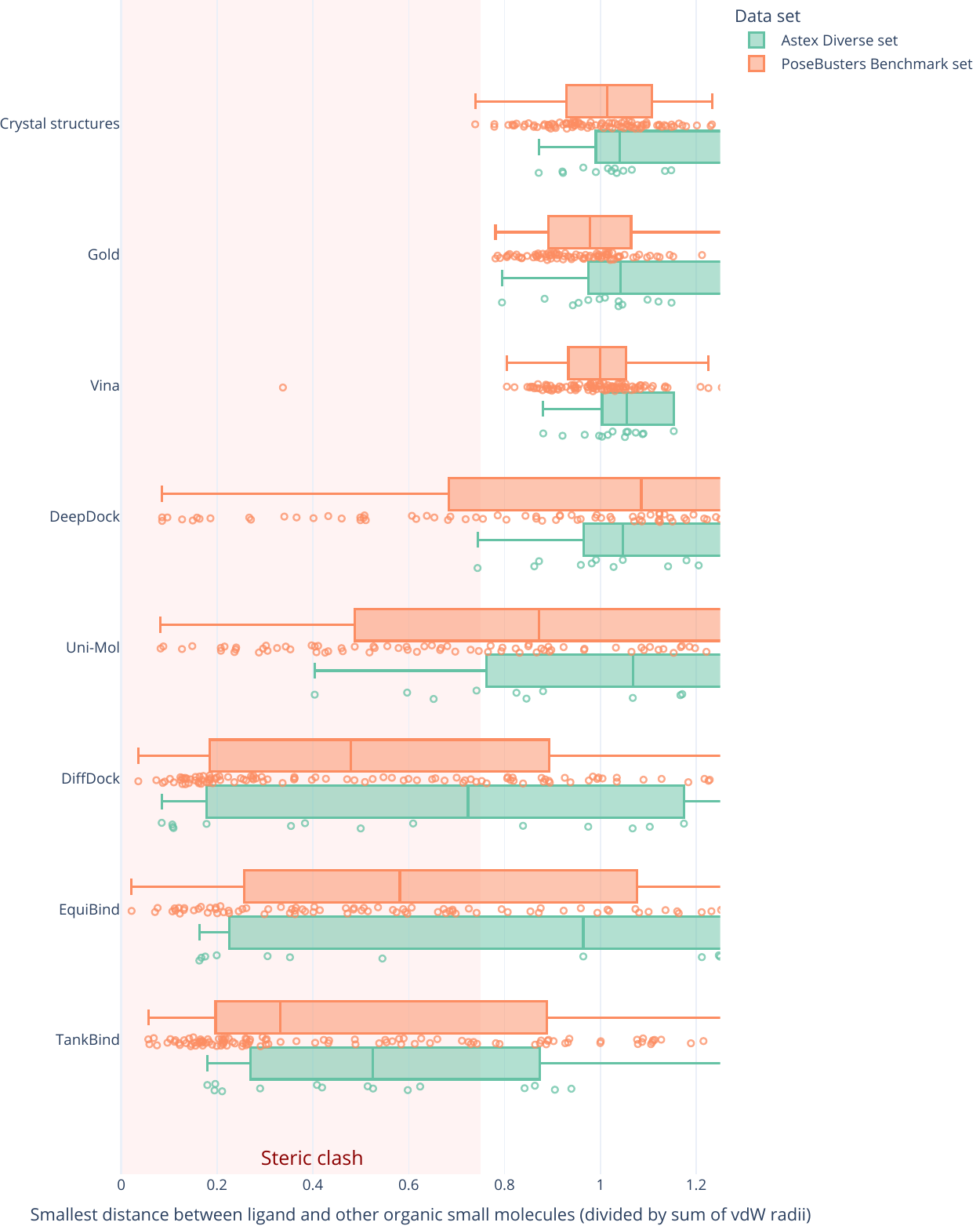}
\caption{
Minimum distances between ligand and organic molecules. Distance is the smallest pairwise distance of heavy atoms of the ligand and organic molecules normalized by their sum of van der Waals radii. The red area highlights the rejection zone below the cutoff of \qty{0.75}.
}
\label{fig:distance_organic}
\end{figure}

\begin{figure}[H]
\centering
\includegraphics[width=\textwidth]{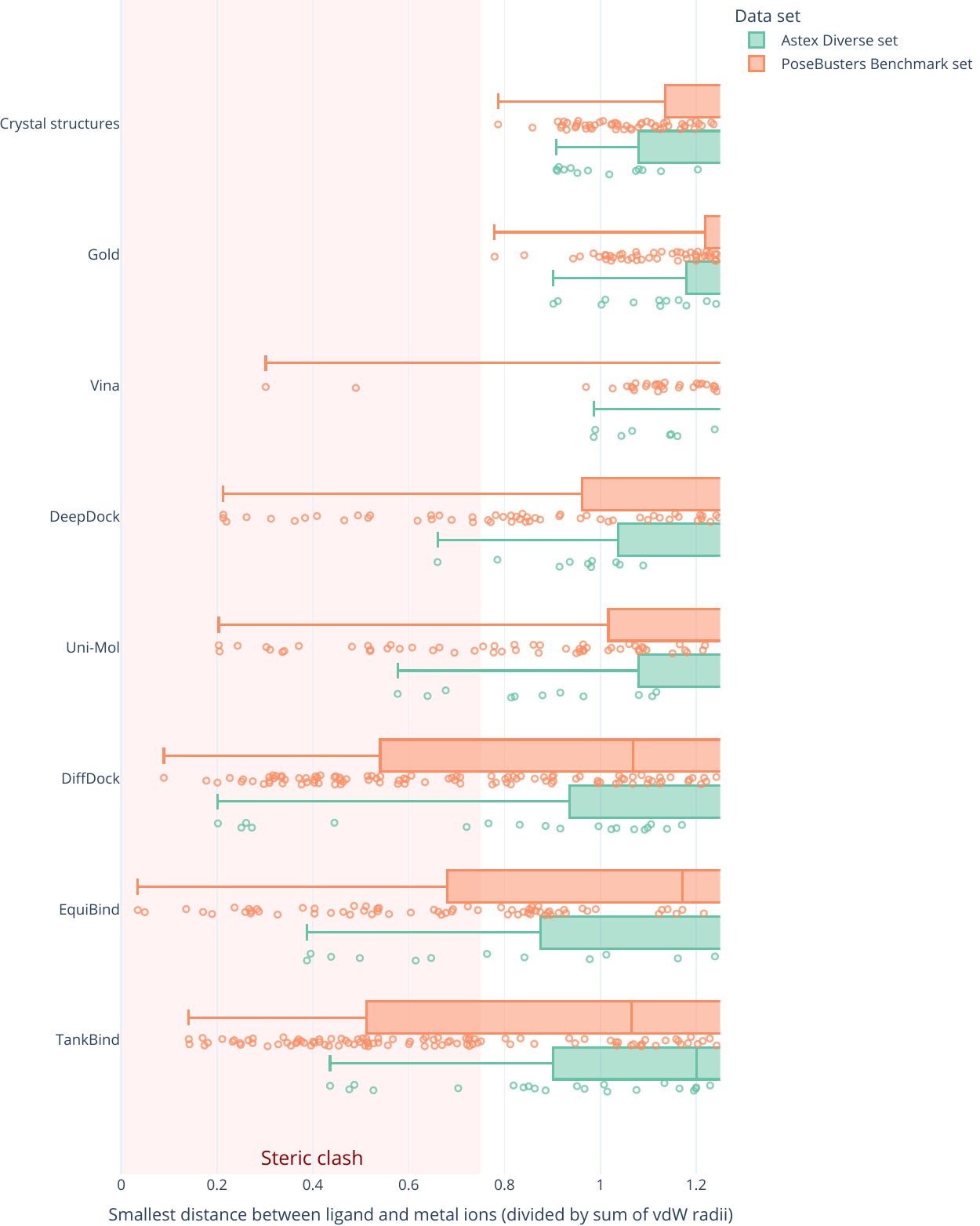}
\caption{
Minimum distances between ligand and inorganic cofactors. Distance is the smallest pairwise distance of heavy atoms of the ligand and inorganic cofactors normalized by their sum of covalent radii. The red area highlights the rejection zone below the cutoff of \qty{0.75}. 
}
\label{fig:distance_inorganic}
\end{figure}

\begin{figure}[H]
\centering
\includegraphics[width=\textwidth]{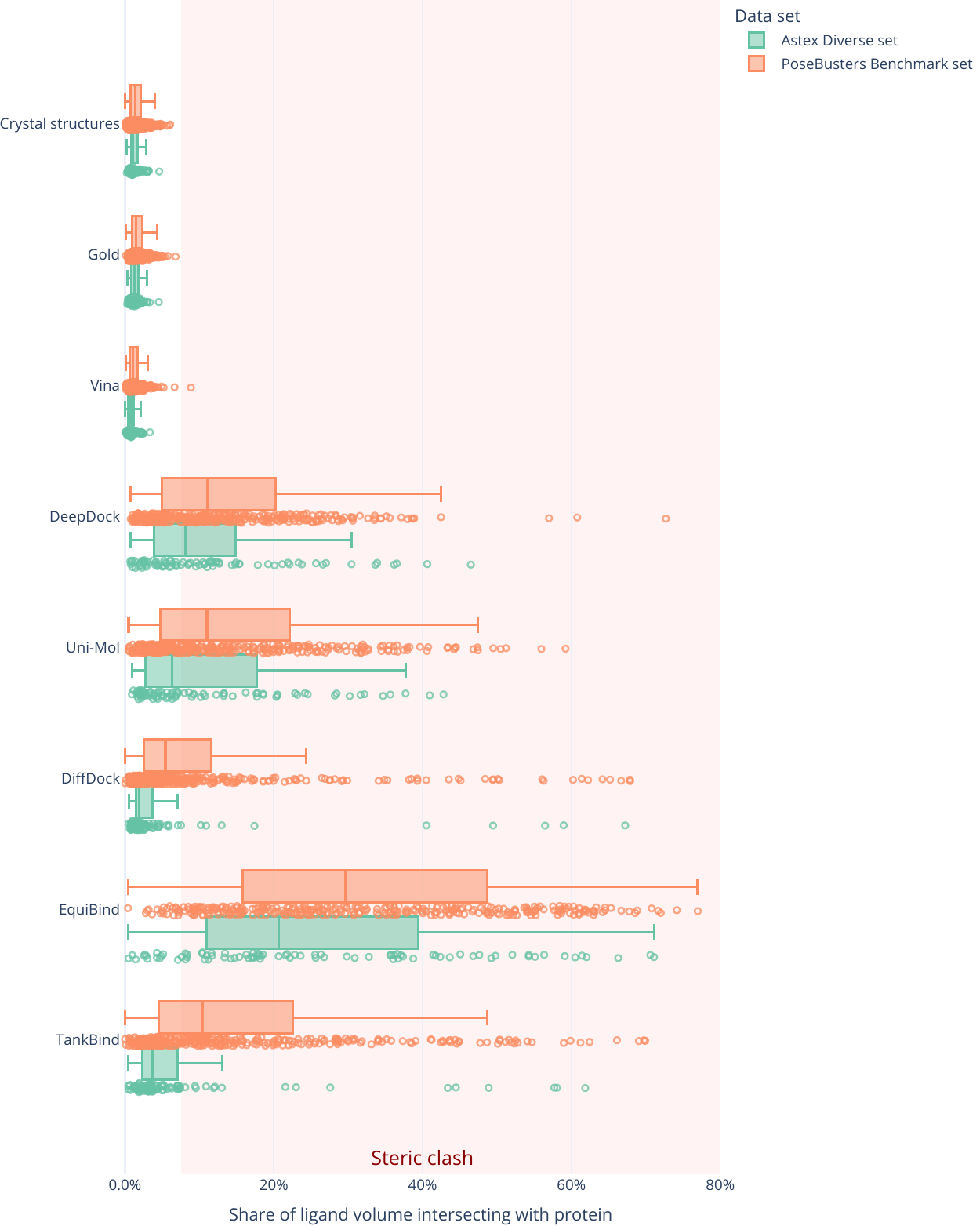}
\caption{
Volume overlap of protein and ligand. Volume overlap is the percentage of the ligand that overlaps with the protein. The volumes are the van der Waals volumes of the heavy atoms of the ligand and protein. The van der Waals radii are scaled by 0.8. The red dashed line shows the used cutoff of 5\%. 
}
\label{fig:volume_overlap_protein}
\end{figure}
\begin{figure}[H]
\centering
\includegraphics[width=\textwidth]{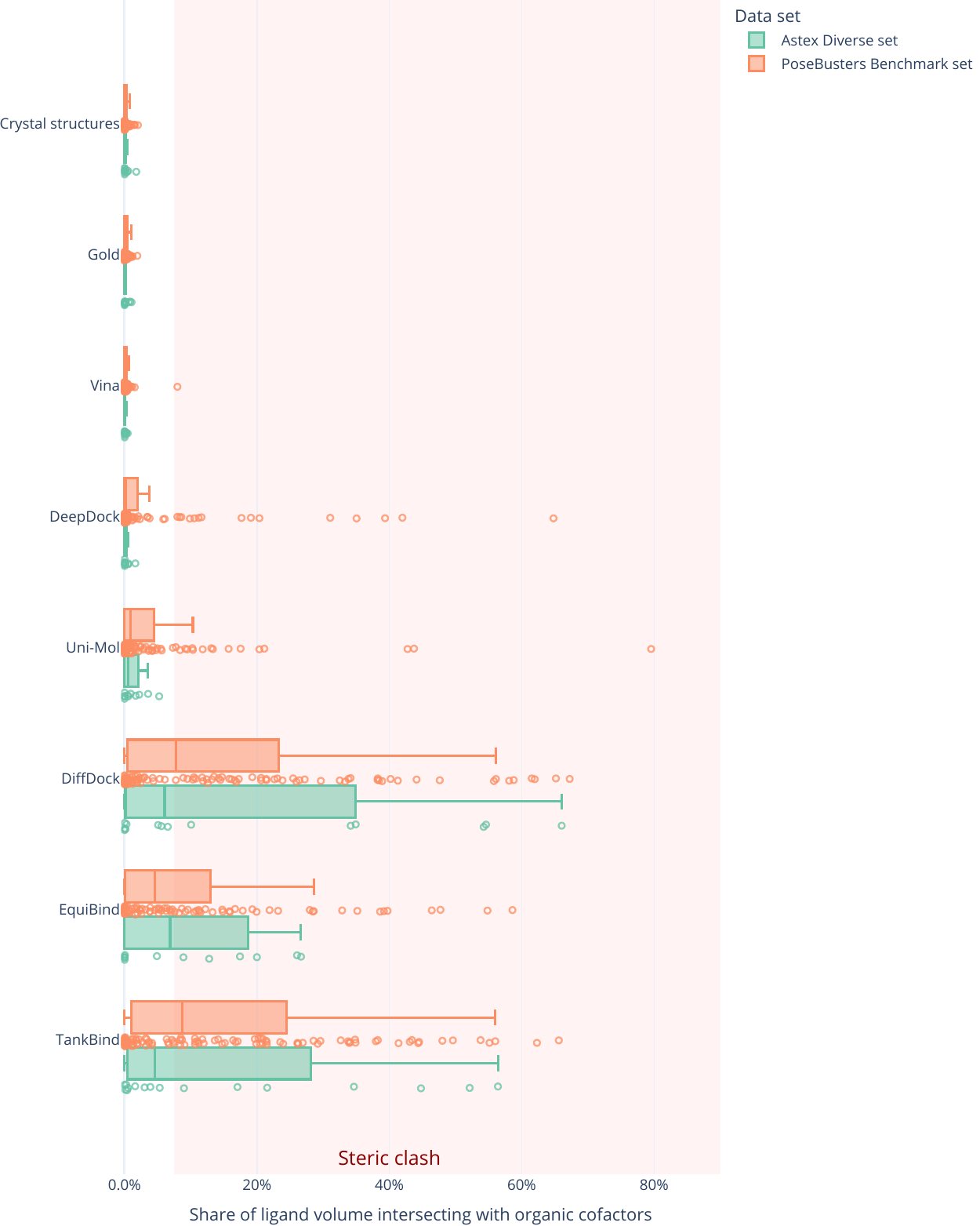}
\caption{
Volume overlap of ligand and organic molecules. Volume overlap is the percentage of the ligand that overlaps with the organic molecules. The volumes are the van der Waals volumes of the heavy atoms of the ligand and organic molecules. The van der Waals radii are scaled by 0.8. The red dashed line shows the used cutoff of 5\%. 
}
\label{fig:volume_overlap_organic}
\end{figure}
\begin{figure}[H]
\centering
\includegraphics[width=\textwidth]{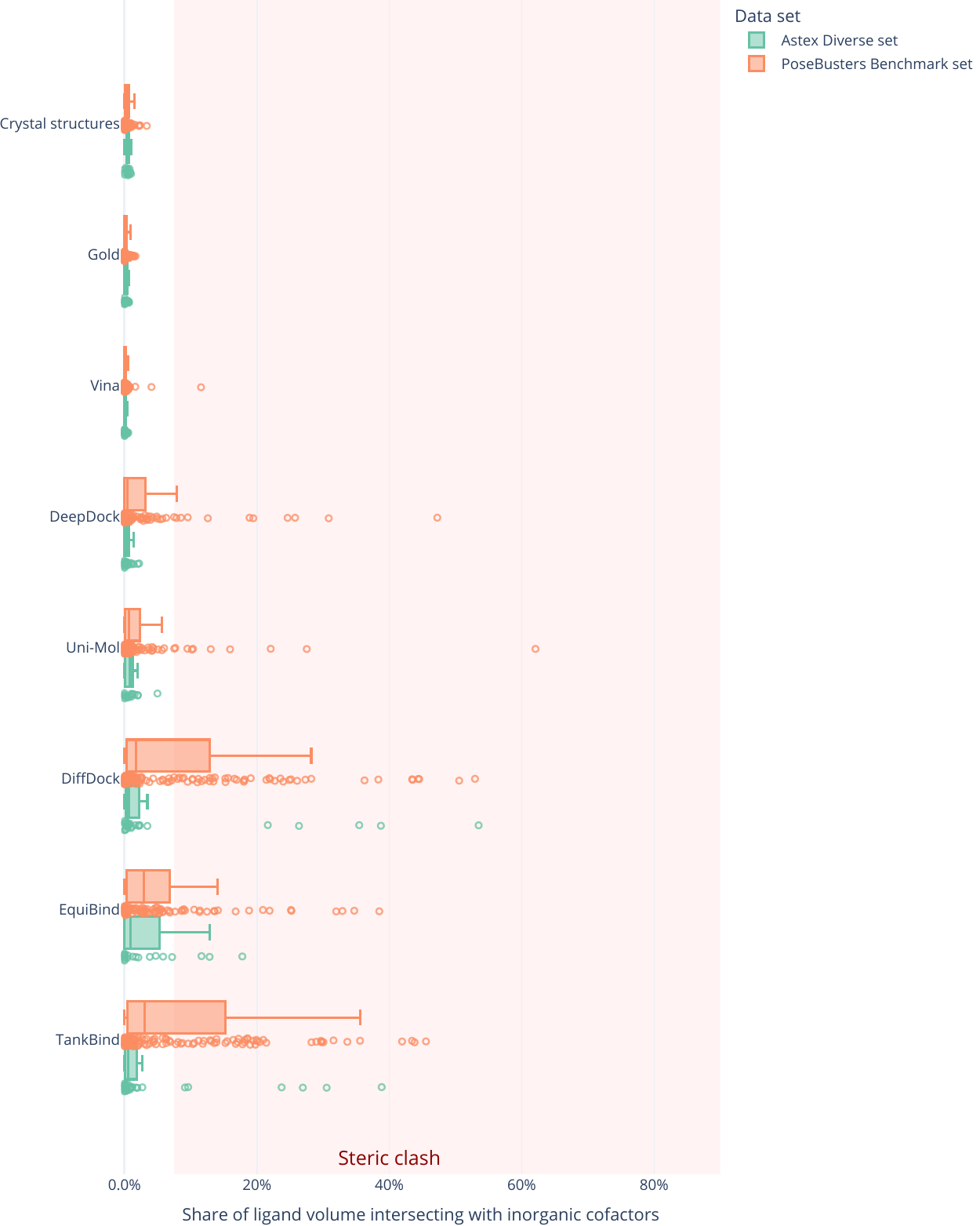}
\caption{
Volume overlap of ligand and metal ions. Volume overlap is the percentage of the ligand that overlaps with the metal ions. The volumes are the van der Waals volumes of the heavy atoms of the ligand and metal ions. The van der Waals radii are scaled by 0.8. The red zone shows the used cutoff of 20\%. 
}
\label{fig:volume_overlap_inorganic}
\end{figure}

\clearpage
\section{Alternative binding site definitions for Uni-Mol}
\comment{entire section added}

This section explores a range of binding site definitions starting from Uni-Mol's preferred definition of all residues with an atom within \qty{6}{\angstrom} of a heavy atom of the crystal ligand. 
The Uni-Mol results shown here were generated using the docking procedure described in Section~\ref{sec:docking_protocols} but with different thresholds for selecting residues around the crystal ligand.
The additional CCDC Gold results were generated using the procedure described in Section~\ref{sec:docking_protocols} but with a binding site selection using \texttt{BindingSiteFromLigand} with different thresholds and always selecting the entire residues.

\begin{figure}[H]
\centering
\includegraphics[width=0.85\columnwidth]{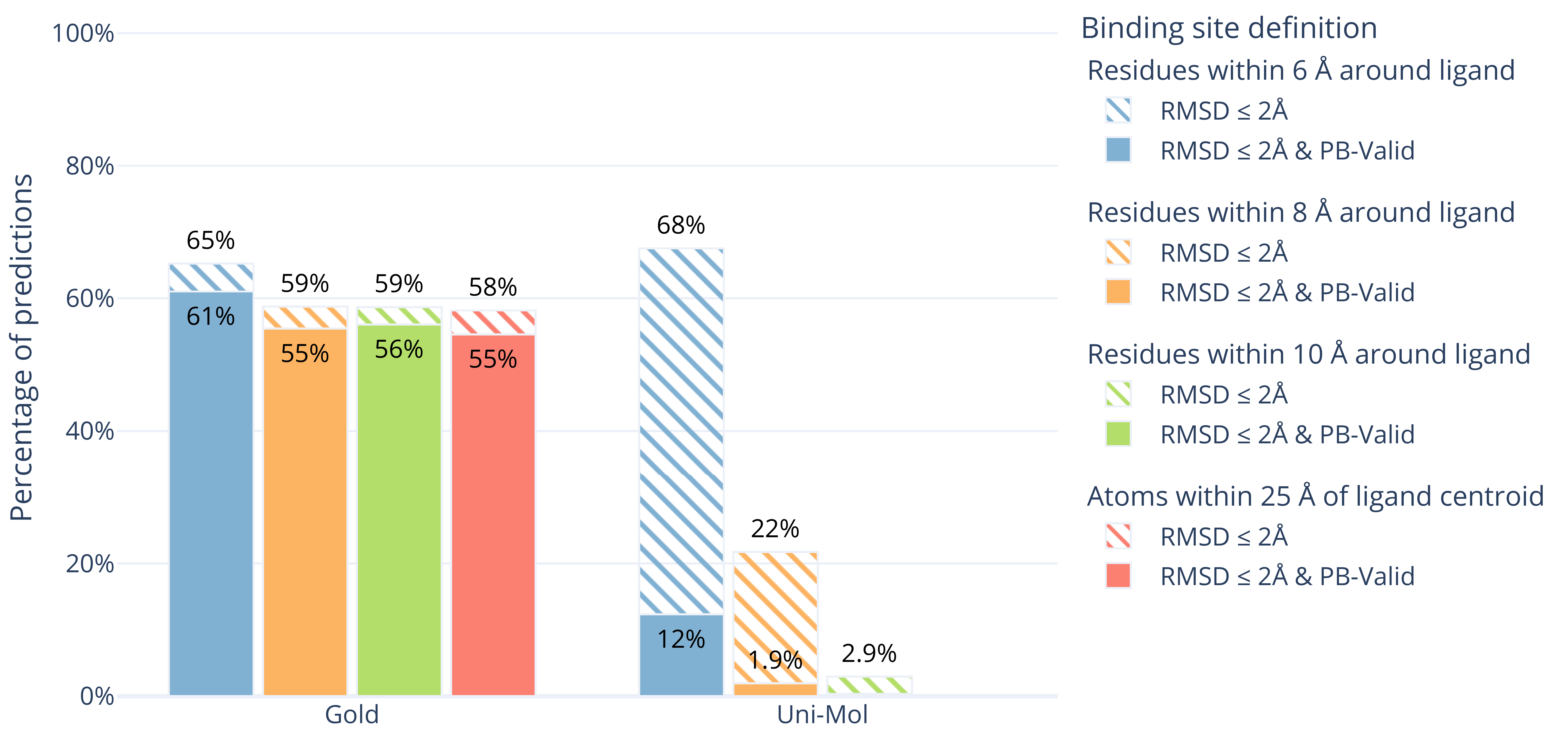}
\caption{
Performance of CCDC Gold and Uni-Mol on the PoseBusters Benchmark set on a range of binding site definitions starting from Uni-Mol's preferred definition of all residues with an atom within \qty{6}{\angstrom} of a heavy atom of the crystal ligand. 
The striped bars show the share of predictions of each method that have an RMSD within \qty{2}{\angstrom} and the solid bars show the subset that in addition have valid geometries and energies, i.e., pass all PoseBuster tests and are therefore `PB-Valid'.
Under the tight pocket definition of \qty{6}{\angstrom} Uni-Mol performs better than any of the blind docking methods.
In the paper (Figure~1), we show the results for Gold shown here in red defining the search space as a sphere with a radius of \qty{25}{\angstrom} centred on the crystal ligand's centroid and for Uni-Mol the results in the paper are the results shown here in orange defining the binding site as all residues with an atom within \qty{8}{\angstrom} of a crystal ligand heavy atom. We selected these search spaces to make both methods more comparable to the blind docking methods.
}
\label{fig:bars_small_search_space_datasets}
\end{figure}
\begin{figure}[H]
\centering
\includegraphics[width=0.85\columnwidth]{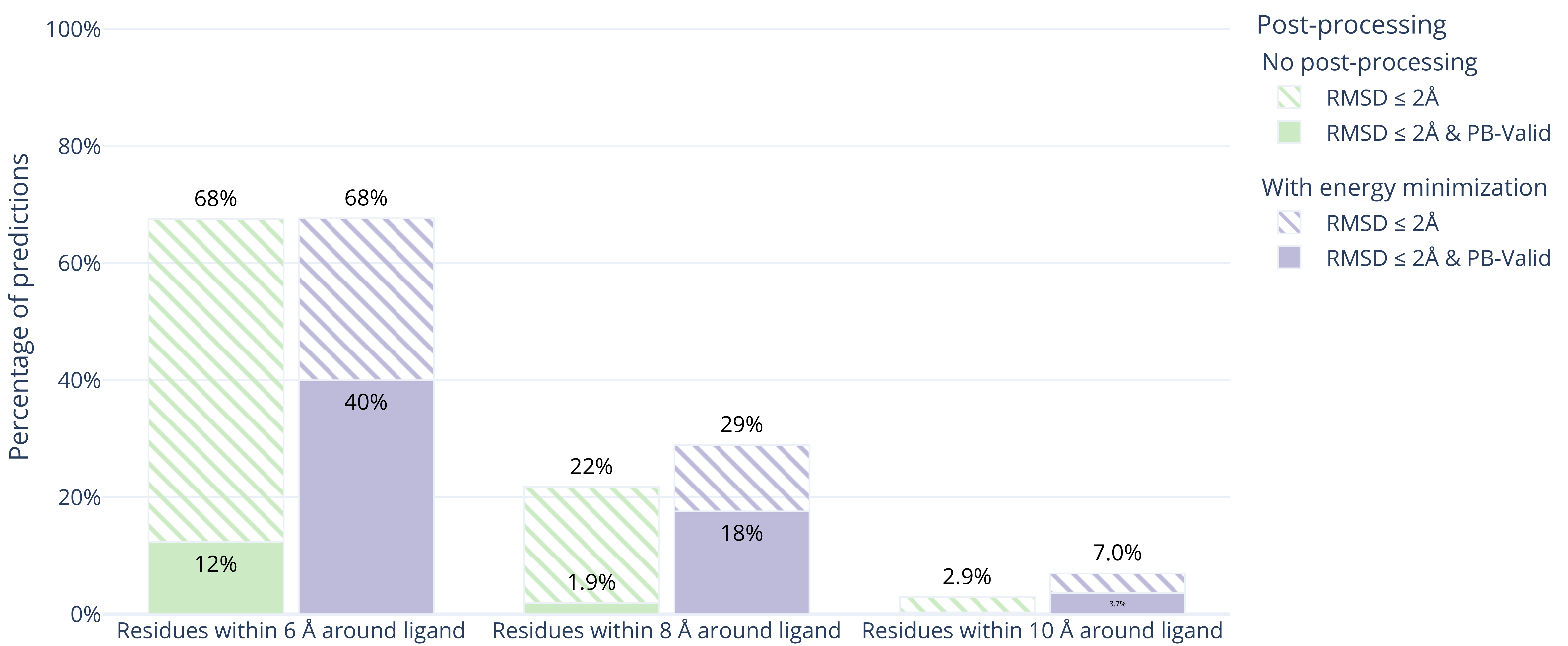}
\caption{
Performance of Uni-Mol on the PoseBusters Benchmark set for a range of binding site definitions with and without energy minimization as an additional post-processing step. 
The striped bars show the share of predictions of each method that have an RMSD within \qty{2}{\angstrom} of the crystal pose and the solid bars show those predictions which in addition pass all PoseBuster tests and are therefore PB-valid.
Under the tight binding site definition the number of poses after energy minimisation that pass the tests increases to about the same level as DiffDock (\qty{35}{\percent}, Figure~5).
}
\label{fig:pb_unimol_em}
\end{figure}

\clearpage
\section*{}
\bibliography{references}
\bibliographystyle{rsc}